\documentclass[
	floatfix,
  reprint,
  aip,
  jcp
	]{revtex4-1}
\let\Twocolumn

\newif\ifTwocolumn
\ifx\Twocolumn\undefined
  \Twocolumnfalse
\else
  \Twocolumntrue
\fi
\ifTwocolumn
\fi
\usepackage{graphicx}
\usepackage{amsmath,amssymb,amsfonts}
\usepackage{gensymb}
\usepackage{bm}
\usepackage{upgreek}
\usepackage{wasysym}
\usepackage{bibentry}
\usepackage[T1]{fontenc}
\usepackage{textcomp}
\usepackage{mathptmx}
\usepackage[scaled=0.92]{helvet}
\usepackage{courier}
\newcommand{\eref}[1]{Eq.~\eqref{#1}}
\newcommand{\fref}[1]{Fig.~\ref{#1}}

\newcommand{\upd}{{\ensuremath{\textrm{d}}}}
\renewcommand*{\vec}[1]{\mathbf{#1}}
\newcommand{\MBC}{\ensuremath{(-)}}
\newcommand{\PBC}{\ensuremath{(+)}}
\newcommand{\Mm}{\ensuremath{(-,-)}}
\newcommand{\Pp}{\ensuremath{(+,+)}}
\newcommand{\Pm}{\ensuremath{(+,-)}}
\newcommand{\Ppm}{\ensuremath{(+,\pm)}}

\newcommand{\Pmm}{\ensuremath{(\pm,-)}}
\newcommand{\Mpm}{\ensuremath{(\mp,-)}}
\newcommand{\Ab}{\ensuremath{(a,b)}}
\newcommand{\Asb}{\ensuremath{ {(a_<,b)}}}
\newcommand{\Alb}{\ensuremath{ {(a_>,b)}}}
\newcommand{\step}{\ensuremath{\textrm{s}}}
\newcommand{\stripe}{\ensuremath{\ell}}
\newcommand{\period}{\ensuremath{\textrm{p}}}
\newcommand{\cyl}{\ensuremath{{ \textrm{cyl}}}}

\newcommand{\fppm}{\ensuremath{f_{\Ppm}}}

\newcommand{\fmpm}{\ensuremath{f_{\Mpm}}}
\newcommand{\fpp}{\ensuremath{f_{\Pp}}}
\newcommand{\fpm}{\ensuremath{f_{\Pm}}}

\newcommand{\Fab}{\ensuremath{F_{\Ab}}}

\newcommand{\Fstep}{\ensuremath{F_{\step}}}
\newcommand{\Fstripe}{\ensuremath{F_{\stripe}}}
\newcommand{\Fperiod}{\ensuremath{F_{\period}}}

\newcommand{\kpm}{\ensuremath{k_{\Pm}}}
\newcommand{\kmm}{\ensuremath{k_{\Mm}}}

\newcommand{\kpmm}{\ensuremath{k_{\Pmm}}}

\newcommand{\kab}{\ensuremath{k_{\Ab}}}

\newcommand{\Kmm}{\ensuremath{K_{\Mm}}}

\newcommand{\Kpmm}{\ensuremath{K_{\Pmm}}}
\newcommand{\Kmpm}{\ensuremath{K_{\Mpm}}}
\newcommand{\Kab}{\ensuremath{K_{\Ab}}}
\newcommand{\Kasb}{\ensuremath{K_{\Asb}}}
\newcommand{\Kalb}{\ensuremath{K_{\Alb}}}
\newcommand{\Kstep}{\ensuremath{K_{\step}}}
\newcommand{\Kstripe}{\ensuremath{K_{\stripe}}}
\newcommand{\Kperiod}{\ensuremath{K_{\period}}}

\newcommand{\Phiab}{\ensuremath{\Phi_{\Ab}}}
\newcommand{\Phistep}{\ensuremath{\Phi_{\step}}}
\newcommand{\Phistripe}{\ensuremath{\Phi_{\stripe}}}
\newcommand{\Phiperiod}{\ensuremath{\Phi_{\period}}}

\newcommand{\psistripe}{\ensuremath{\psi_{\stripe}}}
\newcommand{\psiperiod}{\ensuremath{\psi_{\period}}}

\newcommand{\omegastripe}{\ensuremath{\omega_{\stripe}}}
\newcommand{\omegaperiod}{\ensuremath{\omega_{\period}}}

\newcommand{\varthetapm}{\ensuremath{\vartheta_{\Pm}}}
\newcommand{\varthetamm}{\ensuremath{\vartheta_{\Mm}}}

\newcommand{\varthetapmm}{\ensuremath{\vartheta_{\Pmm}}}
\newcommand{\varthetampm}{\ensuremath{\vartheta_{\Mpm}}}
\newcommand{\varthetaab}{\ensuremath{\vartheta_{\Ab}}}
\newcommand{\varthetastep}{\ensuremath{\vartheta_{\step}}}
\newcommand{\varthetastripe}{\ensuremath{\vartheta_{\stripe}}}
\newcommand{\varthetaperiod}{\ensuremath{\vartheta_{\period}}}
\newcommand{\Dpp}{\ensuremath{\Delta_{\Pp}}}
\newcommand{\Dmm}{\ensuremath{\Delta_{\Mm}}}
\newcommand{\Dpm}{\ensuremath{\Delta_{\Pm}}}

\newcommand{\Dab}{\ensuremath{\Delta_{\Ab}}}

\newcommand{\rdef}{\ensuremath =\mathrel{\mathop:}}
\newcommand{\ddef}{\ensuremath \mathrel{\mathop:}=}

\DeclareMathOperator\sgn{sign}
\DeclareMathOperator\erfc{erfc}
\DeclareMathOperator\erf{erf}
\ifTwocolumn
  \newcommand{\pprule}{\ensuremath{\mbox{\rule[.7mm]{27pt}{0.5pt}}}}
\else
  \newcommand{\pprule}{\ensuremath{\mbox{\rule[1mm]{32pt}{0.5pt}}}}
\fi
\begin{document}
\title{Critical Casimir effect for colloids close to chemically patterned substrates}
\author{M.~Tr{\"o}ndle}
\affiliation{
	Max-Planck-Institut f\"ur Metallforschung,  
	Heisenbergstr.\ 3, D-70569 Stuttgart, Germany}
\affiliation{
	Institut f\"ur Theoretische und Angewandte Physik, 
	Universit\"at Stuttgart, 
	Pfaffenwaldring 57, 
	D-70569 Stuttgart, Germany
}
\author{S.~Kondrat}
\altaffiliation{Present address: Department of Chemistry, Imperial College London, South Kensington
Campus, SW7 2AZ London, U.K.}
\affiliation{
	Max-Planck-Institut f\"ur Metallforschung,  
	Heisenbergstr.\ 3, D-70569 Stuttgart, Germany}
\affiliation{
	Institut f\"ur Theoretische und Angewandte Physik, 
	Universit\"at Stuttgart, 
	Pfaffenwaldring 57, 
	D-70569 Stuttgart, Germany
}
\author{A.~Gambassi}
\affiliation{
  SISSA --- International School for Advanced Studies and INFN, via Bonomea 265, 34136 Trieste, Italy 
  }
\author{L.~Harnau}
\author{S.~Dietrich}
\affiliation{
	Max-Planck-Institut f\"ur Metallforschung,  
	Heisenbergstr.\ 3, D-70569 Stuttgart, Germany}
\affiliation{
	Institut f\"ur Theoretische und Angewandte Physik, 
	Universit\"at Stuttgart, 
	Pfaffenwaldring 57, 
	D-70569 Stuttgart, Germany
}
\date{May, 6 2010}
%
\begin{abstract}
  Colloids immersed in a critical or near-critical binary liquid mixture and close to a chemically
  patterned substrate are subject to normal and lateral critical Casimir 
  forces of dominating strength.
  For a single colloid we calculate these attractive or repulsive forces and the corresponding critical Casimir
  potentials within mean-field theory.
  Within this approach we also discuss  the quality of the Derjaguin approximation and apply
  it to Monte Carlo simulation data available for the system under study.
  We find that the range of validity of the Derjaguin approximation is rather large and that it fails
  only for surface structures which are very small compared to the geometric mean of the size of the colloid and its distance
  from the substrate.
  For certain chemical structures of the substrate the critical Casimir force acting on the colloid can  
  change sign as a function of the distance between the particle and the substrate;  
  this provides a mechanism for stable levitation at a certain distance which can be strongly tuned by temperature, i.e., 
  with a sensitivity of more than $200 \textrm{nm}/\textrm{K}$.
\end{abstract}
\maketitle
\section{Introduction}
%
Since the discovery of the Casimir effect in quantum electrodynamics 
\cite{casimir:1948,kardar:1999} it is well-known that the inherent fluctuations of a medium 
lead to an effective force acting on its confining boundaries.
In soft matter physics, the analogue of the vacuum fluctuations in quantum electrodynamics 
are the thermal fluctuations of the order parameter $\phi$ of a fluid.
These occur on the length scale of the bulk correlation length $\xi$ which is generically of
molecular size.
However, upon approaching a critical point 
at the temperature ${T=T_c}$, the correlation length $\xi$ increases with an algebraic singularity
and attains \emph{macroscopic} values.
The confinement of these long-ranged fluctuations results in the so-called critical Casimir force
acting on a length scale set by $\xi$ \cite{fisher:1978}.
Since the correlation length diverges as ${\xi(T\to T_c)\propto|T-T_c|^{-\nu}}$, where
$\nu$ is a standard bulk critical exponent, the range of the critical Casimir force (and therefore its
strength at a certain distance) can be controlled and tuned by minute temperature changes
(see, e.g., Refs.~\onlinecite{gambassi:2009conf,gambassi:2009news}).
The characteristic energy scale of the critical Casimir effect is given
by $k_B T_c$, which allows for a direct measurement of the critical Casimir forces,
in particular if the critical point is located at ambient thermodynamic conditions 
\cite{hertlein:2008,gambassi:2009}.
\par
The attractive or repulsive character of the critical Casimir force can be controlled
by suitable treatments of the confining surfaces.
Generically, the surfaces which confine a binary liquid mixture preferentially adsorb one 
of its two components (or the gas or liquid phase in the case of a one-component
fluid).
This can be described by effective, symmetry breaking surface fields, which lead to a
preference for either positive $[(+)]$ or negative $[(-)]$ values of the scalar order parameter 
$\phi$,  corresponding to the difference between the local concentrations of the two species
(or the deviation of the density of the one-component fluid from its critical value).
%
\nocite{krech:book}
\nocite{brankov:book}
\nocite{krech:9192all}
\nocite{evans:1994}
\nocite{diehl:2006}
\nocite{zandi:2007}
\nocite{schmidt:2008}
\nocite{mohry:2009}
The critical Casimir force strongly depends on the effective boundary conditions (BC) at the 
walls (see, e.g., Refs.~\onlinecite{krech:9192all,krech:book,brankov:book,evans:1994,
diehl:2006,zandi:2007,schmidt:2008,mohry:2009} and references therein). 
It is attractive for equal symmetry breaking $(\pm,\pm)$ BC and repulsive
for opposing $(\pm,\mp)$ BC.
Inter alia, this latter feature qualifies critical Casimir forces to be a tool to overcome the problem
of ``stiction'' which occurs in micro- and nano-mechanical devices.
(The quantum electrodynamic Casimir force is typically attractive and thus responsible for stiction; turning it to be
repulsive requires a careful choice of the fluid and of the bulk materials of the confinement \cite{Munday:2009}.)
The theoretical description of the critical Casimir forces is particularly challenging due to the non-Gaussian character
of the order parameter fluctuations, which contrasts with the intrinsically Gaussian nature of the low energy fluctuations of
the electromagnetic field; in addition, the critical Casimir effect is also particularly rich as it allows, inter alia, 
symmetry breaking boundary conditions,
which do not occur for electromagnetic fields.
\par
The critical Casimir effect exhibits universality, i.e., the critical Casimir force expressed in terms of suitable 
scaling variables depends only on the universality class of the bulk critical point and on the type of boundary conditions, 
whereas it is independent of the microscopic structure and of the material properties of the specific fluid medium involved.
In our present theoretical analysis we focus on the Ising universality class which encompasses
the experimentally relevant classical binary liquid mixtures and simple fluids.
\par
%
\nocite{garcia:9902all}
\nocite{ganshin:2006}
\nocite{fukuto:2005}
\nocite{rafai:2007}
The existence of the critical Casimir effect has been experimentally confirmed and its strength has been
first measured {indirectly} for wetting films \cite{garcia:9902all,fukuto:2005,ganshin:2006,rafai:2007}.
%
%
%
The first {direct} measurement of this effect has been performed at 
the sub-micrometer scale for a spherical \emph{colloid} immersed in a (near) critical binary liquid mixture close to 
a laterally homogeneous and planar substrate \cite{hertlein:2008,gambassi:2009}. 
\nocite{hucht:2007}
\nocite{vasilyev:2007}
\nocite{vasilyev:2009}
The corresponding Monte Carlo simulation data for the \emph{film} geometry are in 
very good quantitative agreement with all available experimental data
\cite{hertlein:2008,gambassi:2009,hucht:2007,vasilyev:2007,vasilyev:2009,Hasenbusch:2009}.
\nocite{Burkhardt:1995}
\nocite{Eisenriegler:1995}
\nocite{hanke:1998}
\nocite{Schlesener:2003}
\nocite{Eisenriegler:2004}
Theoretical studies of the critical Casimir effect acting on {colloidal} particles involve spherically 
\cite{Burkhardt:1995,Eisenriegler:1995,hanke:1998,Schlesener:2003,Eisenriegler:2004} or
ellipsoidally \cite{kondrat:2009} shaped colloids adjacent to \emph{homogeneous} substrates.
\par
%
\begin{figure} 
  \includegraphics{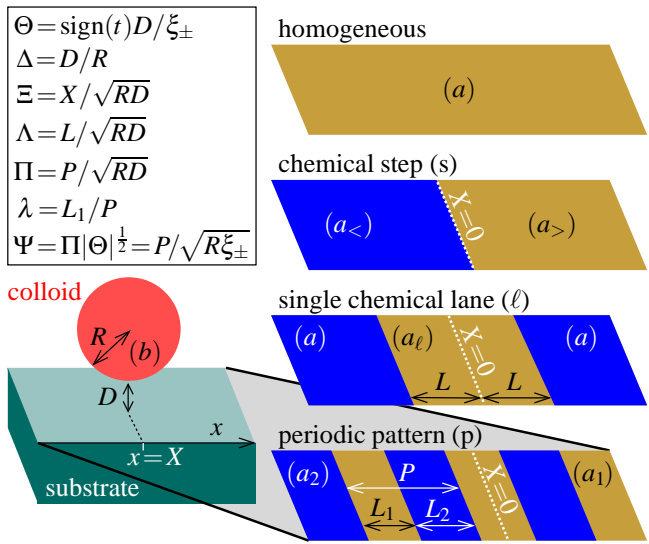}
  \caption{%
    Sketch of a spherical colloid immersed in a near-critical binary liquid mixture (not shown)
    and close to a (patterned) planar substrate.
    The sphere with $(b)$ boundary condition (BC) and radius $R$ is located at a surface-to-surface distance $D$ from
    the substrate and its center has a lateral coordinate $x=X$ with the substrate pattern being translationally invariant in
    all other directions.
    The following four different types of substrate surfaces are considered:
    homogeneous substrate [Sec.~\ref{sec:homog}], a chemical step [$\step$; Sec.~\ref{sec:step}], 
    a single chemical lane [$\stripe$; Sec.~\ref{sec:stripe}], and a periodically patterned substrate 
    [$\period$; Sec.~\ref{sec:period}].
    (Note that for a four-dimensional system, which we also consider, this is a three-dimensional
    cut of the system, which is invariant along the fourth direction; the sphere thus corresponds
    to a hypercylinder in four dimensions.)
    For later reference, the box on the left side summarizes the definitions of the various scaling variables 
    which the scaling functions of the critical Casimir force depend on for the listed geometrical configurations.
    On the right, $(a)$, $(a_\gtrless)$, $(a_{\stripe})$, $(a_1)$, and $(a_2)$ indicate the boundary conditions corresponding to the various chemical patterns.  
  }   
  \label{fig:sketch}
\end{figure}
%
Besides their wide use as model systems in soft matter physics, colloids have applications
at the micro- and nanometer scale.
In this context, they are widely used in micro- and nano-mechanical devices.
Therefore, one may utilize the critical Casimir forces acting on colloids because their strength and their direction can be tuned in a
controlled way.
%
\nocite{troendle:2009}
\nocite{soyka:2008}
\nocite{sprenger:2006}
\nocite{troendle:2008}
Suitably designed chemically or geometrically structured substrates generate 
\emph{lateral} critical Casimir forces acting on colloidal particles \cite{soyka:2008,
troendle:2008,troendle:2009,sprenger:2006}.
Current techniques allow one to endow solid surfaces with precise structures on the nano- and 
micrometer-scale.
Hence, the critical Casimir effect can be used to create laterally confining potentials for a single colloid, 
which can be tuned by temperature \cite{soyka:2008}. 
\par
Recently, the critical Casimir potential of a colloid close to a substrate with a pattern of 
parallel chemical stripes with laterally alternating adsorption preference has been measured 
\cite{soyka:2008}.
In our corresponding theoretical study \cite{troendle:2009}, we have calculated the normal and lateral critical
Casimir forces acting on a colloid close to such a patterned substrate as well as the corresponding potentials.
We have used our theoretical predictions for the universal scaling functions of the critical
Casimir potential in order to interpret the available experimental data in Ref.~\onlinecite{soyka:2008}.
It has turned out that an agreement between theory and experiment can be achieved only if one 
takes into account the geometrical details of 
the chemical substrate pattern.
This demonstrates that the critical Casimir effect is very sensitive to
the details of the imprinted structures and that it can resolve them.
\par
Here we generalize our previous analysis \cite{troendle:2009} to various substrate patterns.
In  particular we study the critical Casimir effect for a \emph{three}-dimensional
sphere close to a homogeneous substrate [Sec.~\ref{sec:homog}], a chemical step
[Sec.~\ref{sec:step}], a single chemical lane [Sec.~\ref{sec:stripe}], and 
periodic patterns of chemical stripes of alternating adsorption preference [Sec.~\ref{sec:period}] 
[see \fref{fig:sketch}].
For completeness, we also consider a \emph{cylinder} which is aligned with the chemical pattern [Sec.~\ref{sec:cylinder}].
We provide quantitative predictions for the scaling functions of the  
critical Casimir forces, pursuing a two-pronged approach:
(i) We calculate the force using the full three-dimensional numerical analysis of the appropriate 
mean-field theory (MFT).
(ii) We use the so-called Derjaguin approximation (DA) based on the scaling functions for the 
critical Casimir force in the film geometry either obtained analytically within MFT \cite{krech:1997}
or obtained from Monte Carlo simulations \cite{vasilyev:2007,vasilyev:2009}, which
allows us to predict the critical Casimir force in the physically relevant three-dimensional case.
Inter alia, we determine the range of validity of the DA within MFT,
which provides guidance concerning its applicability in three spatial dimensions $d=3$.
This is an important information because
presently available Monte Carlo simulations are far from being able to capture complex geometries 
\cite{vasilyev:2007,vasilyev:2009}.
\par
Currently, the possibility of realizing stable levitation of particles by means of the electrodynamic Casimir forces has been the subject of intense theoretical investigation \cite{leonhardt:2007,rodriguez:2008,rodriguez:2009,rahi:2009,rahi:2009a,zhao:2009}.
Our results presented in Secs.~\ref{sec:period} and \ref{sec:cylinder} show that for suitable choices of the 
geometry of the chemical pattern of the substrate, the critical Casimir forces can be used to levitate a colloid above
the substrate at a height which can be tuned by temperature.
This levitation is stable against perturbations because it corresponds to a minimum of the potential of the critical Casimir force acting on the colloid. 
\par
In  Sec.~\ref{sec:background} we briefly introduce the necessary terminology related to
finite-size scaling and we discuss briefly the corresponding MFT.
Section~\ref{sec:homog} is devoted to the well-studied case of a colloid close to a homogeneous substrate.
(In $d=4$, as appropriate for MFT, the three-dimensional colloid is extended to the fourth dimension as a hypercylinder, for which we also present the results of our analysis.)
As mentioned above, the various patterns and setups are considered in Secs.~\ref{sec:step}--\ref{sec:cylinder}.
We conclude and summarize our findings in Sec.~\ref{sec:summary}.
Certain important technical details concerning the calculation of the Derjaguin approximation are presented in the 
Appendices \ref{app:step}--\ref{app:cylinder}.
\section{Theoretical background \label{sec:background}}
\subsection{Finite-size scaling}
%
According to the theory of finite-size scaling, the normal and lateral critical Casimir forces
and the corresponding potentials can be described by \emph{universal} scaling functions,
which are independent of the molecular details of the system but depend only on the
gross features of the system, i.e., on the bulk universality class (see, e.g., Refs.~\onlinecite{krech:book,
brankov:book} and references therein) of the associated critical point.
Here, we focus on the Ising universality class (which is characterized by a scalar order parameter $\phi$)
in spatial dimensions $d=3$ and $d=4$.
In addition, the critical Casimir force depends on the type of effective boundary conditions 
at the walls, which we denote by $(a)$ and $(b)$, and by the geometry of the
confining surfaces \cite{binder:1983,diehl:1986,diehl:1997}.
Note that $(a)$ and $(b)$ can represent the various symmetry preserving fixed-point BC (the so-called
ordinary, special, periodic, or antiperiodic boundary conditions \cite{krech:book,brankov:book}) 
in addition to the symmetry breaking cases $(\pm)$ we are mainly 
interested in, and which describe the adsorption of fluids at the confining walls.
\par
Inspired by the experiments described in Ref.~\onlinecite{soyka:2008} we consider binary liquid
mixtures with their consolute critical point approached by varying the 
temperature $T$ towards $T_c$ at fixed pressure and critical composition.
We first study the \emph{film} geometry in which the fluid undergoing the continuous phase transition
is confined between two parallel, infinitely extended walls at distance $L$.
According to renormalization group theory the normal critical Casimir force $f_{(a,b)}$
per unit area which is acting on the walls scales as \cite{krech:9192all}
\begin{equation} 
  \label{eq:planar-force}
  f_{(a,b)}(L,T)=k_BT \frac{1}{L^d}k_{(a,b)}( \sgn(t)\, L/\xi_\pm),
\end{equation} 
where $(a,b)$ denotes the pair of boundary conditions $(a)$ and $(b)$ characterizing the two walls.
The scaling function $k_{(a,b)}$ depends only on a single scaling variable given by
the sign of the reduced temperature distance $t$ from the critical point ($\pm$ for $t\gtrless0$) 
and the film thickness $L$ in units of the bulk correlation length $\xi_\pm(t\to0^\pm)=\xi_0^\pm|t|^{-\nu}$,
where $\nu\simeq0.63$  in $d=3$ and $\nu=1/2$ in $d=4$ \cite{pelissetto:2002}.
(Clearly, one has $f_{(a,b)}(L,T)=f_{(b,a)}(L,T)$.)
Positive values of $t$, $t>0$, correspond to the disordered (homogeneous) phase of the 
fluid, whereas negative values of $t$, $t<0$, correspond to the ordered (inhomogeneous) phase,
where phase separation occurs.
Typically, the homogeneous phase is found at high temperatures, and one has $t=(T-T_c)/T_c$.
However,  many experimentally relevant binary liquid mixtures exhibit a \emph{lower} critical point, for which
the homogeneous phase corresponds to the low-temperature
phase and one has $t=-(T-T_c)/T_c$ \cite{hertlein:2008,gambassi:2009}.
The two non-universal amplitudes $\xi_0^\pm$ of the correlation length are of
molecular size and characterized
by the universal ratio $\xi_0^+/\xi_0^-\simeq1.9$ in $d=3$ \cite{pelissetto:2002,privman:1991}
and $\xi_0^+/\xi_0^-=\sqrt{2}$ in $d=4$ \cite{tarko:all};
$\xi_\pm$ is determined by the exponential spatial decay of the two-point correlation function of the order parameter $\phi$ in the bulk.
\par
At the critical point $T=T_c$, the correlation length diverges, $\xi_\pm\to\infty$, and the scaling
function of the critical Casimir force acting on the two planar walls attains a universal 
constant value referred to as the critical Casimir amplitude \cite{krech:book,brankov:book}:
\begin{equation} 
  \label{eq:delta-ab}
  k_{(a,b)}(L/\xi_\pm=0)=\Dab.
\end{equation} 
\par
Away from criticality, the critical Casimir force 
decays exponentially as a function of $L/\xi_\pm$.
For the specific case of symmetry breaking BC  $a,b\in\{+,-\}$
and for $t>0$ one expects for ${L/\xi_+\gg1}$ a \emph{pure} exponential decay of $\fppm$ (see, e.g., Refs.~\onlinecite{evans:1994,
krech:1997,borjan:2008} and footnote $3$ in Ref.~\onlinecite{troendle:2009}, i.e., a decay without an algebraic prefactor to the 
exponential and without a numerical prefactor to $L/\xi_+$ in the argument of the exponential) corresponding to
\begin{equation} 
  \label{eq:exponential-decay}
  k_{(+,\pm)}(L/\xi_+\gg1)=A_\pm \left(\frac{L}{\xi_+}\right)^d \exp(-L/\xi_+),
\end{equation}
where $A_\pm$ are universal constants \cite{gambassi:2009}.
Note that, in the absence of symmetry-breaking fields inside the film, the scaling functions for $(+,+)$ BC are the same as for $(-,-)$ BC.
\subsection{Mean-field theory \label{sec:MFT}}
%
The standard Landau-Ginzburg-Wilson fixed-point effective Hamiltonian describing critical phenomena
of the Ising universality class is given by \cite{binder:1983,diehl:1986}
\begin{equation} 
  \label{eq:hamiltonian}
   \mathcal{H}[\phi]=\int_V\,\upd^d\vec{r}\,\left\{
        \frac{1}{2}(\nabla\phi)^2
       +\frac{\tau}{2}\phi^2
       +\frac{u}{4!}\phi^4
			 \right\},
\end{equation} 
where $\phi(\vec{r})$ is the order parameter describing the fluid, which completely fills the
volume $V$ in $d$-dimensional space.
The first term in the integral in \eref{eq:hamiltonian} penalizes local fluctuations of
the order parameter.
The parameter $\tau$ in \eref{eq:hamiltonian} is proportional to $t$, and the coupling constant 
$u$ is positive and provides stability of the Hamiltonian for $t<0$.
The mean-field order parameter profile $m\ddef u^{1/2}\langle\phi\rangle$ minimizes
the Hamiltonian, i.e., $\updelta \mathcal{H}[\phi]/\updelta\phi|_{\phi=u^{-1/2}m}=0$.
In the bulk the mean-field order parameter is spatially constant and attains the values
$\langle\phi\rangle=\pm a|t|^\beta$ for $t<0$ and $\langle\phi\rangle=0$ for $t>0$, where, besides $\xi_0^+$,
$a$ is the only additional independent non-universal amplitude appearing 
in the description of bulk critical phenomena \cite{binder:1983,diehl:1986}, and $\beta(d=4)=1/2$ 
is a standard critical exponent.
Within MFT $\tau=t (\xi_0^+)^{-2}$ and $u=6a^2(\xi_0^+)^{-2}$.
In a finite-size system the bulk Hamiltonian $\mathcal{H}[\phi]$ is supplemented by appropriate 
surface and curvature (edge) contributions \cite{binder:1983,diehl:1986}.
In the strong adsorption limit \cite{burkhardt:1994,diehl:1993}, these contributions 
generate boundary conditions for the order parameter such that 
$\phi\big|_{\text{surface}}=\pm\infty$.
For binary liquid mixtures these fixed-point $(\pm)$ BC are the experimentally relevant ones.
(Note that a \emph{weak} adsorption preference might lead to a crossover between various kinds of 
effective boundary conditions for the order parameter $\phi$
\cite{mohry:2009,schmidt:2008,gambassi:2009}.)

\par
We have minimized numerically $\mathcal{H}[\phi]$ using a $3d$ finite element method in order to obtain
the (spatially inhomogeneous) profile $m(\vec{r})$ for the geometries under consideration 
[see \fref{fig:sketch}].
The normal and the lateral critical Casimir forces are calculated directly from these mean-field order parameter
profiles using the stress tensor \cite{krech:1997,kondrat:2009}.
This allows one to infer the universal scaling functions of the critical Casimir forces at the 
upper critical dimension $d=4$ up to an overall prefactor $\propto u^{-1}$ and up to logarithmic
corrections.
The corresponding critical Casimir potential is obtained by the appropriate integration of
the normal or of the lateral critical Casimir forces.
\par
In the  case of planar walls the MFT scaling functions for the critical Casimir force can be determined analytically 
\cite{krech:1997} and one finds [see \eref{eq:delta-ab}] for the case of symmetry breaking boundary conditions 
the following critical Casimir amplitudes: $\Dpp=\Dmm=24[K(1/\sqrt{2})]^4/u\simeq-283.61\times u^{-1}$, 
where $K$ is the complete elliptic integral of the first kind, and 
$\Dpm=-4\Dpp$ [see Ref.~\onlinecite{krech:1997} and Eq.~(27) and Ref.~[49] in Ref.~\onlinecite{vasilyev:2009}].
\par
In ${d=4}$ (corresponding to MFT) the three-dimensional sphere is a hypercylinder and the 
physical properties are invariant along the fourth dimension.
Accordingly, the MFT results for the force and the potential given below are those per length 
along this additional direction.
%
%
\section{Homogeneous substrate \label{sec:homog}}
We first consider a three-dimensional sphere of radius $R$ with $(b)$ BC facing a 
chemically \emph{homogeneous} substrate with $(a)$ BC at a surface-to-surface distance $D$ 
as shown in \fref{fig:sketch}, denoting this combination by $(a,b)$. 
The critical Casimir force $\Fab(D,R,T)$ \emph{normal} to the substrate surface 
and the corresponding critical Casimir potential $\Phiab(D,R,T)=\int_D^\infty \upd z\; \Fab(z,R,T)$ 
take the scaling forms \cite{hanke:1998,hertlein:2008,troendle:2009,gambassi:2009}
\begin{align}
  \label{eq:force-homog}
  \Fab(D,R,T)&=k_BT\frac{R}{D^{d-1}}\Kab(\Theta,\Delta)\\
\intertext{and}
  \label{eq:potential-homog}
\Phiab(D,R,T)& =k_BT\frac{R}{D^{d-2}}\varthetaab(\Theta,\Delta),
\end{align}
where $\Delta={D}/{R}$ and $\Theta=\sgn(t)\,{D}/{\xi_\pm}$ (for $t\gtrless 0$)
are the scaling variables corresponding to the distance $D$
in units of the radius $R$ of the colloid and of the correlation length $\xi_\pm$, respectively.
The case $d=4$ corresponds to the MFT solution up to logarithmic corrections, which we shall neglect
here.
Equations~\eqref{eq:force-homog} and \eqref{eq:potential-homog} describe a force and an energy, respectively,
per $D^{d-3}$, which for $d=4$ corresponds to considering $\Fab$ and $\Phiab$ per length $L_4$ of the extra 
translationally invariant direction of the hypercylinder.
%
\subsection{Derjaguin approximation \label{sec:homog-da}}
%
The Derjaguin approximation (DA) is based on the idea of decomposing the surface of the spherical colloid into infinitely thin
circular rings of radius $\rho$ and area $\upd S (\rho)= 2\pi\rho\upd\rho$ which are parallel to the 
opposing substrate surface \cite{derjaguin:1934,hanke:1998,hertlein:2008,gambassi:2009,troendle:2009}.
(Here we do not multiply $2\pi\rho\upd\rho$ by the linear extension $L_4$ of the hypercylinder along its axis in the fourth dimension, because
the critical Casimir force is eventually expressed in units of $L_4$, which therefore drops out from the final expressions.)
The distance $L$ of a ring with radius $\rho$ from the substrate is given by
\begin{equation} 
  \label{eq:da-L}
  L(\rho)= D+R\left(1-\sqrt{1-{\rho^2}/{R^2}}\right).
\end{equation}
Assuming \emph{additivity} of the forces and neglecting edge effects, 
the normal critical Casimir forces $\upd F(\rho)$ acting
on these rings can be expressed in terms of the force acting on parallel plates [\eref{eq:planar-force}]: 
\begin{equation} 
  \label{eq:da-dF}
  \frac{{\upd F(\rho)}}{k_BT} =  \frac{\upd S}{\left[L(\rho)\right]^{d}} \kab(\sgn(t)\, L(\rho)/\xi_\pm).
\end{equation}
Finally, in order to calculate the total force $\Fab$ acting on the colloid, one sums up 
the contributions of the rings, which yields
\begin{equation}   
  \label{eq:da-def}
  \frac{\Fab(D,R,T)}{k_BT}\simeq2\pi   \int_0^R \upd\rho \rho \left[L(\rho)\right]^{-d} \kab(\sgn(t)\, L(\rho)/\xi_\pm).
\end{equation} 
(For $d=3$, $\Fab$ is the force on a sphere whereas in $d=4$ it is the force on a hypercylinder per length
of its axis.)
\begin{figure} 
  \includegraphics{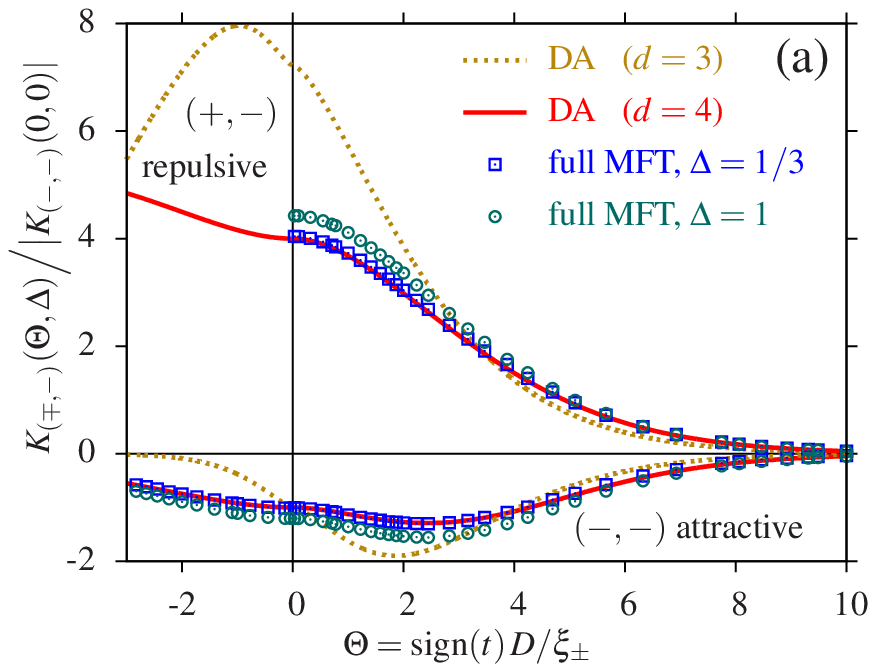}\\ 
  \includegraphics{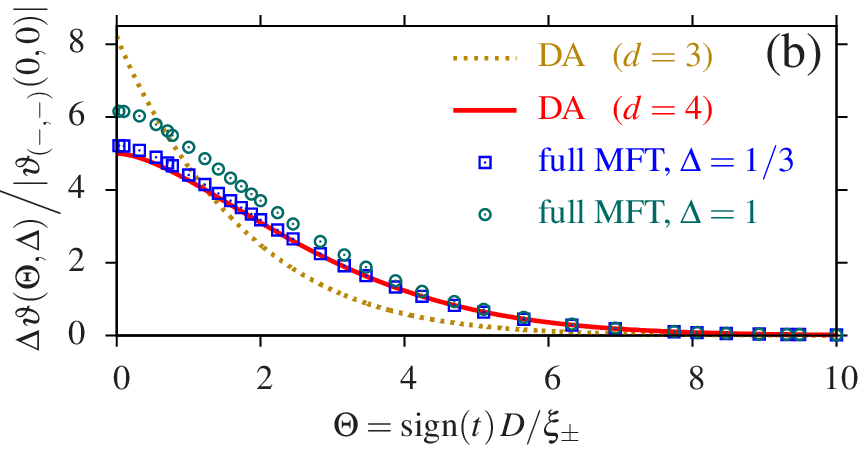} 
  \ifTwocolumn
  \else
  \linespread{1.2}
  \selectfont
  \fi
  \caption{
    (a)
    Scaling functions $\Kmpm$ for the normal critical Casimir force [\eref{eq:force-homog}] 
		acting on a three-dimensional sphere with $(b)=\MBC$ BC close to a homogeneous substrate 
    with $(a)=(\mp)$ BC [\fref{fig:sketch}].
    The suitably normalized scaling functions $\Kmpm$ are shown as a function of the 
    scaling variable $\Theta=\sgn(t) D/\xi_\pm$ for $t\gtrless0$,
    where $t$ is the reduced deviation from the critical temperature
		and $\Kmm(0,0)$ is the value of the critical Casimir force
    scaling function within the DA at $T=T_c$ for $(-,-)$ BC.
		The solid lines correspond to the Derjaguin approximation (DA, $\Delta=D/R\to0$) within mean-field theory (MFT, $d=4$)
    whereas the dotted lines correspond to the DA obtained by using Monte Carlo (MC) results for films in $d=3$
    the systematic uncertainties of which are not indicated \cite{mcdata}.
    The normalization implies that at $\Theta=0$ both the solid and dotted lines pass through $-1$ for
    $\Mm$ BC whereas the solid line passes through $4$ for $\Pm$ BC.
    The symbols correspond to the full numerical MFT results obtained for $\Delta=1/3$ and $\Delta=1$, 
		the size of which indicates the estimated numerical error.
		(For $\Pm$ BC and $t<0$ we have not been able to calculate the corresponding scaling functions with
	  adequate precision due to severe numerical difficulties in obtaining the full three-dimensional order parameter 
		profile in the presence of two ``competing'' bulk values.)
    Since within the DA the dependence of $\Kmpm$ on $\Delta$ drops out, the difference between the symbols
    $\boxdot$ and $\odot$ and the solid lines measures the accuracy of the DA in $d=4$.
    (b)
		Difference $\Delta\vartheta=\varthetapm-\varthetamm$  of the scaling functions for the Casimir potentials
    [\eref{eq:potential-homog}] for $\Pm$ and $\Mm$ BC, suitably normalized 
    by $\varthetamm(0,0)$.
		The solid line corresponds to the DA within MFT and the symbols correspond to the full MFT results 
    for $\Delta=1/3$ and $\Delta=1$; the dotted line is the DA for $d=3$.
    Due to the normalization the solid line reaches $5$ for $\Theta=0$.
    } 
  \label{fig:homog}
\end{figure}
\par
One expects the DA to describe the actual behavior accurately if the colloid is very close to
the substrate, i.e., for $\Delta=D/R\to0$.
In this limit, \eref{eq:da-L} can be approximated by $L(\rho)=D\alpha$ where $\alpha=1+\rho^2/(2RD)$, 
so that one finds for the scaling function of the force \cite{hanke:1998,gambassi:2009}
\begin{equation} 
  \label{eq:da-force}
  \Kab(\Theta,\Delta\to0)=2\pi\int_{1}^{\infty}\upd \alpha\alpha^{-d}\kab(\alpha\Theta),
\end{equation} 
and, accordingly, for the scaling function of the potential \cite{hertlein:2008,troendle:2009,gambassi:2009}
\begin{equation}
  \varthetaab(\Theta,\Delta\to0)=\\
  2\pi\int_{1}^{\infty}\upd \beta\left( \frac{1}{\beta^{d-1}}-\frac{1}{\beta^d} \right)\kab(\beta\Theta).
  \label{eq:derjaguinpot}
\end{equation}
At the bulk critical point, using \eref{eq:delta-ab}, one finds the well known values
${\Kab(0,0)} = {2\pi\Dab/(d-1)}$ and  ${\varthetaab(0,0)} = {2\pi\Dab/[(d-2)(d-1)]}$.
We note that the DA implies that the dependence of $\Fab$ and $\Phiab$ on the size $R$
of the sphere reduces to the proportionality $\propto R$ indicated explicitly in
Eqs.~\eqref{eq:force-homog} and \eqref{eq:potential-homog}.
\par
Before proceeding further one first has to assess the accuracy of the DA, which will carried out below within MFT ($d=4$).
We expect the range of validity of the DA to be similar for $d=3$, so that within that range
one can use the DA based on scaling functions for the film geometry obtained
from Monte Carlo simulations \cite{mcdata} in order to calculate the critical Casimir force acting on a colloid in $d=3$.
%
\subsection{Scaling functions for the normal critical Casimir force and the potential}
%
The expressions obtained above within the DA hold for general boundary conditions $(a)$ and $(b)$ and
are valid beyond the cases we consider in the following, i.e., $a\in\{+,-\}$ and $b=-$.
Figure~\ref{fig:homog}(a) shows the full numerical MFT ($d=4$) results for the scaling functions $\Kpmm$ with $\Delta=1$ and $\frac{1}{3}$
compared with the corresponding DA results based on the suitable numerical integration 
[\eref{eq:da-force}] of the analytic (MFT) 
expression for $\kpmm$ \cite{krech:1997}.
Moreover, in \fref{fig:homog}, the corresponding DA results for $d=3$ are shown;
they are obtained from the film scaling functions determined by MC simulations \cite{mcdata} 
and by using the corresponding ratio of the correlation lengths above and below $T_c$ \cite{pelissetto:2002}.
In \fref{fig:homog}(b) we report the difference  
$\Delta\vartheta(\Theta,\Delta)\ddef\varthetapm(\Theta,\Delta)-\varthetamm(\Theta,\Delta)$
computed for the various cases reported in \fref{fig:homog}(a), which will be useful
for describing the case of a chemically patterned substrate.
The scaling functions in $d=4$ are reasonably well reproduced by the DA for 
$\Delta\lesssim0.4$ and we expect this to hold for $d=3$ as well.
The fact that for increasing values of $\Delta$ the magnitude of the actual scaling functions becomes larger 
compared with those within the DA (corresponding to $\Delta\to0$) is in agreement with earlier results obtained 
for a $d$-dimensional hypersphere (see, e.g., Ref.~\onlinecite{hanke:1998}).
\par
It has been shown that the scaling functions obtained within the DA 
for $d=3$ agree very well -- within the experimental accuracy -- with the ones obtained from direct measurements of the critical Casimir 
potential \cite{hertlein:2008,gambassi:2009} corresponding to $\Delta\lesssim0.35$ (see also
Ref.~[48] in Ref.~\onlinecite{vasilyev:2009}).
%
%
\section{Chemical step ($\step$) \label{sec:step}} 
The basic building block of a chemically patterned substrate of the type we consider here, i.e., 
with translational invariance in all directions but one ($x$), is a chemical step ($\step$) realized by
a substrate with $(a_\gtrless)$ BC for $x\gtrless0$ at its surface.
In this section we analyze the critical Casimir force if such a substrate is approached by
a colloid with $(b)$ BC with its center located at the lateral position $x=X$ (see \fref{fig:sketch}
and Ref.~\onlinecite{soyka:2008} for experimental realizations). 
We denote this configuration by $(a_<|a_>,b)$.
The normal critical Casimir force $\Fstep$ is described by the scaling form \cite{troendle:2009}
\begin{equation} 
  \label{eq:step-force}
  \Fstep (X,D,R,T)= k_BT \frac{R}{D^{d-1}} \; \Kstep(\Xi,\Theta,\Delta),
\end{equation}
where $\Xi=X/\sqrt{RD}$ is the scaling variable corresponding to the lateral position of the colloid.
It is useful to write the scaling function $\Kstep$ as
\ifTwocolumn
\begin{multline}
  \label{eq:step-K}
  \Kstep (\Xi,\Theta,\Delta) =  
                                \frac{K_\Asb+K_\Alb}{2}
                                \\
                                +\frac{K_\Asb-K_\Alb}{2}
                                    \psi_{(a_<|a_>,b)}(\Xi,\Theta,\Delta)
                                ,%
\end{multline}
\else
\begin{equation}
  \label{eq:step-K}
  \Kstep (\Xi,\Theta,\Delta) =  
                                \frac{K_\Asb+K_\Alb}{2}
                                +\frac{K_\Asb-K_\Alb}{2}
                                    \psi_{(a_<|a_>,b)}(\Xi,\Theta,\Delta)
                                ,%
\end{equation}
\fi
where the scaling functions of the laterally homogeneous substrates $K_{(a_\gtrless,b)}$ depend on $\Theta$ and $\Delta$ only [\eref{eq:force-homog}],
and the scaling function $\psi_{(a_<|a_>,b)}$ varies from $+1$ at $\Xi\to-\infty$ to $-1$
at $\Xi\to+\infty$, such that the laterally homogeneous cases are recovered far from the step.
Accordingly, the corresponding critical Casimir potential $\Phistep(X,D,R,T)=\int_D^\infty \upd z\;\Fstep(X,z,R,T)$
can be cast in the form \cite{troendle:2009}
\begin{equation} 
  \label{eq:step-pot}
  \Phistep(X,D,R,T)=k_BT\frac{R}{D^{d-2}}\varthetastep(\Xi,\Theta,\Delta),
\end{equation}
and
\ifTwocolumn
\begin{multline}
  \label{eq:step-vartheta}
  \varthetastep(\Xi,\Theta,\Delta)=\frac{\vartheta_\Asb+\vartheta_\Alb}{2}
                                \\
                                +\frac{\vartheta_\Asb-\vartheta_\Alb}{2}
                                    \omega_{(a_<|a_>,b)}(\Xi,\Theta,\Delta)
                                ,
\end{multline}
\else
\begin{equation}
  \label{eq:step-vartheta}
  \varthetastep(\Xi,\Theta,\Delta)=\frac{\vartheta_\Asb+\vartheta_\Alb}{2}
                                +\frac{\vartheta_\Asb-\vartheta_\Alb}{2}
                                    \omega_{(a_<|a_>,b)}(\Xi,\Theta,\Delta)
                                ,
\end{equation}
\fi
where $\vartheta_{(a_\gtrless,b)}$ depend on $\Theta$ and $\Delta$ only [\eref{eq:potential-homog}], 
and $\omega_{(a_<|a_>,b)}(\Xi=\pm\infty,\Theta,\Delta)=\mp1$.
Note that the scaling functions $\psi_{(a_<|a_>,b)}$ and $\omega_{(a_<|a_>,b)}$ are independent
of the common prefactor $\propto u^{-1}$ [see Sec.~\ref{sec:MFT}], which is left undetermined
by the analytical and numerical mean-field calculation of $\Kstep$ and $\varthetastep$.
\subsection{Derjaguin approximation}
%
If the sphere is close to the substrate, i.e., $\Delta\to0$, the DA can be applied,
and one finds for the scaling function of the critical Casimir force [see Appendix~\ref{app:step}]
\ifTwocolumn
\begin{multline}
    \label{eq:step-psi-da}
      \psi_{(a_<|a_>,b)}(\Xi\gtrless0,\Theta,\Delta\to0)=
      \mp 1 \\
      \pm\frac{
                4
                \int_{1+\Xi^2/2}^{\infty}\upd\alpha\;
                  \alpha^{-d}\arccos\left({|\Xi|}(2\alpha-2)^{-1/2}\right)
                  \Delta k( \alpha\Theta)} 
              {
                K_{(a_<,b)}(\Theta,\Delta\to0)-K_{(a_>,b)}(\Theta,\Delta\to0)
              },
\end{multline}
\else
\begin{equation}
    \label{eq:step-psi-da}
      \psi_{(a_<|a_>,b)}(\Xi\gtrless0,\Theta,\Delta\to0)=
      \mp 1 
      \pm\frac{
                4
                \int_{1+\Xi^2/2}^{\infty}\upd\alpha\;
                  \alpha^{-d}\arccos\left({|\Xi|}(2\alpha-2)^{-1/2}\right)
                  \Delta k( \alpha\Theta)} 
              {
                K_{(a_<,b)}(\Theta,\Delta\to0)-K_{(a_>,b)}(\Theta,\Delta\to0)
              },
\end{equation}
\fi
where $\Delta k(\Theta) = k_\Asb(\Theta) - k_\Alb(\Theta)$ is the difference between the scaling functions
for the critical Casimir forces acting on two planar walls with $\Asb$ and with $\Alb$ boundary conditions, respectively.
We note that according to Eqs.~\eqref{eq:step-psi-da} and \eqref{eq:da-force} within the DA $\psi_{(a_<|a_>,b)}$ can be
determined from the knowledge of the film scaling functions $\kab(\Theta)$ [\eref{eq:planar-force}] only.
Due to the assumption of additivity which underlies the DA, (i) $\psi_{(a_<|a_>,b)}$ vanishes at 
$\Xi=0$ for all $\Theta$ and it is an antisymmetric function of $\Xi$ and (ii) $\psi_{(a_<|a_>,b)}=\psi_{(a_>|a_<,b)}$;
within the DA both of these properties are valid irrespective of the type of
boundary conditions on both sides of the chemical step.
(However, the actual scaling function $\psi_{(a_<|a_>,b)}$ as, e.g., obtained from full numerical
MFT calculations may violate this symmetry because the actual critical Casimir forces are non-additive.)
At the bulk critical point one has $\Theta=0$ so that [see Appendix~\ref{app:step-crit}],
\ifTwocolumn
\begin{multline} 
  \label{eq:step-psi-da-crit}
  \psi_{(a_<|a_>,b)}(\Xi,\Theta=0,\Delta\to0)=\\
  \Xi^{2d-7}\left(\tfrac{15}{2}(3-d)+(3-2d)\Xi^2-\Xi^4\right)\left(2+\Xi^2\right)^{-(d-\frac{3}{2})}
\end{multline}
\else
\begin{equation} 
  \label{eq:step-psi-da-crit}
  \psi_{(a_<|a_>,b)}(\Xi,\Theta=0,\Delta\to0)=\\
  \Xi^{2d-7}\left(\tfrac{15}{2}(3-d)+(3-2d)\Xi^2-\Xi^4\right)\left(2+\Xi^2\right)^{-(d-\frac{3}{2})}
\end{equation} 
\fi
independent of $k_{(a_\gtrless,b)}$.
Similarly, within the DA one finds for the scaling function $\omega$ 
of the critical Casimir potential [see Appendix~\ref{app:step} and Ref.~\onlinecite{troendle:2009}]
\ifTwocolumn
\begin{multline}
    \label{eq:step-omega-da}
      \omega_{(a_<|a_>,b)}(\Xi\gtrless0,\Theta,\Delta\to0)=\mp1
    \\
      \pm  
      \frac{\Xi^4
    \int_{1}^{\infty}\upd s
    \frac{
    s\arccos\left( s^{-1/2}\right)-\sqrt{s-1}}
    {(1+\Xi^2s/2)^{d}}
    \Delta k
    \left( \Theta[1+\Xi^2 s/2] \right)}
    {
    \vartheta_{(a_<,b)}(\Theta,\Delta\to0)-\vartheta_{(a_>,b)}(\Theta,\Delta\to0)
    }.
\end{multline}
\else
\begin{equation}
    \label{eq:step-omega-da}
      \omega_{(a_<|a_>,b)}(\Xi\gtrless0,\Theta,\Delta\to0)=
      \mp1\pm  
      \frac{\Xi^4
    \int_{1}^{\infty}\upd s
    \frac{
    s\arccos\left( s^{-1/2}\right)-\sqrt{s-1}}
    {(1+\Xi^2s/2)^{d}}
    \Delta k
    \left( \Theta[1+\Xi^2 s/2] \right)}
    {
    \vartheta_{(a_<,b)}(\Theta,\Delta\to0)-\vartheta_{(a_>,b)}(\Theta,\Delta\to0)
    }.
\end{equation}
\fi
This yields $\omega_{(a_<|a_>,b)}(\Xi=0,\Theta,\Delta\to0)=0$, as expected from the underlying 
assumption of additivity; within full MFT this only holds in the limit $\Delta\to0$.
At the critical point we find [see Appendix~\ref{app:step-crit}]
\begin{equation}
      \label{eq:step-omega-da-crit}
        \omega_{(a_<|a_>,b)}(\Xi,\Theta=0,\Delta\to0) =
          {\Xi\left(1-d-\Xi^2\right)}{\left(\Xi^2+2\right)^{-3/2}}.
\end{equation}%
%
\begin{figure} 
  \ifTwocolumn
    \includegraphics[width=8cm]{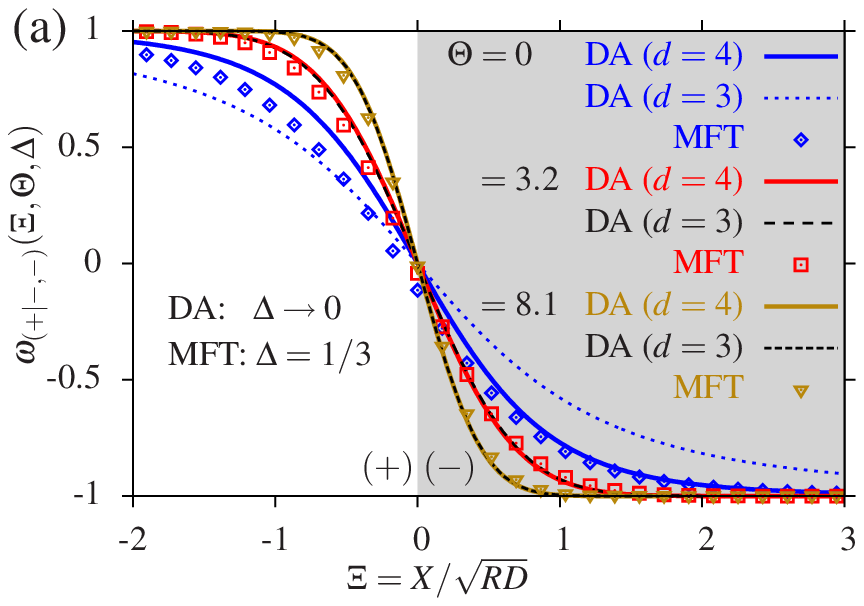}\\
    \includegraphics[width=8cm]{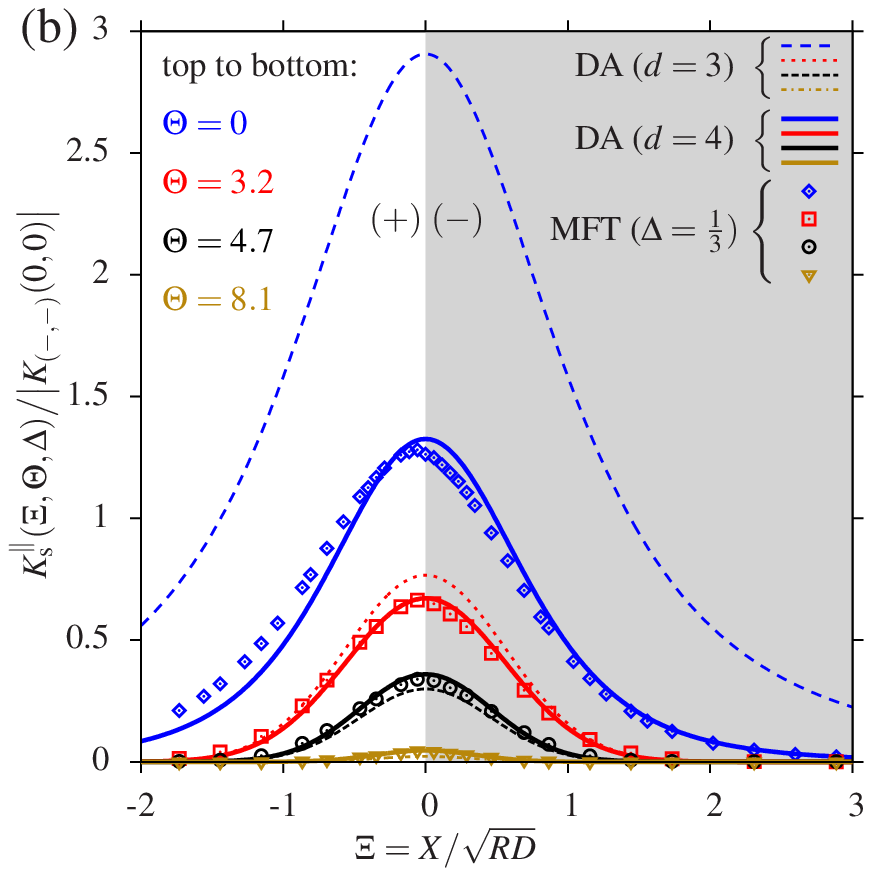}
  \else
  \includegraphics{step_temp}\\
  \includegraphics{lateral_step}
    \linespread{1}\selectfont
  \fi
  \caption{%
  (a) Scaling function $\omega_{(+|-,-)}$ [\eref{eq:step-vartheta}] for the  critical Casimir potential of a
  spherical colloid with $\MBC$ BC across a chemical step $(+|-)$ as a function of $\Xi \equiv X/\sqrt{R D}$ 
  for various (positive) values of $\Theta=D/\xi_+$ \cite{troendle:2009}.
  Within the DA  $\omega_{(+|-,-)}$ is an antisymmetric function of $\Xi$ [\eref{eq:step-omega-da}]
  whereas within full MFT this antisymmetry is slightly violated, in particular for small $\Theta$.
  (b) Corresponding scaling function $\Kstep^\parallel$ [\eref{eq:step-lateral-force}] of 
  the \emph{lateral} critical Casimir force, normalized by the 
  amplitude $\Kmm(0,0)=2\pi\Dmm/(d-1)$  of the normal critical 
  Casimir force at $T=T_c$ acting on a colloid with $\MBC$ BC close to a homogeneous substrate with $\MBC$ BC
  within the DA [Sec.~\ref{sec:homog-da}].
  For both (a) and (b) the full numerical MFT results obtained for $\Delta=1/3$ are shown as 
  symbols (the symbol size represents the estimated numerical error) whereas the lines show 
  the corresponding results obtained within the DA (i.e., $\Delta\to0$); the dotted lines refer to $d=3$
  and are obtained by using Monte Carlo simulation data \cite{mcdata} and the solid lines refer to $d=4$.
  The lines for $\Theta=0$ are obtained by using \eref{eq:step-omega-da-crit} and 
  \eref{eq:step-lateral-force-crit}, respectively; 
  for $\Theta=3.2,4.7,8.1$ the DA lines de facto coincide with the asymptotic results 
  obtained for symmetry breaking BC and
  $\Theta\gg1$ [\eref{eq:erf} and \eref{eq:step-lateral-force-erf}, respectively]
  and thus are indeed independent of $d$.  
  The DA ($d=4$) provides a good approximation for the full numerical MFT data, in particular for $\Theta\gtrsim3$.
  $\Kstep^\parallel>0$ implies that the colloid moves to the right where it enjoys an attractive potential versus 
  a repulsive one for $\Xi<0$.
  Within the DA $\Kstep^\parallel$ is a symmetric function of $\Xi$ [Eqs.~\eqref{eq:step-omega-da} and \eqref{eq:step-22new}]
  whereas within full MFT this symmetry is slightly violated, in particular for small $\Theta$.
  }   
  \label{fig:lateral}
\end{figure}
\par
For symmetry breaking $(\mp,-)$ BC and $\Theta\gg1$ the critical Casimir force $\fmpm(D,T)$ acting on two planar 
walls at a distance $D$ decays $\propto\exp(-\Theta)$ [Eqs.~\eqref{eq:exponential-decay} and \eqref{eq:planar-force}], 
which within the DA leads to the 
same \emph{$d$-independent} result for the scaling functions $\psi_{(+|-,-)}$ and 
$\omega_{(+|-,-)}$ [see Appendix~\ref{app:step-gg}]:
\ifTwocolumn
\begin{multline}
  \label{eq:erf}
  \psi_{(+|-,-)}(\Xi,\Theta\gg1,\Delta\to0) =\\
  \omega_{(+|-,-)}(\Xi,\Theta\gg1,\Delta\to0) =\\
  - \erf\left( \sqrt{\Theta/2}\;\Xi\right),
\end{multline}
\else
\begin{equation}
  \label{eq:erf}
  \psi_{(+|-,-)}(\Xi,\Theta\gg1,\Delta\to0) =
  \omega_{(+|-,-)}(\Xi,\Theta\gg1,\Delta\to0) =
  - \erf\left( \sqrt{\Theta/2}\;\Xi\right),
\end{equation}
\fi
where $\erf$ is the error function.
\par
Figure~\ref{fig:lateral}(a) compares the scaling function $\omega_{(a_<|a_>,b)}$
for the critical Casimir potential of a sphere with $\MBC$ BC in front of a $(+|-)$ step, as
obtained within the DA for $d=4$ [\eref{eq:step-omega-da}], with the one obtained numerically
within full MFT for $\Delta=1/3$.
For $\Delta\lesssim1/3$ the DA captures the scaling function very well, 
in particular for $\Theta\gtrsim3$ \cite{troendle:2009}.
The scaling function $\omega_{(a_<|a_>,b)}$ obtained within the DA ($d=3$)
on the basis of the Monte Carlo data of Ref.~\onlinecite{vasilyev:2009}, which is also shown in \fref{fig:lateral}(a),
has been used successfully in order to
interpret the experimental data of Ref.~\onlinecite{soyka:2008}, for
which the analysis in terms of separate, independent, and consecutive chemical steps turned out to be
accurate.
Moreover, the critical Casimir forces turned out to be a sensitive probe of the chemical pattern and its geometric
design \cite{troendle:2009}.
\subsection{Lateral critical Casimir force \label{sec:lateral}}
%
The \emph{lateral} critical Casimir force is given by $\Fstep^\parallel=-\partial_X\Phistep$ and
can be cast in the scaling form
\begin{equation} 
  \label{eq:step-lateral-force}
  \Fstep^\parallel(X,D,R,T)=k_BT\,\frac{R}{D^{d-1}}\left(\frac{D}{R}\right)^{1/2}\,\Kstep^\parallel(\Xi,\Theta,\Delta),
\end{equation} 
where $\Kstep^\parallel$ is a universal scaling function.
$\Fstep^\parallel$ and $\Kstep^\parallel$ vanish far from the chemical
step, i.e., for $|\Xi|\to\infty$.
In \eref{eq:step-lateral-force} the prefactors  in terms of $R$ and $D$ and their exponents are 
chosen such that $\Kstep^\parallel$ is regular and non-vanishing for $\Delta\to0$.
We note that the same holds for the normal critical Casimir forces and the corresponding potentials 
[see Eqs.~\eqref{eq:force-homog}, \eqref{eq:potential-homog}, \eqref{eq:step-force}, \eqref{eq:step-pot}, and 
the considerations following below].
\par
Within the DA $\Kstep^\parallel$ can be calculated from Eqs.~\eqref{eq:step-vartheta} and \eqref{eq:step-omega-da}:
\ifTwocolumn
\begin{multline} 
  \label{eq:step-22new}
  \Kstep^\parallel(\Xi,\Theta,\Delta\to0)=\\-\frac{1}{2}
    \left[\vartheta_\Asb(\Theta,\Delta\to0)-\vartheta_\Alb(\Theta,\Delta\to0)\right]\times\\
    \partial_\Xi\omega_{(a_<|a_>,b)}(\Xi,\Theta,\Delta\to0).
\end{multline} 
\else
\begin{equation} 
  \label{eq:step-22new}
  \Kstep^\parallel(\Xi,\Theta,\Delta\to0)=-\frac{1}{2}
    \left[\vartheta_\Asb(\Theta,\Delta\to0)-\vartheta_\Alb(\Theta,\Delta\to0)\right]
    \partial_\Xi\omega_{(a_<|a_>,b)}(\Xi,\Theta,\Delta\to0).
\end{equation} 
\fi
At bulk criticality $\Theta=0$ one finds with \eref{eq:step-omega-da-crit} [see \eref{eq:step-psi-da}]
\begin{equation} 
  \label{eq:step-lateral-force-crit}
  \Kstep^\parallel(\Xi,\Theta=0,\Delta\to0)=\pi\Delta k(0)\left(2+\Xi^2\right)^{-(d-\frac{3}{2})}.
\end{equation} 
For $(\mp,-)$ BC and $\Theta\gg1$ Eqs.~\eqref{eq:step-pot}, \eqref{eq:step-vartheta}, and \eqref{eq:erf} lead to
\ifTwocolumn
\begin{multline} 
  \label{eq:step-lateral-force-erf}
  \Kstep^\parallel(\Xi,\Theta\gg1,\Delta\to0)=\\
  \left[\varthetapm(\Theta,\Delta)-\varthetamm(\Theta,\Delta)\right]\sqrt{\frac{\Theta}{2\pi}}\;\exp\left\{-\frac{\Theta\Xi^2}{2}\right\},
\end{multline} 
\else
\begin{equation} 
  \label{eq:step-lateral-force-erf}
  \Kstep^\parallel(\Xi,\Theta\gg1,\Delta\to0)=\\
  \left[\varthetapm(\Theta,\Delta)-\varthetamm(\Theta,\Delta)\right]\sqrt{\frac{\Theta}{2\pi}}\;\exp\left\{-\frac{\Theta\Xi^2}{2}\right\},
\end{equation} 
\fi
for \emph{both} $d=3$ and $d=4$.
[The prefactor $\Delta\vartheta(\Theta,\Delta)=\varthetapm(\Theta,\Delta)-\varthetamm(\Theta,\Delta)$ in \eref{eq:step-lateral-force-erf} is
shown in \fref{fig:homog}(b).]
\par
Figure~\ref{fig:lateral}(b) shows the comparison between the normalized 
lateral critical Casimir force obtained within the DA (solid lines)
and the full MFT data obtained for $\Delta=1/3$ (symbols).
We infer that not only the shape of $\Kstep^\parallel$ as a function of $\Xi$
but also its \emph{amplitude} is described well by the DA [Eqs.~\eqref{eq:step-lateral-force-crit} and
\eqref{eq:step-lateral-force-erf}] for $\Delta\lesssim1/3$, and in particular for $\Theta\gtrsim3$.
We expect this feature to hold in $d=3$, too, as well as for the normal critical Casimir force
and the critical Casimir potential.
The lateral critical Casimir forces for $d=3$ obtained within the DA on the basis of Monte Carlo simulation
data for the film geometry \cite{mcdata} are shown in \fref{fig:lateral}(b) as dashed lines. 
Compared with the previous curves, these ones have similar shapes but their overall amplitudes 
in units of the normal critical Casimir force at $\Theta = 0$ are significantly different 
for $\Theta = 0$ and $\Theta = 3.2$. This difference reflects the analogous one observed in the normalized 
difference between the corresponding critical Casimir potentials for $(+,-)$ and $(-,-)$ BC, reported in \fref{fig:homog}(b).
%
\section{Single chemical lane ($\stripe$)\label{sec:stripe}}
In this section we consider the case of a colloid with $(b)$ BC close to a substrate with a 
single chemical lane ($\stripe$) with $(a_{\stripe})$ BC and width $2L$ in the lateral $x$ direction and 
which is invariant along the other lateral direction(s).
The remaining parts of the substrate are two semi-infinite planes at $|x|>L$ with $(a)$ BC [see \fref{fig:sketch}].
The lateral coordinate $X$ of the center of mass of the sphere along the $x$ direction is chosen to vanish
in the center of the chemical lane.
One expects that for ``broad'' lanes a description in terms of two subsequent chemical steps is 
appropriate [Sec.~\ref{sec:step} and Ref.~\onlinecite{troendle:2009}], whereas for ``narrow'' lanes the effects of the two
subsequent chemical steps interfere.
We find that in addition to the variables characterizing the chemical step [\eref{eq:step-force}], a further 
scaling variable $\Lambda=L/\sqrt{RD}$ emerges naturally, which corresponds to the width of the lane.
Accordingly,
the normal critical Casimir force $\Fstripe$ acting on the colloid can be cast in the form
\begin{equation} 
  \label{eq:stripe-force}
  \Fstripe(L,X,D,R,T)=k_BT \frac{R}{D^{d-1}} \Kstripe\left(\Lambda,\Xi,\Theta,\Delta\right),
\end{equation} 
where $\Kstripe$ is the corresponding universal scaling function.
The critical Casimir potential scales as
\begin{equation} 
  \label{eq:stripe-pot}
  \Phistripe(L,X,D,R,T)=k_BT \frac{R}{D^{d-2}} \varthetastripe\left(\Lambda,\Xi,\Theta,\Delta\right),
\end{equation} 
with $\varthetastripe$ as the universal scaling function for the potential of a sphere close to
a single chemical lane.
Analogously to Eqs.~\eqref{eq:step-K} and \eqref{eq:step-vartheta}  we define $\psistripe$ and
$\omegastripe$ according to
\ifTwocolumn
\begin{multline} 
  \label{eq:stripe-psi}
  \Kstripe(\Lambda,\Xi,\Theta,\Delta)=
    \frac{K_{(a,b)}+K_{(a_{\stripe},b)}}{2}\\
    +\frac{K_{(a,b)}-K_{(a_{\stripe},b)}}{2}     
    \psistripe(\Lambda,\Xi,\Theta,\Delta),
\end{multline}
\else
\begin{equation} 
  \label{eq:stripe-psi}
  \Kstripe(\Lambda,\Xi,\Theta,\Delta)=
    \frac{K_{(a,b)}+K_{(a_{\stripe},b)}}{2}+
    \frac{K_{(a,b)}-K_{(a_{\stripe},b)}}{2}     
    \psistripe(\Lambda,\Xi,\Theta,\Delta),
\end{equation}
\fi
and
\ifTwocolumn
\begin{multline} 
  \label{eq:stripe-omega}
  \varthetastripe(\Lambda,\Xi,\Theta,\Delta)=
    \frac{\vartheta_{(a,b)}+\vartheta_{(a_{\stripe},b)}}{2}\\
    +\frac{\vartheta_{(a,b)}-\vartheta_{(a_{\stripe},b)}}{2}     \omegastripe(\Lambda,\Xi,\Theta,\Delta),
\end{multline}
\else
\begin{equation} 
  \label{eq:stripe-omega}
  \varthetastripe(\Lambda,\Xi,\Theta,\Delta)=
    \frac{\vartheta_{(a,b)}+\vartheta_{(a_{\stripe},b)}}{2}+\\
    \frac{\vartheta_{(a,b)}-\vartheta_{(a_{\stripe},b)}}{2}     \omegastripe(\Lambda,\Xi,\Theta,\Delta),
\end{equation}
\fi
so that far from the lane 
$\psistripe\left(\Lambda,|\Xi|\gg\Lambda,\Theta,\Delta \right)=
\omegastripe\left(\Lambda,|\Xi|\gg\Lambda,\Theta,\Delta \right)=1$.
On the other hand, only for a ``broad'' lane the scaling functions
at the center of the chemical lane approach their limiting value
$\psistripe\left(\Lambda\to\infty,\Xi=0,\Theta,\Delta \right)=-1=
\omegastripe\left(\Lambda\to\infty,\Xi=0,\Theta,\Delta \right)$, corresponding
to the homogeneous case with $(a_{\stripe},b)$ BC.
\subsection{Derjaguin approximation}
%
Using the underlying assumption of additivity of the forces, within the DA ($\Delta\to0$) we find for the scaling functions
of the critical Casimir force and of the critical Casimir potential 
[see Appendix~\ref{app:stripe}]
\ifTwocolumn
\begin{multline}
  \label{eq:stripe-psi-da}
  \psistripe(\Lambda,\Xi,\Theta,\Delta\to0)=\\
  1+\psi_{(a_{\stripe}|a,b)}(\Xi+\Lambda,\Theta,\Delta\to0)-\psi_{(a_{\stripe}|a,b)}(\Xi-\Lambda,\Theta,\Delta\to0)
\end{multline} 
\else
\begin{equation}
  \label{eq:stripe-psi-da}
  \psistripe(\Lambda,\Xi,\Theta,\Delta\to0)=
  1+\psi_{(a_{\stripe}|a,b)}(\Xi+\Lambda,\Theta,\Delta\to0)-\psi_{(a_{\stripe}|a,b)}(\Xi-\Lambda,\Theta,\Delta\to0)
\end{equation} 
\fi
and
\ifTwocolumn
\begin{multline}
  \label{eq:stripe-omega-da}
  \omegastripe(\Lambda,\Xi,\Theta,\Delta\to0)=\\
  1+\omega_{(a_{\stripe}|a,b)}(\Xi+\Lambda,\Theta,\Delta\to0)-\omega_{(a_{\stripe}|a,b)}(\Xi-\Lambda,\Theta,\Delta\to0),
\end{multline}
\else
\begin{equation}
  \label{eq:stripe-omega-da}
  \omegastripe(\Lambda,\Xi,\Theta,\Delta\to0)=
  1+\omega_{(a_{\stripe}|a,b)}(\Xi+\Lambda,\Theta,\Delta\to0)-\omega_{(a_{\stripe}|a,b)}(\Xi-\Lambda,\Theta,\Delta\to0),
\end{equation} 
\fi
respectively.
Thus, within the DA, from the knowledge of the scaling functions $\psi_{(a_{\stripe}|a,b)}$ [\eref{eq:step-psi-da}]
and $\omega_{(a_{\stripe}|a,b)}$ [\eref{eq:step-omega-da}] for the chemical step with the appropriate BC, one can 
directly calculate the corresponding scaling functions for the chemical lane configuration.
Accordingly, in the limit $\Delta \rightarrow 0$ and for symmetry breaking BC, $\psistripe$ and $\omegastripe$ 
can be analytically calculated on the basis of Eqs.~\eqref{eq:stripe-psi-da} and~\eqref{eq:stripe-omega-da} 
by taking advantage of Eqs.~\eqref{eq:step-psi-da-crit}, \eqref{eq:step-omega-da-crit}, and \eqref{eq:erf}.
\subsection{Scaling function for the critical Casimir potential}
\begin{figure} 
  \includegraphics[width=8.4cm]{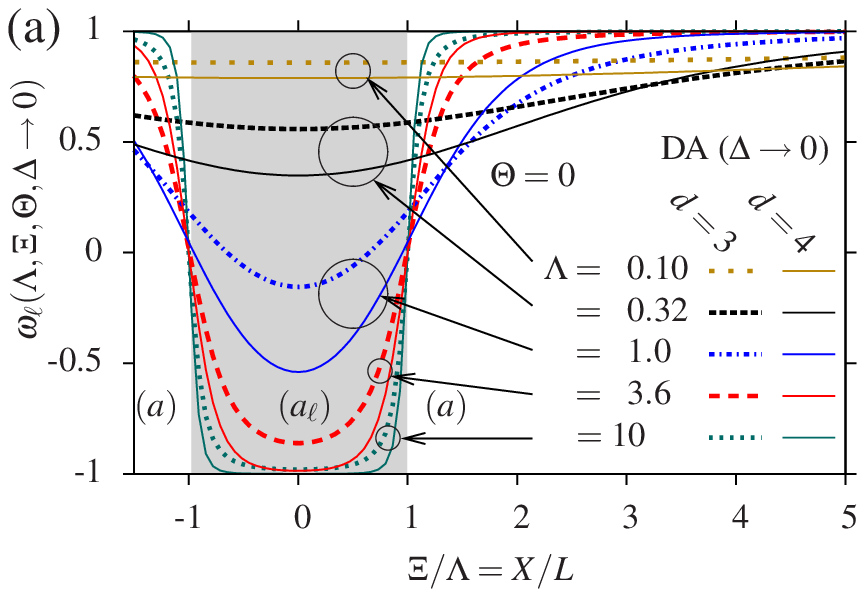}\\ 
  \includegraphics[width=8.4cm]{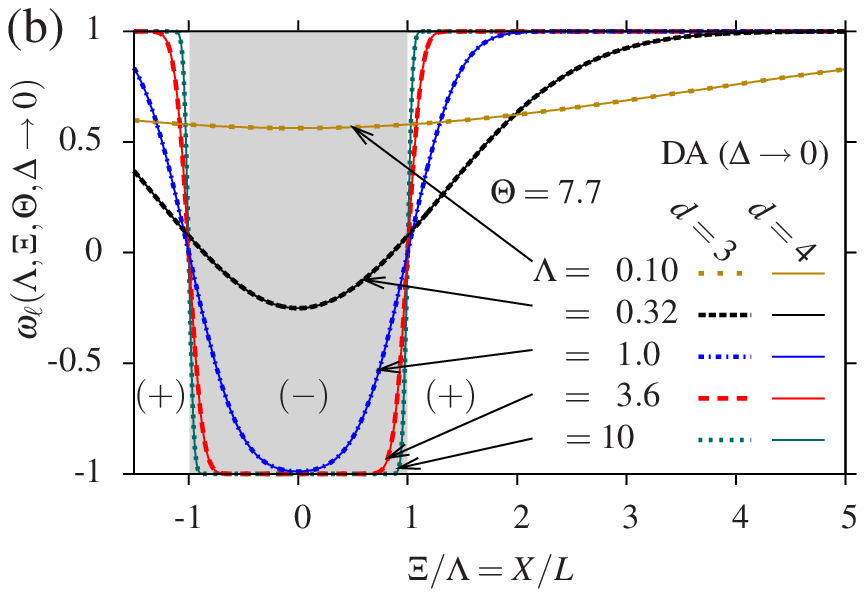} 
  \caption{%
    Scaling function $\omegastripe$ [\eref{eq:stripe-omega}] describing the lateral variation 
    of the critical Casimir potential of a colloid across a single chemical lane of width $2L$
    as a function of the lateral position $X$ of the colloid in units of the half width
    of the lane [see \fref{fig:sketch}; $\Xi=X/\sqrt{RD}$, $\Lambda=L/\sqrt{RD}$, $\Theta=D/\xi_+$].
    Here, $\omegastripe$ has been obtained within the DA ($\Delta \rightarrow 0$) 
    in $d=3$ and $4$ [\eref{eq:stripe-omega-da}].
    In (a) the curves correspond to $\Theta=0$ [\eref{eq:step-omega-da-crit}], whereas in (b) they correspond 
    to $\Theta=7.7$ and $a,a_{\stripe},b\in\{+,-\}$ BC [\fref{fig:sketch}].
    For $\Theta\gg1$ [(b)] the corresponding scaling functions obtained from Monte Carlo simulation
    data \cite{mcdata} in $d=3$ and from analytic MFT results \cite{krech:1997} in $d=4$ de
    facto coincide and their asymptotic expressions are given by Eqs.~\eqref{eq:erf} and
    \eqref{eq:stripe-omega-da}.
    $\omegastripe=1$ corresponds to the laterally homogeneous critical Casimir potential for $(a,b)$ BC
    outside the chemical lane, whereas $\omegastripe=-1$ corresponds to the value of the 
    critical Casimir potential for the homogeneous case with $(a_{\stripe},b)$ BC as within the chemical lane.
    For large values of $\Lambda$ the critical Casimir potential is the same as for two
    independent chemical steps, and $\omegastripe$ reaches its limiting value $-1$ in
    the center of the lane at $\Xi=0$ [see the main text].
    In (b), for $\Theta\gg1$, $\omegastripe$ attains $-1$ in the center of the chemical lane already 
    for smaller values of $\Lambda$ due to the exponential decay of the critical Casimir force.
    We note that the DA results for $\Theta=0$ (i.e., at the critical point) are independent of the actual
    boundary conditions which, accordingly, were not specified in (a).
    }   
    \label{fig:stripe}
\end{figure}
%
In \fref{fig:stripe}(a) we show the scaling function $\omegastripe$ for the critical
Casimir potential obtained within the DA for $d=3$ and $d=4$ (MFT) at the bulk critical point $T=T_c$
[Eqs.~\eqref{eq:stripe-omega-da} and \eqref{eq:step-omega-da-crit}]
for various values of $\Lambda=L/\sqrt{RD}$ as a function of the lateral coordinate of the colloid.
One can infer from \fref{fig:stripe} that, at bulk criticality,  the critical Casimir
potential varies less pronounced in $d=3$ than in $d=4$.
As expected, for small values of $\Lambda$ (i.e., ``narrow'' chemical lanes), the potential does not
reach the limiting homogeneous value $-1$ in the 
center of the chemical lane.
On the other hand for large values of the scaling variable $\Lambda$ (i.e., ``broad''
chemical lanes), $\omegastripe$ does attain the value $-1$ in the center of the 
chemical lane and the critical Casimir potential flattens.
In this case the potential is adequately described by two independent
chemical steps.
However, the criterion for being a sufficiently ``broad'' lane depends sensitively on $\Theta$ and $d$.
Indeed, from Eqs.~\eqref{eq:stripe-omega-da} and \eqref{eq:step-omega-da-crit} we find that at criticality ($\Theta=0$) the 
critical Casimir potential at the center of the chemical lane ($\Xi=0$) reaches the limiting value corresponding to the colloid facing a homogeneous substrate 
by up to $1\%$ for $\Lambda\gtrsim3.3$ in $d=4$ and for $\Lambda\gtrsim10$ in $d=3$.
We note that the curves in \fref{fig:stripe}(a) as well as these bounds 
are \emph{independent} of the actual
boundary conditions because for all kinds of BC the scaling function of the normal critical Casimir force is constant
at the critical point [see \eref{eq:delta-ab}].
\par
Below we shall discuss some properties which are specific for  BC with $a,a_{\stripe},b\in\{+,-\}$, 
which exhibit the feature that the normal critical Casimir force $\fmpm$ acting on two planar walls
decays \emph{purely} exponentially [see the text preceding \eref{eq:exponential-decay}] as a function of their distance 
expressed in units of the bulk correlation length
[see Eqs.~\eqref{eq:planar-force} and \eqref{eq:exponential-decay}].
In \fref{fig:stripe}(b) the scaling functions $\omegastripe$ in $d=3$ and $d=4$ obtained from Monte Carlo simulation data \cite{mcdata}
and analytic MFT results \cite{krech:1997}, respectively, within the DA 
[see Eqs.~\eqref{eq:stripe-omega-da} and \eqref{eq:erf}] are shown  for the same values of $\Lambda$ as in \fref{fig:stripe}(a)
but off criticality.
For $\Theta=7.7$ the curves for $d=3$ and $d=4$ are indistinguishable from each other and from their common
asymptotic expression given in \eref{eq:erf} [see also Ref.~\onlinecite{troendle:2009}].
For $\Theta\gg1$, the critical Casimir potential attains its limiting 
homogeneous value in the center of the lane for values of $\Lambda$ which are smaller than the ones for
$\Theta=0$ due to the shorter range of the forces.
That is, for both $d=3$ and $d=4$ the single chemical lane is almost equally well approximated by 
two independent chemical steps for $\Lambda\gtrsim1.5$ at $\Theta=3.3$ (data not shown) and for $\Lambda\gtrsim1.0$ at 
$\Theta=7.7$ [\fref{fig:stripe}(b)].
\par
\begin{figure} 
  \includegraphics{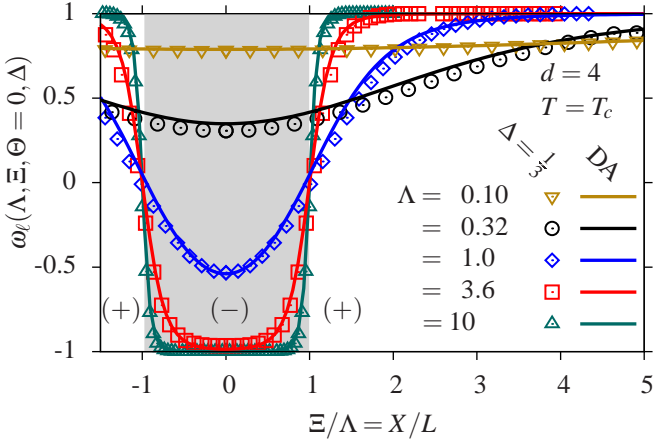} 
  \caption{%
    Test of the performance of the DA for the scaling function $\omegastripe$ [\eref{eq:stripe-omega}] 
    of the critical Casimir potential for a sphere with $\MBC$ BC close to
    a single chemical lane with $\MBC$ BC embedded in a substrate with $\PBC$ BC.
    The MFT $\omegastripe$ is evaluated at bulk criticality $\Theta=0$ in $d=4$ 
    both on the basis of the DA (lines, $\Delta\to0$) and of the full numerical MFT (symbols, $\Delta=1/3$).
    There is good agreement between the DA and the full MFT results, even for
    small values of $\Lambda=L/\sqrt{RD}$.
    Nonlinear effects, which are inherently present in the theory, do not strongly affect the potential.
    For $\Delta\to0$ the assumption of additivity of the critical Casimir forces underlying
    the DA is reliable even for small $\Lambda$.
    }   
  \label{fig:omega_cl}
\end{figure}
In \fref{fig:omega_cl} we compare the MFT $\omegastripe$ obtained within the DA ($\Delta\to0$) at $\Theta=0$ 
[Eqs.~\eqref{eq:stripe-omega-da} and \eqref{eq:step-omega-da-crit}] with the scaling function 
obtained from the full numerical MFT calculations for $\Delta=1/3$.
We find a rather good agreement even for small values of $\Lambda$ (i.e., ``narrow'' chemical lanes).
This shows that for the geometry of a colloid close to a single chemical lane, nonlinearities, which
are actually present in the critical Casimir effect and potentially invalidate the assumption of additivity 
underlying the DA, do not affect the resulting potential for small values of $\Delta$.
We expect this property to hold beyond MFT in $d=3$ as well, in particular off criticality,
i.e., for $\Theta\ne0$.
%
%
\section{Periodic chemical patterns ($\period$) \label{sec:period}}
In this section we consider a pattern of chemical stripes which are alternating \emph{periodically} 
along the $x$ direction.
The pattern consists of stripes of width $L_1$ with $(a_1)$ BC joined with stripes of width
$L_2$ with $(a_2)$ BC, such that the periodicity is given by $P=L_1+L_2$.
Thus, the geometry of the substrate pattern is characterized by the two variables $L_1$ and $P$ 
[see \fref{fig:sketch}].
The coordinate system is chosen such that the lateral coordinate $X$ of the center of the 
sphere is zero at the center of a $(a_1)$ stripe.
The normal critical Casimir force $F_{\period}$ acting on the colloidal particle and its corresponding potential 
$\Phi_{\period}$ take on the following scaling forms:
\begin{align} 
  \label{eq:period-force}
  \Fperiod(L_1,P,X,D,R,T)=&k_BT \frac{R}{D^{d-1}} \Kperiod(\lambda,\Pi,\Xi,\Theta,\Delta)\\
  \intertext{and}
  \label{eq:period-pot}
  \Phiperiod(L_1,P,X,D,R,T)=&k_BT\frac{R}{D^{d-2}} \varthetaperiod(\lambda,\Pi,\Xi,\Theta,\Delta),
\end{align} 
where $\Pi=P/\sqrt{RD}$ is the scaling variable characterizing the periodicity of the pattern and $\lambda=L_1/P$ is
the scaling variable chosen to correspond to the relative width of the stripe with $(a_1)$ BC.
$\Kperiod$ and $\varthetaperiod$ are universal scaling functions for the normal critical Casimir force and the
critical Casimir potential, respectively.
For $\lambda=1$ or $0$ the force and the potential correspond to the homogeneous cases with $(a_1,b)$ BC or $(a_2,b)$ BC,
respectively [see Sec.~\ref{sec:homog}].
As before it is useful to define scaling functions $\psiperiod$ and $\omegaperiod$ which 
vary for $\lambda\in[0,1]$ within the range $[-1,1]$ and describe the lateral behavior of the critical Casimir effect:
\ifTwocolumn
\begin{multline} 
  \label{eq:period-psi}
  \Kperiod(\lambda,\Pi,\Xi,\Theta,\Delta)=
    \frac{K_{(a_2,b)}+K_{(a_1,b)}}{2}\\
    +\frac{K_{(a_2,b)}-K_{(a_1,b)}}{2}     
    \psiperiod(\lambda,\Pi,\Xi,\Theta,\Delta)
\end{multline}
\else
\begin{equation} 
  \label{eq:period-psi}
  \Kperiod(\lambda,\Pi,\Xi,\Theta,\Delta)=
    \frac{K_{(a_2,b)}+K_{(a_1,b)}}{2}+
    \frac{K_{(a_2,b)}-K_{(a_1,b)}}{2}     
    \psiperiod(\lambda,\Pi,\Xi,\Theta,\Delta)
\end{equation}
\fi
and
\ifTwocolumn
\begin{multline} 
  \label{eq:period-omega}
  \varthetaperiod(\lambda,\Pi,\Xi,\Theta,\Delta)=
    \frac{\vartheta_{(a_2,b)}+\vartheta_{(a_1,b)}}{2}\\
    +\frac{\vartheta_{(a_2,b)}-\vartheta_{(a_1,b)}}{2}     
    \omegaperiod(\lambda,\Pi,\Xi,\Theta,\Delta).
\end{multline}
\else
\begin{equation} 
  \label{eq:period-omega}
  \varthetaperiod(\lambda,\Pi,\Xi,\Theta,\Delta)=
    \frac{\vartheta_{(a_2,b)}+\vartheta_{(a_1,b)}}{2}+
    \frac{\vartheta_{(a_2,b)}-\vartheta_{(a_1,b)}}{2}     
    \omegaperiod(\lambda,\Pi,\Xi,\Theta,\Delta).
\end{equation}
\fi
%
\subsection{Derjaguin approximation\label{sec:period-da}}
%
Taking advantage of the assumption of additivity of the forces  underlying the DA, one finds
for the scaling function of the normal critical Casimir force  in the limit $\Delta\to0$ [see Appendix~\ref{app:period}]
\ifTwocolumn
\begin{multline}
  \label{eq:period-psi-da}
  \psiperiod(\lambda,\Pi,\Xi,\Theta,\Delta\to0)=\\
  1+\sum_{n=-\infty}^{\infty}
  \left\{\psi_{(a_1|a_2,b)}(\Xi+\Pi(n+\tfrac{\lambda}{2}),\Theta,\Delta\to0)\right.\\
  \left.-\psi_{(a_1|a_2,b)}(\Xi+\Pi(n-\tfrac{\lambda}{2}),\Theta,\Delta\to0)\right\}.
\end{multline} 
\else
\begin{multline}
  \label{eq:period-psi-da}
  \psiperiod(\lambda,\Pi,\Xi,\Theta,\Delta\to0)=\\
  1+\sum_{n=-\infty}^{\infty}
  \left\{\psi_{(a_1|a_2,b)}(\Xi+\Pi(n+\tfrac{\lambda}{2}),\Theta,\Delta\to0)
  -\psi_{(a_1|a_2,b)}(\Xi+\Pi(n-\tfrac{\lambda}{2}),\Theta,\Delta\to0)\right\}.
\end{multline} 
\fi
Thus, the knowledge of the scaling function $\psi_{(a_1|a_2,b)}$ for a single chemical step with
the appropriate BC [Sec.~\ref{sec:step}] is sufficient to calculate directly the corresponding scaling 
function of the critical Casimir force acting on a colloid close to a periodic pattern of chemical stripes.
As expected, from \eref{eq:period-psi-da} one recovers the values 
${\psiperiod(\lambda=0,\Pi,\Xi,\Theta,\Delta)=1}$ 
and
${\psiperiod(\lambda=1,\Pi,\Xi,\Theta,\Delta)=-1}$, i.e., the cases of 
a colloid with $(b)$ BC facing a homogeneous substrate
with $(a_2)$ BC and $(a_1)$ BC, respectively [see Appendix~\ref{app:period}].
\par
In the limit $\Pi\to0$, i.e., for a pattern with a very fine structure compared to the size of the colloid,
the sum in \eref{eq:period-psi-da} turns into an integral [see Appendix~\ref{app:period}] and, as expected, $\psiperiod$ 
becomes independent of $\Xi$, i.e., of the lateral position of the colloid:
\begin{equation} 
  \label{eq:small-period}
  \psiperiod(\lambda,\Pi\to0,\Xi,\Theta,\Delta\to0)=1-2\lambda.
\end{equation} 
Accordingly, in the limit $\Pi\to0$ the force acting on the colloid -- within the DA -- is the average 
of the ones corresponding to the two boundary conditions weighted by the corresponding relative stripe width
[see Eqs.~\eqref{eq:small-period} and \eqref{eq:period-psi}]:
\ifTwocolumn
\begin{multline} 
  \label{eq:small-period-K}
  \Kperiod(\lambda,\Pi\to0,\Xi,\Theta,\Delta\to0)=\\
  \frac{L_1}{L_1+L_2} K_{(a_1,b)}(\Theta,\Delta\to0)+\frac{L_2}{L_1+L_2}K_{(a_2,b)}(\Theta,\Delta\to0).
\end{multline} 
\else
\begin{equation} 
  \label{eq:small-period-K}
  \Kperiod(\lambda,\Pi\to0,\Xi,\Theta,\Delta\to0)=
  \frac{L_1}{L_1+L_2} K_{(a_1,b)}(\Theta,\Delta\to0)+\frac{L_2}{L_1+L_2}K_{(a_2,b)}(\Theta,\Delta\to0).
\end{equation} 
\fi
\par
For the scaling function of the critical Casimir potential the results 
are completely analogous to Eqs.~\eqref{eq:period-psi-da}--\eqref{eq:small-period-K}
[see Appendix~\ref{app:period}].
%
\subsection{Scaling function for the normal critical Casimir force}
%
\begin{figure} 
  \includegraphics{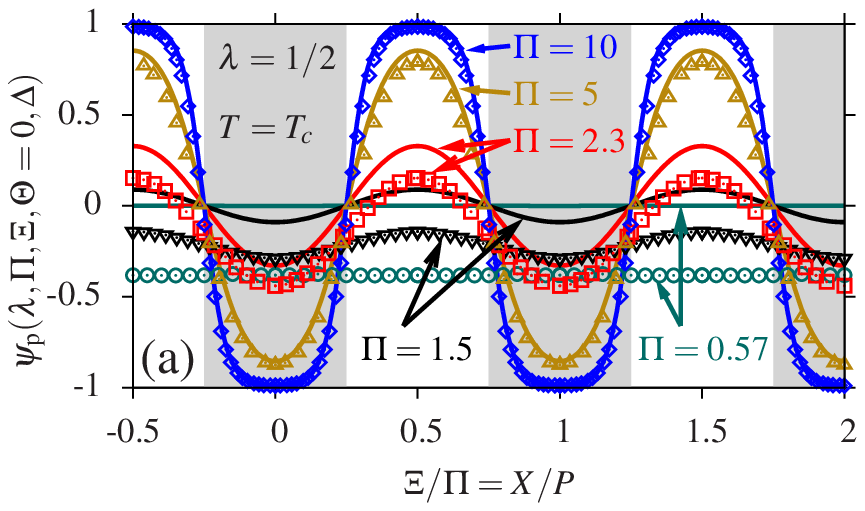}\\
  \includegraphics{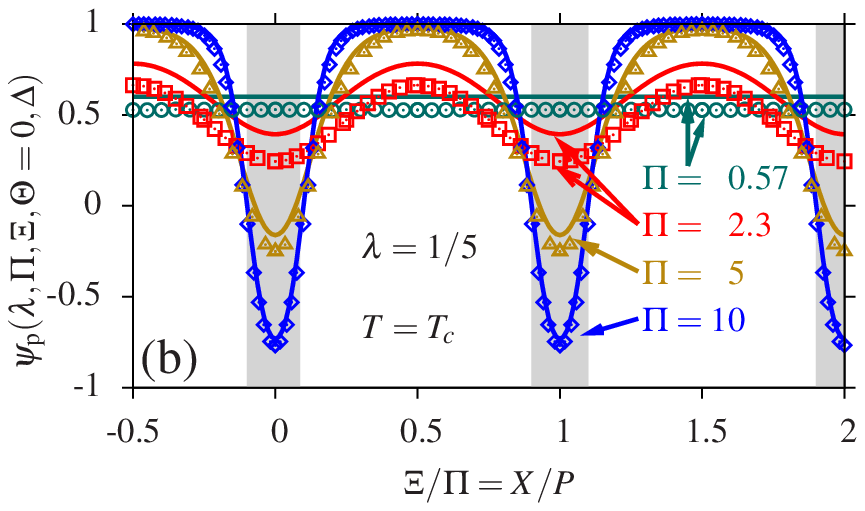}
  \caption{%
  MFT ($d=4$) scaling function $\psiperiod$ [\eref{eq:period-psi}] of the normal critical Casimir
  force acting on a colloidal sphere with $(b)=\MBC$ BC which is close to a periodically patterned
  substrate [\fref{fig:sketch}] with $(a_1)=\MBC$ BC on one kind of stripes [shaded areas]
  and $(a_2)=\PBC$ BC on the other kind of stripes.
  Due to this choice of the BC the colloid is attracted by the shaded stripes and repelled by the others.
  $\psiperiod$ is shown as a function of the lateral position of the colloid $X/P$ with $P=L_1+L_2$
  and at the bulk critical point $\Theta=0$.
  The geometry of the pattern is characterized by $\Pi=P/\sqrt{RD}$ and $\lambda=L_1/P$, for which
  we have chosen the values (a) $\lambda=0.5$ and (b) $\lambda=0.2$. 
  The lines are the results for $\psiperiod$ as obtained within the DA for $d=4$
  [Eqs.~\eqref{eq:period-psi-da} and \eqref{eq:step-psi-da-crit}], whereas the symbols represent the full
  numerical data obtained within MFT for $\Delta=1/3$ for various values of $\Pi$.
  For patterns which are finely structured on the scale of the colloid size, i.e., $\Pi\lesssim2$, 
  the actual results   deviate from the approximate ones obtained within the DA due to the strong 
  influence (in this context) of the inherent nonlinear effects.
  }   
  \label{fig:period_normal}
\end{figure}
%
\begin{figure} 
  \includegraphics{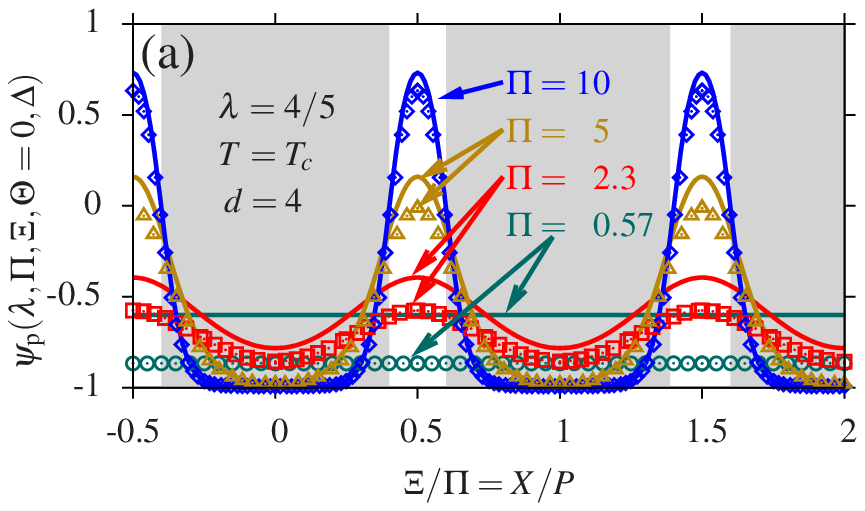}\\
  \includegraphics{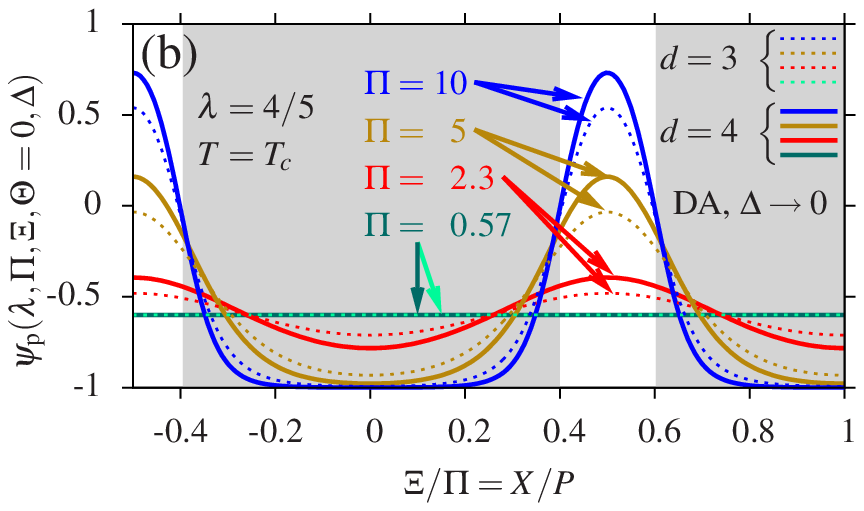}
  \caption{%
    (a) The same as in \fref{fig:period_normal}, but for $\lambda=0.8$.
    Also in this case, the DA turns out to be accurate for $\Pi\gtrsim2$ while it fails to describe 
    quantitatively the full numerical data for smaller values of $\Pi$.
    (b) Comparison between the scaling functions $\psiperiod$ in $d=3$ (dotted lines) and $d=4$
    (solid lines), at $T=T_c$, for $\lambda=0.8$, and within the DA. 
    At the critical point the expression for this scaling function $\psiperiod$ is known analytically
    [see Eqs.~\eqref{eq:period-psi-da} and \eqref{eq:step-psi-da-crit}],
    and the corresponding plot presented here shows that the lateral variation of the normal critical Casimir 
    force is less pronounced in $d=3$ than in $d=4$.
    (We note that for $\Pi\to0$ we expect that also in $d=3$ the DA fails to describe quantitatively 
    the actual behavior; however, we nonetheless present the curve for $\Pi=0.57$ in order to show that the 
    critical Casimir force obtained within the DA practically does not change laterally for such small values of $\Pi$.)
  }   
  \label{fig:normal_3d}
\end{figure}
Figure~\ref{fig:period_normal} shows the scaling function $\psiperiod$ [\eref{eq:period-psi}] as a function of $\Xi/\Pi=X/P$,
describing the lateral variation of the normal critical Casimir force at $\Theta=0$ as obtained within the DA 
for $d=4$ [\eref{eq:period-psi-da} with \eref{eq:step-psi-da-crit}; solid lines] compared with the one obtained from the full 
numerical MFT calculation [$\Delta=1/3$; symbols] for symmetry breaking boundary conditions 
$(a_1)=\MBC$, $(a_2)=\PBC$, and $(b)=\MBC$ [\fref{fig:sketch}].
From this comparison for $\lambda=0.5$ [\fref{fig:period_normal}(a)] and $\lambda=0.2$ [\fref{fig:period_normal}(b)] 
and for various values of $\Pi$ one
can infer that for $\Delta\to0$ and $\Pi\gg1$, i.e., $L_1+L_2\gg\sqrt{RD}$ the DA describes well the actual 
behavior of the scaling function, even if the force scaling function does not attain its limiting
homogeneous values $\psiperiod=\pm1$
in the center of the stripes.
However, for $\Pi\lesssim2$ (in $d=4$ at $T=T_c$) the DA does not quantitatively describe
the actual behavior and the scaling function $\psiperiod$ obtained from the full numerical MFT calculations
deviates from the one obtained within the DA.
Within both the DA and the full numerical MFT calculation, for $\Pi\to0$ the normal critical Casimir force 
loses its lateral dependence on $\Xi$.
But from the full numerical calculation we find that the corresponding constant 
value which is attained by $\psiperiod$ differs from the one obtained within DA [\eref{eq:small-period}].
This shows that for small periodicities $P\lesssim\sqrt{RD}$ nonlinearities inherent in the 
critical Casimir effect strongly affect the resulting scaling functions of the force and the potential,
so that in this respect the assumption of additivity of the force and thus the use of the DA are not justified.
\par
Figure \ref{fig:normal_3d}(a) shows the same comparison as  \fref{fig:period_normal} but for $\lambda=0.8$, which 
corresponds to an areal occupation of 80\% of the substrate surface with $\MBC$ BC and 20\% with $\PBC$ BC.
Due to the fact that at the critical point $\psi_{(a_1|a_2,b)}(\Xi,\Theta=0,\Delta\to 0)$ is actually 
independent of the BC, $\psiperiod(\lambda=0.8,\Pi,\Xi,\Theta=0,\Delta\to 0)$ in \fref{fig:normal_3d}(a) is, 
within the DA, complementary to the one for $\lambda = 0.2$ in \fref{fig:period_normal}(b), i.e., it is obtained from the 
latter by a reflection with respect to $\psiperiod=0$ followed by a shift in $\Xi/\Pi$ of $0.5$.
Instead, the full numerical data in \fref{fig:normal_3d}(a) and \fref{fig:period_normal}(b) show a different 
behavior as they clearly tend to assume the value $-1$ corresponding to the homogeneous case with $(-,-)$ BC.
By contrast, for the case $\lambda=0.2$ shown in \fref{fig:period_normal}(b), 
the full numerical data do not reach as closely the value $+1$ corresponding to $(+,-)$ BC, although the 
substrate area is covered by 80\% with $\PBC$ BC.
This feature is addressed in more detail in Sec.~\ref{sec:cylinder}.
Figure \ref{fig:normal_3d}(b) compares the scaling function $\psiperiod$ of the normal critical Casimir force
at $T=T_c$ and for $\lambda=0.2$ as obtained within the DA for $d=4$ (solid lines) with the corresponding one for $d=3$
(dotted lines).
At $T=T_c$, $\psiperiod$ is determined by Eqs.~\eqref{eq:period-psi-da} and \eqref{eq:step-psi-da-crit} 
from which one can infer that the lateral variation of the normal Casimir force is less pronounced for $d=3$ than for $d=4$.
This qualitative feature holds for all values of $\lambda$ (not shown).
However, off criticality, $\Theta\gg1$, [according to Eqs.~\eqref{eq:period-psi-da} and \eqref{eq:erf}] the  DA
scaling functions both for $d=3$ as obtained from MC simulation data and for $d=4$ as obtained from MFT de facto coincide
(not shown), similarly to the case of a single chemical lane in \fref{fig:stripe}(b).
\par
\begin{figure} 
  \includegraphics{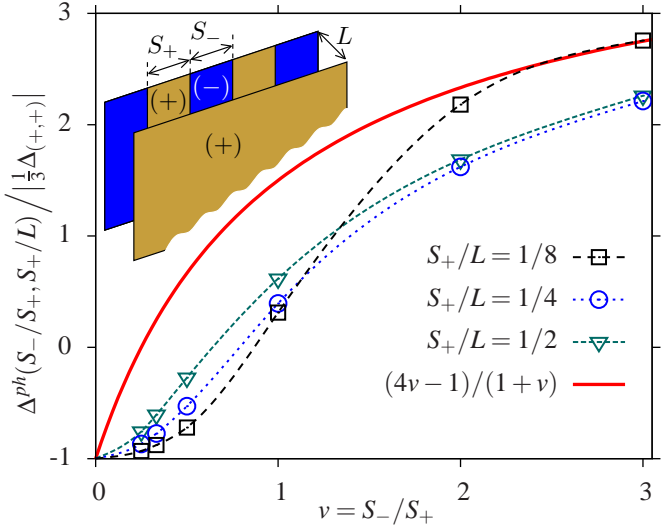}
  \caption{
    Normalized scaling function $\Delta^{ph}$ of the critical Casimir force at criticality acting on a homogeneous \emph{planar}
    wall with $(+)$ BC opposite to a periodically patterned planar substrate with stripes of alternating $(+)$ and $(-)$ 
    BC as a function of $v=S_-/S_+$, where $S_+$ and $S_-$ are the respective widths.  
    The symbols correspond to the MFT ($d=4$) data presented in Fig.~12 of Ref.~\onlinecite{sprenger:2006} for various
    values of $S_+/L$ (note that $\Delta_0^{++}=\Dpp/(d-1)$ in Fig.~12 of Ref.~\onlinecite{sprenger:2006}).
    The dashed and dotted lines which join the data points are a guide to the eye.
    The solid line corresponds to the DA result given in \eref{eq:kraft-da-delta-fin} which assumes additivity of the forces and
    turns out to be independent of the ratio $S_+/L$.
    One can immediately infer from the graph that here the assumption of additivity is not justified,
    which is the limiting configuration of the sphere-wall geometry for $\Pi\to0$.
  }
  \label{fig:kraft}
\end{figure}
Although one would expect the DA to be valid for large radii $R$, the lateral variation of the boundary 
conditions at the surface of the patterned substrate on a scale $P \lesssim \sqrt{R D}$ -- corresponding to the limit $\Pi\to0$ --  
renders the DA less accurate, as it clearly emerges from the numerical data presented in 
Figs.~\ref{fig:period_normal} and \ref{fig:normal_3d}.
The fact that a large colloid radius $R$ does not guarantee the validity of the DA can be understood by
noting that such a discrepancy between the full numerical calculation and the result of the DA approximation 
already emerges in the \emph{film} geometry (formally corresponding to the limit $R\to\infty$),
i.e., for a chemically {\it p}atterned wall opposite to a laterally {\it h}omogeneous flat wall.
This ``$ph$'' configuration has been studied in Ref.~\onlinecite{sprenger:2006} within MFT for laterally 
alternating chemical stripes of width $L_1=S_+$ and $L_2=S_-$ with $(+)$ and $(-)$ BC, respectively, 
opposite to a homogeneous substrate with $(+)$ BC a distance $L$ apart
[see \fref{fig:sketch} and the inset of \fref{fig:kraft}].
Indeed, by using the assumption of additivity of the critical Casimir forces underlying the DA and 
neglecting edge effects, the normal critical Casimir
force $f^{ph}_{\rm (DA)}(S_+,S_-,L,T)$ per unit area acting on the walls is predicted to be given by
\ifTwocolumn
\begin{multline} 
  \label{eq:kraft-da}
  f^{ph}_{\rm (DA)}(S_+,S_-,L,T)=\\
  \frac{S_+}{S_+ + S_-}\fpp(L,T)+\frac{S_-}{S_+ + S_-}\fpm(L,T),
\end{multline} 
\else
\begin{equation} 
  \label{eq:kraft-da}
  f^{ph}_{\rm (DA)}(S_+,S_-,L,T)=\frac{S_+}{S_+ + S_-}\fpp(L,T)+\frac{S_-}{S_+ + S_-}\fpm(L,T),
\end{equation} 
\fi
where $f_{(+,\pm)}$ refer to homogeneous parallel walls, as in \eref{eq:planar-force}.
At the bulk critical point the critical Casimir force is given in general by \cite{sprenger:2006}
\begin{equation} 
  \label{eq:kraft-def}
  f^{ph}(S_+,S_-,L,T=T_c)=k_B T_c \frac{d-1}{L^d}\Delta^{ph}\left(v=\frac{S_-}{S_+},\frac{S_+}{L}\right).
\end{equation} 
Using \eref{eq:kraft-da} together with Eqs.~\eqref{eq:planar-force} and \eqref{eq:delta-ab} one finds within the DA that
\begin{equation} 
  \label{eq:kraft-da-delta}
  (d-1)\Delta^{ph}_{\rm (DA)}\left(v,\frac{S_+}{L}\right)=\frac{v\Dpm+\Dpp}{1+v},
\end{equation} 
which renders the rhs of \eref{eq:kraft-da-delta} to be independent of the scaling variable $S_+/L$.
Within MFT as studied in Ref.~\onlinecite{sprenger:2006} ($d=4$), one has $\Dpm=-4\Dpp>0$ [see the end of Sec.~\ref{sec:MFT}] 
so that 
\begin{equation} 
  \label{eq:kraft-da-delta-fin}
  \Delta^{ph}_{\rm (DA)}\left(v,\frac{S_+}{L}\right)=\frac{|\Dpp|}{3}\;\frac{4v-1}{1+v}.
\end{equation} 
In \fref{fig:kraft} we show the comparison between the actual scaling function $\Delta^{ph}$ 
(data points, obtained numerically as reported in Fig.~12 of Ref.~\onlinecite{sprenger:2006}) 
and $\Delta^{ph}_{\rm (DA)}$ (\eref{eq:kraft-da-delta-fin}, solid line) derived by assuming 
additivity of the forces and neglecting edge effects. 
Figure \ref{fig:kraft} clearly shows that the actual behavior of the critical Casimir force
in the film geometry is not properly predicted within these assumptions.
This is expected to be due to the presence of nonlinear effects and of edge effects in this context.
This explains why in the limit $\Pi\to0$ the DA ($R\gg D$) used here does not capture the 
behavior of the critical Casimir force acting on a colloid close to periodically patterned substrate.
\begin{figure} 
  \ifTwocolumn
    \includegraphics[width=7.8cm]{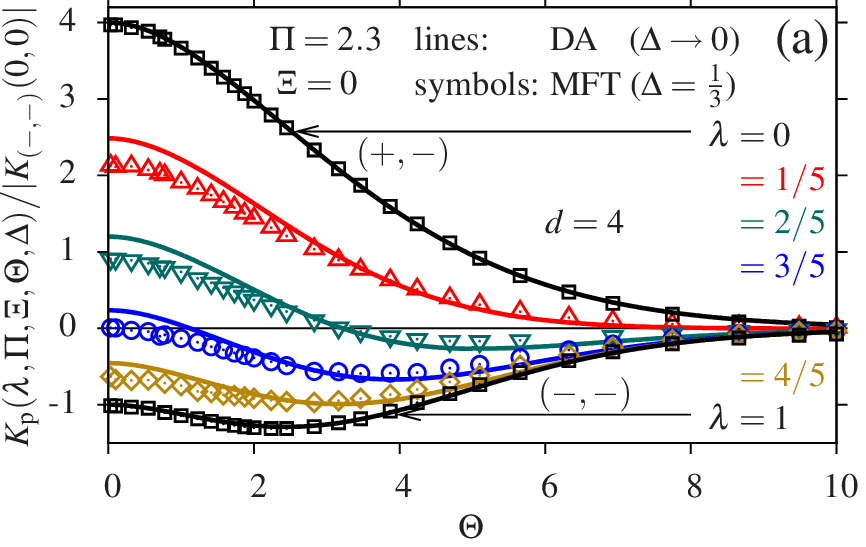}\\
    \includegraphics[width=7.8cm]{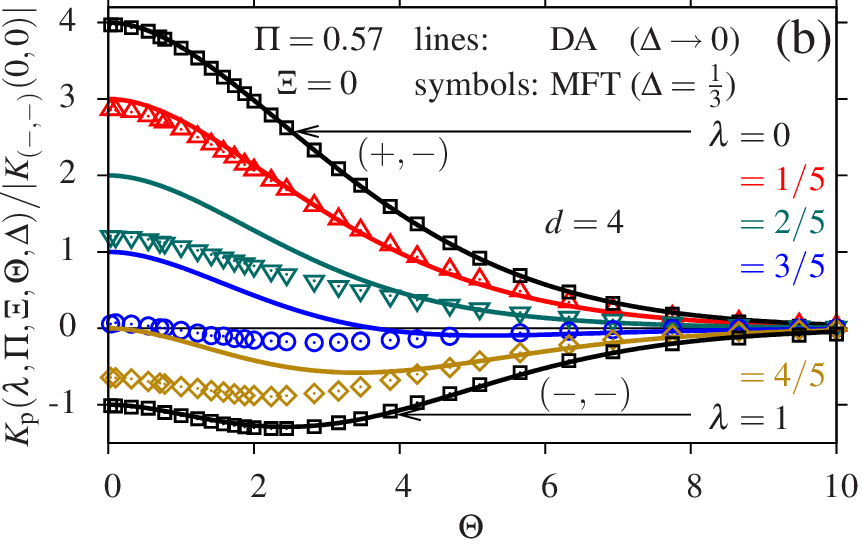}\\
    \includegraphics[width=7.8cm]{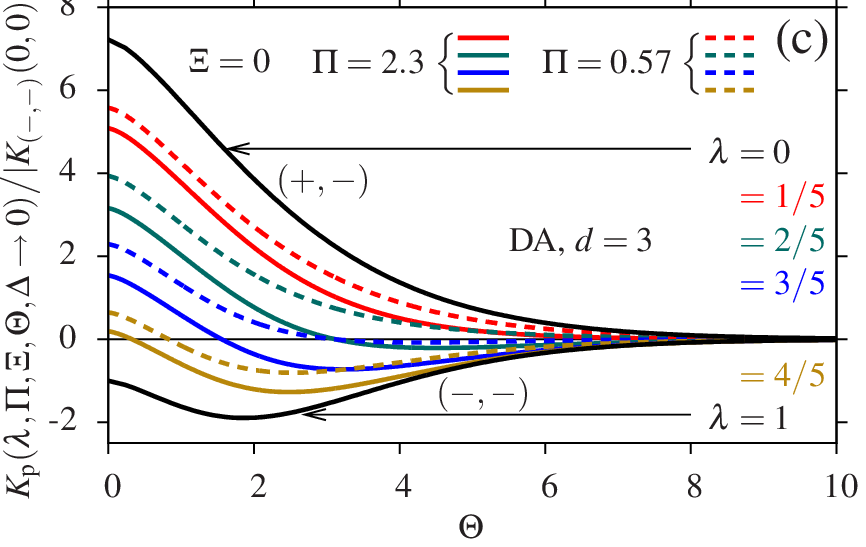}
  \else
  \hspace*{-1.6cm}
  \parbox{18.4cm}{
    \parbox{9.7cm}{
      \includegraphics{comparesystem_b}\\
      \includegraphics{comparesystem_d}}
    \parbox{8.5cm}{
      \includegraphics{period_3d}
      }
      }
  \fi
  \caption{%
    Scaling function $\Kperiod$ [\eref{eq:period-force}] of the normal critical Casimir force acting on a 
    spherical colloid with $\MBC$ BC located at $X=0$  ($\Xi=X/\sqrt{RD}$) close to a periodically chemically patterned 
    substrate [see \fref{fig:sketch}].
    $\Kperiod$ is suitably normalized by the absolute value of the force scaling
    function $\Kmm(0,0)=2\pi\Dmm/(d-1)$ for the homogeneous $(-,-)$ case at criticality and within the DA [Sec.~\ref{sec:homog-da}].
    The lateral position of the center of the colloid is fixed at the center of a stripe with $(a_1)=(-)$ BC
    and width $L_1=\lambda P$, which it is attracted to, in contrast to the second type of stripes with 
    $(a_2)=(+)$ BC and width $L_2=(1-\lambda)P$, which it is repelled from.
    The scaling variable corresponding to the periodicity of the substrate pattern is (a) $\Pi=P/\sqrt{RD}=2.7$ 
    and (b) $\Pi=0.57$, whereas the relative area fraction of the $\MBC$ stripes changes from
    $\lambda=L_1/(L_1+L_2)=0$ to $\lambda=1$ (top to bottom: fully repulsive to fully attractive).
    In (a) and (b) the lines represent the result for the MFT critical Casimir force within the DA 
    [$\Delta\to0$, $d=4$, see \eref{eq:period-psi-da}], whereas the symbols represent the full numerical 
    MFT data obtained for $\Delta=1/3$.
    The DA agrees reasonably well with the full data
    for $\Pi=2.3$ [(a)] and $\Theta\gtrsim1$, but for $\Pi=0.57$ [(b)] it fails to describe the actual behavior
    within the ranges $\Theta\lesssim4$ and $0.3\lesssim\lambda\lesssim0.9$ where the nonlinear 
    effects strongly affect the resulting scaling function.
    In (c) $\Kperiod$ is shown for $\Pi=0.57$ and $2.3$, as obtained for $d=3$ within the DA on the basis of the 
    Monte Carlo simulation data for the film geometry \cite{mcdata}.
    (We note, however, that we do not expect that the curves shown for $\Pi=0.57$ are quantitatively reliable.)
    }   
  \label{fig:comparesystem}
\end{figure}
\par
In \fref{fig:comparesystem} we show the behavior of scaling function $\Kperiod$ [\eref{eq:period-force}]
of the normal critical Casimir force acting on the colloid in $d=4$ with $(b)=\MBC$ BC as a function of 
$\Theta=D/\xi_+$ (i.e., as a function of the normal distance of the colloid from the substrate in units of the 
bulk correlation length) and for various values of $\lambda$ and $\Pi$.
In \fref{fig:comparesystem} the scaling function 
$\Kperiod$ is evaluated at $X=0$ [see \fref{fig:sketch}] which corresponds to 
the most preferred lateral position of the colloid in which the normal force is least repulsive or most attractive
[see \fref{fig:period_normal}].
From \fref{fig:comparesystem} one can infer that the DA does not provide an accurate estimate of
$\Kperiod$ in the whole range of $\Theta$ for $\Pi=0.57$ [panel (b)], whereas it does so for $\Pi=2.3$ [panel (a)].
Indeed, for $\Pi=0.57$ the discrepancy between the DA and the numerical data is already significant for $\Theta\lesssim4$ and $0.3\lesssim\lambda\lesssim0.9$, whereas for $\Pi=2.3$ agreement is found
 for all values of $\lambda$ except for $\Theta\lesssim1$ [\fref{fig:comparesystem}(a)].
This fact suggests that for relatively small periodicities $\Pi\lesssim2$ non-additive and edge effects become important.
On the other hand, for large values of $\Theta\gg1$ the DA describes the behavior of $\Kperiod$ rather well 
for all values of $\Pi$ due to the exponential decay of the critical Casimir force for $\Theta\gg1$ 
[\eref{eq:exponential-decay}].
Figure~\ref{fig:comparesystem}(c) shows the scaling function $\Kperiod$ for $d=3$ within the DA
as obtained from Monte Carlo simulation data for the film geometry \cite{mcdata}.
The qualitative features of the behavior of $\Kperiod$ in $d=3$ and $d=4$ are similar.
\par
From our analysis in $d=4$ we conclude that the DA describes quantitatively well the behavior of the actual critical
Casimir force for $\Pi\gtrsim2$ for all values of $\Theta$.
For smaller values of $\Pi$, the DA is only quantitatively reliable for large values of $\Theta$ (at which
the force decays exponentially).
For example, for $\Pi\gtrsim0.5$ the DA result is quantitatively correct for $\Theta\gtrsim4$. 
We expect these properties to be carried over to $d=3$.
%
\subsection{Critical Casimir levitation \label{sec:levitation}}
%
Rather remarkably, within a certain range of values of $\lambda$, $\Kperiod$ \emph{changes sign} as a function of 
$\Theta = D/\xi_+$ [\fref{fig:comparesystem}].
In this context it is convenient to introduce for later purposes another scaling variable 
$\Psi=\Pi|\Theta|^{1/2}=P/\sqrt{R\xi_\pm}$ which is independent of $D$ and therefore does not vanish in the DA limit 
$D\ll R$ (i.e., $\Delta\to0$).
Due to this change of sign of $\Kperiod$, there exists a certain value $\Theta = \Theta_0(\Psi,\lambda,\Xi,\Delta)$ 
at which the normal critical Casimir force $\Fperiod$ acting on the colloid vanishes. 
This implies that in the absence of additional forces the colloid levitates at a height $D_0$ determined by 
$\Theta_0$ and $\xi_+$, which can be tuned by changing the temperature.
Since for fixed geometrical parameters $R$, $X$, and $P$ the scaling variables $\Theta$, $\Pi$, $\Xi$, and $\Delta$ depend on $D$,  
one has to consider the behavior of $\Fperiod$ as a function of $D$ near $D_0$ in order to assess whether the levitation is stable
against perturbations of $D$ or not. 
Stability requires $\partial_D \Fperiod|_{D=D_0}<0$ (so that for $D<D_0$ the colloid is repelled from the
patterned substrate, whereas for for $D>D_0$ it is attracted).
According to \eref{eq:period-force} one has
\ifTwocolumn
\begin{multline} 
  \label{eq:levitation-1}
  \partial_D \Fperiod=
  k_BT \frac{R}{D^d}\times\left\{-(d-1)\right.\\\left.-\tfrac{1}{2}\Pi\partial_\Pi-\tfrac{1}{2}\Xi\partial_\Xi+\Theta\partial_\Theta
  +\Delta\partial_\Delta\right\}\Kperiod(\lambda,\Pi,\Xi,\Theta,\Delta).
\end{multline} 
\else
\begin{equation} 
  \label{eq:levitation-1}
  \partial_D \Fperiod=k_BT \frac{R}{D^d}\left\{-(d-1)-\tfrac{1}{2}\Pi\partial_\Pi-\tfrac{1}{2}\Xi\partial_\Xi+\Theta\partial_\Theta
  +\Delta\partial_\Delta\right\}\Kperiod(\lambda,\Pi,\Xi,\Theta,\Delta).
\end{equation} 
\fi
The laterally preferred position is always at $X=X_0=0$, corresponding to $\Xi=\Xi_0=0$, 
so that within the DA ($\Delta\to0$) one has 
\ifTwocolumn
\begin{multline} 
  \label{eq:levitation-2}
  \sgn\left(\partial_D \Fperiod\big|_{D=D_0,X=X_0,{\rm DA}}\right)=\\
  \sgn\left(\left\{-\tfrac{1}{2}\Pi\partial_{\Pi}+\Theta\partial_\Theta\right\}
      \Kperiod(\lambda,\Pi,\Xi=0,\Theta,\Delta\to 0)\big|_{\Theta=\Theta_0}\right),
\end{multline} 
\else
\begin{equation} 
  \label{eq:levitation-2}
  \sgn\left(\partial_D \Fperiod\big|_{D=D_0,X=X_0,{\rm DA}}\right)=
  \sgn\left(\left\{-\tfrac{1}{2}\Pi\partial_{\Pi}+\Theta\partial_\Theta\right\}
      \Kperiod(\lambda,\Pi,\Xi=0,\Theta,\Delta\to 0)\big|_{\Theta=\Theta_0}\right),
\end{equation} 
\fi
where we have used the implicit equation $\Fperiod|_{D=D_0}=0$ so that $\Kperiod|_{D=D_0}=0$.
(Equation~\eqref{eq:levitation-2} assumes that $\partial_\Delta\Kperiod$ does not diverge $\propto\Delta^{-1}$ for $\Delta\to0$.)
In the following we only consider $\Theta\ge0$ and BC $(a_1)=(-)$, $(a_2)=(+)$, and $(b)=(-)$.
\par
Within the DA we find that both $\partial_{\Pi}\Kperiod|_{\Theta=\Theta_0,\Xi=\Xi_0}$ and 
$\partial_{\Theta}\Kperiod|_{\Theta=\Theta_0, \Xi=\Xi_0}$ 
are negative, so that according to \eref{eq:levitation-2} the sign of $\partial_D\Fperiod|_{D=D_0,X=X_0,{\rm DA}}$ 
can vary and depends on their values as well as on $\Theta_0$ and $\Pi$.
However, at criticality ($\Theta=0$) the second term of the rhs of \eref{eq:levitation-2} vanishes.
Thus, at the bulk critical point $T=T_c$ the derivative $\partial_D\Fperiod$ evaluated at $D=D_0$ and $X=X_0=0$
is always positive so that one cannot achieve stable levitation.
On the other hand, for ${\Theta>0}$ it is always possible to find geometrical configurations for which the colloid
exhibits stable levitation, as described in the following.
%
%
\begin{figure} 
  \includegraphics{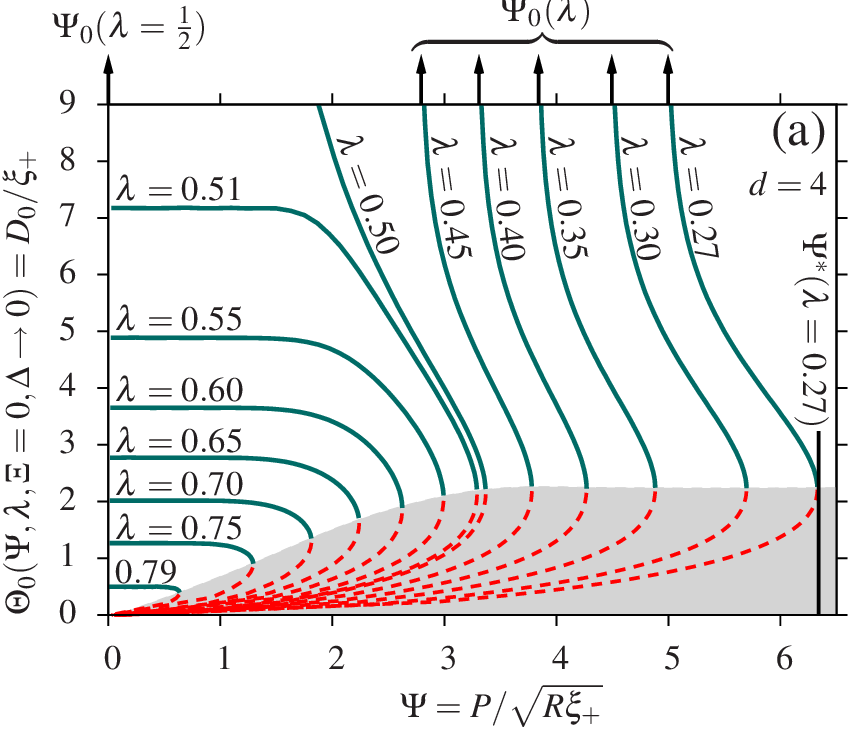}\\[2mm]
  \includegraphics{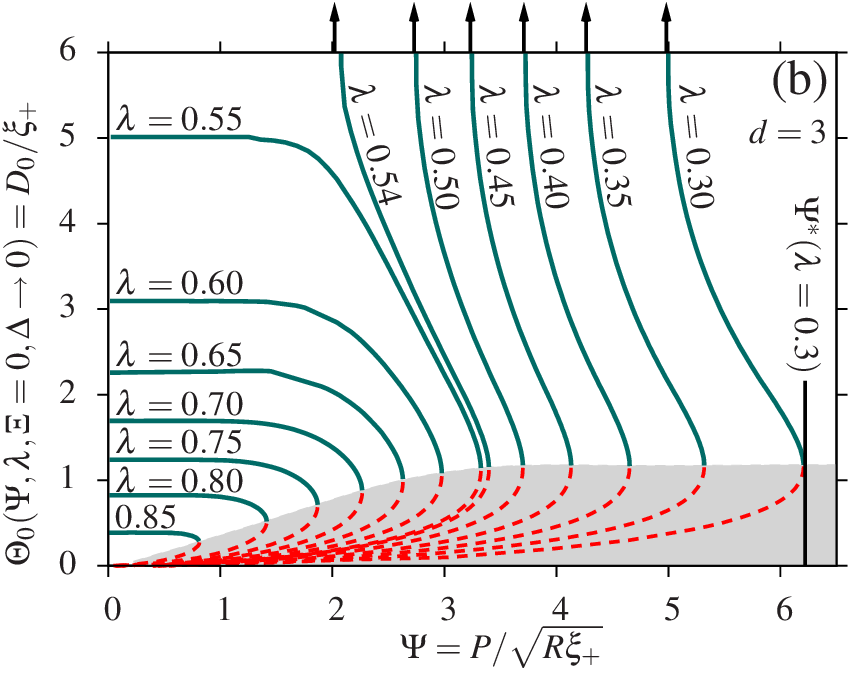}
  \caption{%
    Values of the scaling variable $\Theta_0$ at which within the DA ($\Delta\to0$) the normal critical Casimir force $\Kperiod$ 
    shown in  \fref{fig:comparesystem} vanishes as a function 
    of $\Psi$ for (a) $d=4$ and (b) $d=3$ on the basis of Monte Carlo simulation data \cite{mcdata} 
    and for various values of $\lambda=L_1/P$.
    The solid lines correspond to values of $\Theta_0$ for which the levitation of the colloid at a height $D_0$ above the substrate is 
    \emph{stable} against perturbations of $D$ [$\partial_D\Fperiod|_{D=D_0}<0$,
    see \eref{eq:levitation-2}].
    The shaded region and the dashed lines indicate those values of $\Theta_0$ for which 
    $\partial_D\Fperiod|_{D=D_0}>0$ and thus do \emph{not} correspond to stable levitation.
    For $\lambda>\lambda_0$ with $\lambda_0(d=4)=4/5$ and $\lambda_0(d=3)\simeq0.88$, $\Theta_0$ ceases to
    exist, i.e., $\Kperiod$ does not exhibit a zero. 
    For $\lambda<\lambda_1$ with $\lambda_1(d=4)=1/2$ and $\lambda_1(d=3)\simeq0.545$, 
    $\Theta_0(\Psi\searrow\Psi_0(\lambda))$ diverges.
    (The values for $\Psi_0(\lambda)$ are indicated by upward arrows.)
    For any $\lambda<\lambda_0$, $\Theta_0$ exists for $\Psi<\Psi^*(\lambda)$.
    (From the analysis in \fref{fig:comparesystem} we expect the DA to be quantitatively reliable only 
    for $\Psi\gtrsim2\sqrt{\Theta_0}$ for $\Theta_0\lesssim4$
    and for $\Psi\gtrsim0.5\sqrt{\Theta_0}$ for $\Theta_0\gtrsim4$, which implies
    $\lambda\lesssim0.7$ in $d=3$ and $\lambda\lesssim0.6$ in $d=4$.)
  }   
  \label{fig:levitation}
\end{figure}
%
%
%
\par
\begin{figure} 
  \ifTwocolumn
    \includegraphics[width=6.5cm]{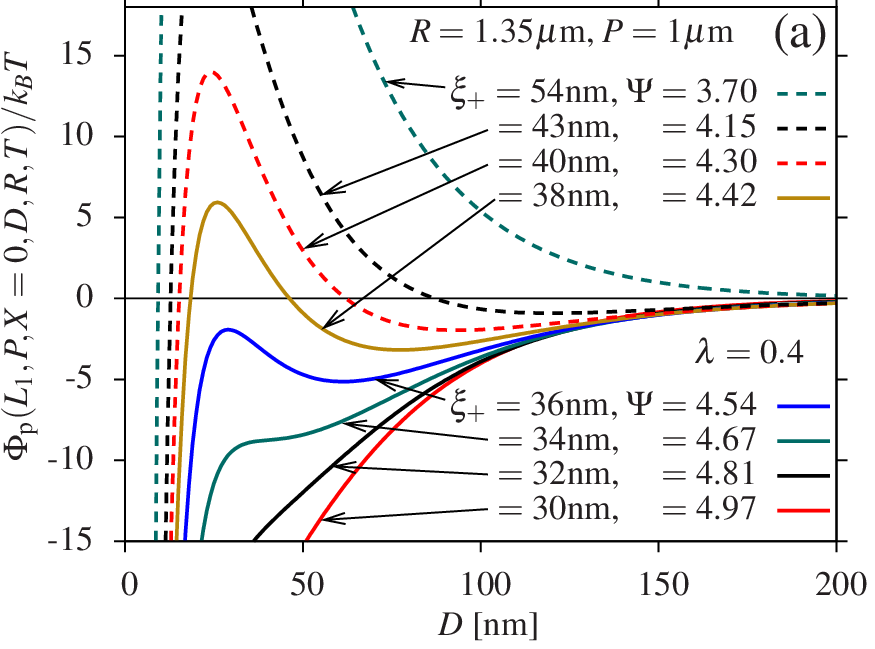}\\[1mm]
    \includegraphics[width=6.5cm]{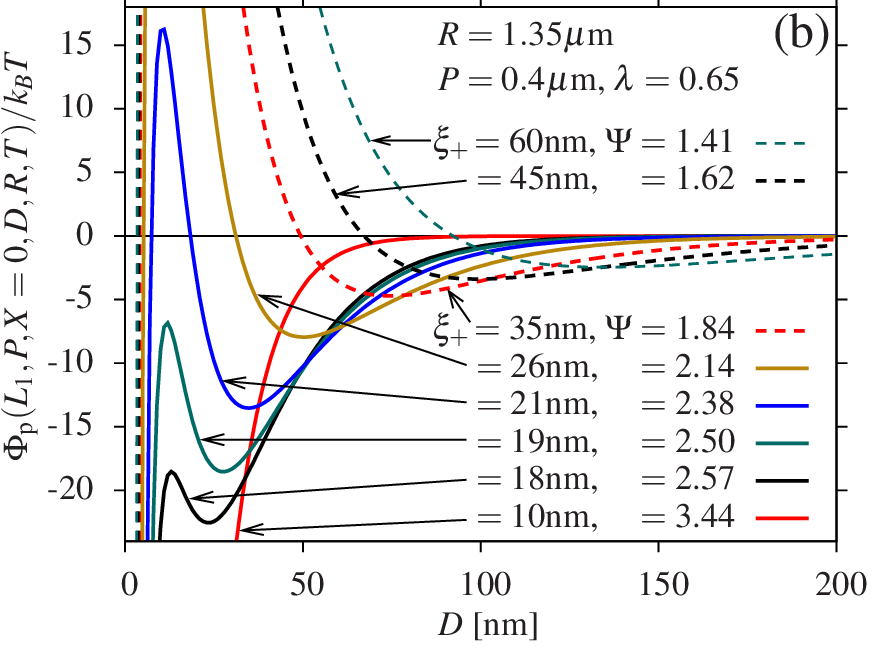}\\
    \includegraphics[width=6.5cm]{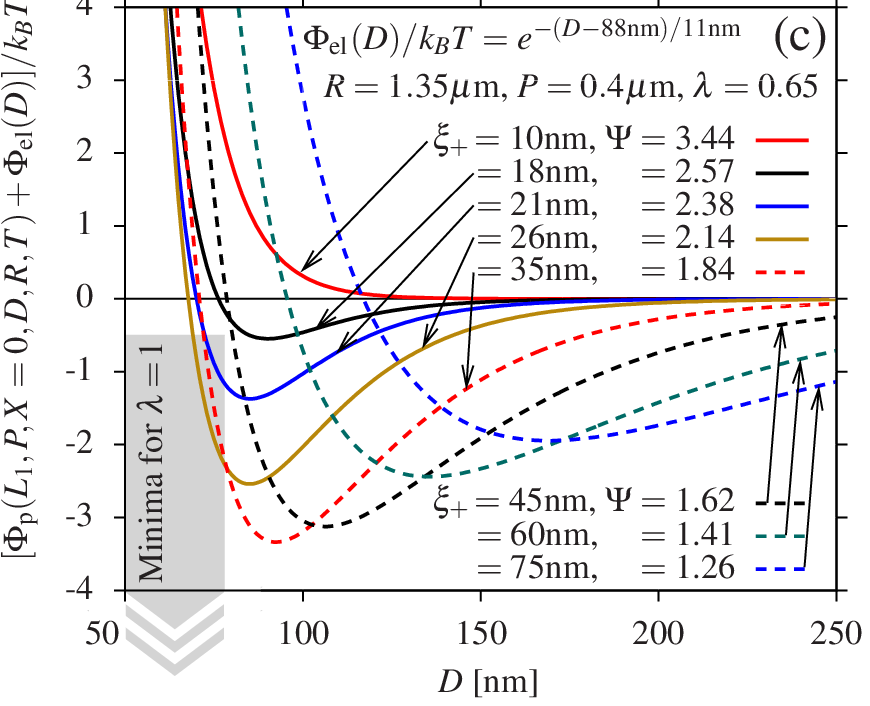}\\[-2mm]
  \else
    \hspace*{-1.6cm}
    \parbox{18.4cm}{%
      \parbox{9.7cm}{%
        \includegraphics{lev_ex_1}\\[1mm]
        \includegraphics{lev_ex_2}
        }
      \parbox{8.5cm}{%
        \includegraphics{lev_ex_2_el}
        }
      }
    \linespread{1.2}
  \fi
  \caption{%
  \ifTwocolumn
  \else
  \fi
  Critical Casimir potential $\Phiperiod$ [\eref{eq:period-pot}] in $d=3$ of a colloid of radius $R=1.35\mu$m close
  to a periodically patterned substrate as a function of $D$ and for various values of $\xi_+$ for $P=1\mu$m with
  $\lambda=0.4$ in (a) and $P=0.4\mu$m with $\lambda=0.65$ in (b) and (c).
  The values of $P$, $\lambda$, and $\xi_+$ are chosen as to be experimentally accessible in a colloidal
  suspension exhibiting critical Casimir forces \cite{nellen:2009,hertlein:2008,gambassi:2009,soyka:2008}.
  The critical Casimir potential for the colloid close to a patterned substrate may exhibit -- depending on the
  value of $\xi_+$, and, thus, on the temperature -- a local minimum corresponding to stable levitation.
  In (c) an electrostatic potential $\Phi_{\textrm{el}}$ [\eref{eq:phitot}] 
  is added to $\Phiperiod$, which refers to actual experimental data \cite{nellen:2009}.
  The shaded area indicates the ranges of the positions and the depths of the local minima of the total potential occurring
  if the substrate is laterally homogeneous and purely attractive, i.e., for $\lambda=1$ ($(-,-)$ BC) for the range
  $14$nm $<\xi_+<75$nm leading to potential depths between $0.5k_BT$ and $70k_BT$ (indicated
  by the shaded arrow); for $\lambda=1$ the preferred colloid position is dictated by the electrostatic
  repulsion and restricted to the range of $50$nm to $75$nm, whereas the colloid position $D_{0,s}=\Theta_{0,s}\xi_+$ due to critical 
  Casimir levitation can be much larger and tuned by temperature.
  Moreover, whereas for $\lambda=1$ and upon approaching $T_c$ the minima monotonically become deeper, the levitation
  minima first deepen and move to smaller values of $D$ followed by a decrease of the depth, by becoming more shallow, and moving to
  larger values of $D$.
  Reducing the range and strength of the electrostatic repulsion by adding salt to the solvent
  is expected to provide access to even more details of the critical Casimir levitation potential $\Phiperiod$ shown in (b).
  }   
\label{fig:levitation_example}
\end{figure}

Figure~\ref{fig:levitation} shows the values of $\Theta_0$ at which the normal critical Casimir 
force acting on a colloid vanishes as a function of the new scaling variable $\Psi$ introduced at the beginning
of this subsection, for various $\lambda$, for $\Xi=0$, 
and within the DA ($\Delta\to0$) for (a) $d=4$ and (b) $d=3$.
The corresponding sign of $\partial_D \Fperiod\big|_{D=D_0}$ [according to \eref{eq:levitation-2}] is also indicated: 
$\Theta_0$ drawn as a solid line indicates $\partial_D \Fperiod\big|_{D=D_0} < 0$,
i.e., stable levitation of the colloid; a dashed line, instead, indicates $\partial_D \Fperiod\big|_{D=D_0}>0$ and therefore 
a local maximum of the critical Casimir potential
with respect to $D$, which occurs within the shaded regions in \fref{fig:levitation}.
For a given value of $\lambda$ (with $\lambda_1<\lambda<\lambda_0$ as we shall discuss in detail further below), e.g., 
$\lambda =0.60$ in \fref{fig:levitation}(a), the corresponding curve for $\Theta_0$ shows a bifurcation at 
$\Psi = \Psi^*(\lambda)$ such that a vertical line drawn in \fref{fig:levitation} at a certain $\Psi$ intersects this curve in two
points $\Theta_{0,u}$ and $\Theta_{0,s}>\Theta_{0,u}$ if $\Psi < \Psi^*(\lambda)$, whereas it has no intersection 
for $\Psi  >\Psi^*(\lambda)$. 
In the former case $\Theta_{0,u}$ and $\Theta_{0,s}$ correspond to a local maximum and to a local minimum of
the critical Casimir potential at distances $D_{0,u} = \xi_+ \Theta_{0,u}$ and $D_{0,s} = \xi_+ \Theta_{0,s}$, respectively, i.e.,
to an {\emph u}nstable and a {\emph s}table levitation point for the colloid, respectively.
Instead, for $\Psi>\Psi^*(\lambda)$, the critical Casimir force has no zero at any finite value of $D$.
We note that $D=0$ (stiction) and thus $\Theta=0$ always corresponds to the global minimum of the potential because
for $D\to 0$  the critical Casimir potential is strongly attractive.
The corresponding geometrical configuration into which the colloid is finally attracted by the substrate 
[due to $(a_1) =(-)$, $(b)=(-)$, and $X=0$, see Fig.1] is stabilized by the steric repulsion of the wall. 
We note that within the DA the critical Casimir potential for $X=0$ is attractive at sufficiently small distances, even if 
the major part of the substrate is characterized by $(+)$ BC, i.e., even if $0\ne\lambda \ll 1$. 
Indeed, in this case the potential of the colloid at $X=0$ and close to a periodically patterned substrate can be 
approximated by the one due to a single chemical lane centered at $X=0$, which has been discussed in Sec.~\ref{sec:stripe}. 
For given colloid radius $R$ and width $L_1=\lambda P>0$ of the attractive stripe, the scaling variable 
$\Lambda = L_1/(2\sqrt{RD})$ diverges as $D\to 0$, so that the scaling function $\omegastripe(\Lambda,\Xi,\Theta,\Delta)$  
which characterizes the potential of the lane [see \eref{eq:stripe-omega}] attains the value $-1$ corresponding to the case of homogeneous, 
attractive $(-,-)$ BC [see \fref{fig:stripe}]. 
Within this approximation and for $D\ll\xi_\pm$ the critical Casimir force becomes attractive if 
$\varthetaperiod \simeq \varthetastripe < 0$ which, due to Eqs.~\eqref{eq:stripe-omega}, \eqref{eq:derjaguinpot}, and \eqref{eq:delta-ab},
yields the condition $\omegastripe(\Lambda,\Xi=0,\Theta\to0,\Delta\to0)<1-2\Dpm/(\Dpm-\Dpp)$, 
i.e., $\omegastripe<-0.6$ in $d=4$ \cite{krech:1997} and $\omegastripe \lesssim-0.76$ in $d=3$ \cite{mcdata}; 
this occurs for $\Lambda > \Lambda_0 = 1.1$ in $d = 4$, and $\Lambda>\Lambda_0=2.7$ in $d = 3$, respectively 
[see also \fref{fig:stripe}(a)].  
Accordingly, at distances $D<\lambda^2P^2/(4R\Lambda_0^2)$ (together with $D\ll\xi_\pm$) the critical Casimir potential $\Phiperiod$ is
negative and diverges to $-\infty$ for $D \to 0$. 
(However, for very small values of $\lambda$ this would occur at distances of microscopic scale such that the scaling limit and 
thus the form of $\Phiperiod$ do no longer hold). 
Thus the bifurcation of $\Theta_0$ at $\Psi^*(\lambda)$ corresponds to a transition from (metastable) levitation at 
$D=D_{0,s}$ for $\Psi<\Psi^*(\lambda)$ to stiction at $D=0$ for $\Psi>\Psi^*(\lambda)$. 
For $\Psi<\Psi^*(\lambda)$ the metastable levitation minimum at $D_{0,s}$ is shielded from the global minimum at $D=0$
by a potential barrier the height of which vanishes for $\Psi\nearrow\Psi^*(\lambda)$ [see \fref{fig:levitation_example}].
Experimentally, one typically varies the value of $\xi_+$ by changing the temperature 
\cite{hertlein:2008,gambassi:2009,nellen:2009,soyka:2008} and leaves the geometry ($\lambda$, $P$, and $R$) 
unchanged, which results in a change of $\Psi$ via varying $T$.
Thus, experimentally, the transition at $\Psi^*(\lambda)$ corresponds to a de facto irreversible transition from separation to 
stiction of the colloid as a function of temperature.
\par
Moreover, from \fref{fig:levitation} one can infer that for both $d=3$ and $d=4$ there is a $\lambda_0$ such that, 
for $1\ge\lambda>\lambda_0$, $\Kperiod$ has no zero for any choice of $\Psi$ (i.e., there is no solution $\Theta_0$)
and the critical Casimir force is attractive at all distances.
Within the DA, $\lambda_0=\Dpm/(\Dpm-\Dmm)$ [see also \eref{eq:small-period-K}], which renders the values $\lambda_0=0.80$ in
$d=4$ \cite{krech:1997} and $\lambda_0\simeq0.88$ in $d=3$ \cite{mcdata}.
In addition, from \fref{fig:levitation} one can infer that for $\lambda_0>\lambda>\lambda_1 \simeq 0.5$ 
and $\Psi \lesssim 1$, $\Theta_{0,s}$ effectively does no longer depend on $\Psi$ but solely on $\lambda$. 
Accordingly, the distance $D_{0,s} \propto \xi_+$ at which the colloid stably levitates can be tuned by 
temperature upon approaching criticality. 
However, for $\lambda < \lambda_1 \simeq 0.5$, $\Theta_{0,s}$ diverges at $\Psi = \Psi_0(\lambda) < \Psi^*(\lambda)$ 
such that for $\Psi_0(\lambda)<\Psi<\Psi^*(\lambda)$ the colloid exhibits critical Casimir levitation at a local 
minimum of the potential, whereas within this range of $\lambda$ values for $\Psi<\Psi_0(\lambda)$ the critical Casimir potential has only a local 
(positive) maximum at $D_{0,u}$; it is repulsive for $D>D_{0,u}$ and therefore for large values of $D$ (i.e., 
$\Theta \gg 1$ and $\Pi \ll 1$) it approaches zero from positive values. 
This qualitative change in the behavior of the critical Casimir potential occurs at $\lambda = \lambda_1$. 
The value of $\lambda_1$ is close to 0.5 because the repulsive and attractive forces for $(+,-)$ and $(-,-)$ BC, 
respectively, have similar strengths but opposite signs for $\Theta \gg 1$, i.e., $\kpm(\Theta \gg 1)\simeq-\kmm(\Theta\gg 1)$ 
for both $d=3$ and $d=4$ [see \eref{eq:exponential-decay}, where $|A_-/A_+|\simeq 1.2$ in $d=3$ \cite{gambassi:2009}
and $|A_-/A_+|=1$ in $d=4$ \cite{krech:1997}].
Accordingly, depending on $\lambda$ being larger or smaller than $\lambda_1\simeq 0.5$, the area covered by one of the two BC 
prevails and the resulting force is asymptotically (i.e., $\Theta\gg1$) attractive or repulsive, respectively  
[see the remark at the end of Sec.~\ref{sec:period-da} and Eqs.~\eqref{eq:small-period-K} and \eqref{eq:app-step-K-homog}]. 
Taking into account the slight difference in the strength of the asymptotic forces for $(+,-)$ and $(-,-)$ BC one 
finds $\lambda_1 = (1-A_+/A_-)^{-1} $ which renders $\lambda_1=1/2$ in $d=4$ and $\lambda_1\simeq 0.545$ in $d=3$. 
The asymptotic behavior of the force at large distances can be inferred from the asymptotic behavior of 
$\Kperiod(\lambda,\Pi =\Psi\Theta^{-1/2},\Xi=0,\Theta\gg1,\Delta\to 0) \simeq {\mathcal A}(\Psi,\lambda)\; \Theta^{d-1} e^{-\Theta}$,
which can be obtained from Eqs.~\eqref{eq:period-psi-da}, \eqref{eq:period-psi}, \eqref{eq:app-step-K-homog}, 
\eqref{eq:erf}, and \eqref{eq:exponential-decay} .
Accordingly, the value $\Psi_0(\lambda)$ at which $\Theta_{0,s}$ diverges is characterized by the fact that 
${\mathcal A}(\Psi \lessgtr \Psi_0(\lambda),\lambda)\gtrless 0$ so that the force approaches zero from above or 
from below depending on having $\Psi<\Psi_0(\lambda)$ or $\Psi>\Psi_0(\lambda)$, respectively.
The condition ${\mathcal A}(\Psi_0(\lambda),\lambda) = 0$ yields the following implicit equation for $\Psi_0(\lambda)$:
%
\begin{equation} 
  \label{eq:psi0}
  2\lambda_1=\sum_{n=-\infty}^{\infty}
  \erf\left\{\tfrac{\Psi_0(\lambda)}{\sqrt{2}}(n+\tfrac{\lambda}{2})\right\}
  -\erf\left\{\tfrac{\Psi_0(\lambda)}{\sqrt{2}}(n-\tfrac{\lambda}{2})\right\}.
\end{equation} 
For $\lambda\ll1$ the sum on the rhs of \eref{eq:psi0} can be approximated by the term $n=0$ alone and one finds
$\Psi_0(\lambda\ll1)\simeq2^{3/2}\lambda^{-1}\erf^{-1}(\lambda_1)$, where $\erf^{-1}$ is the inverse error function,
which yields the relations $\Psi_0(\lambda\ll1)\simeq1.49/\lambda$ for $d=3$ and $\Psi_0(\lambda\ll1)\simeq1.35/\lambda$ 
for $d=4$. 
On the other hand, in the marginal case one expects $\Psi_0(\lambda=\lambda_1)=0$.
However, as argued above, \emph{at} the critical point ($\Theta=0$) the colloid does not exhibit stable levitation for any
geometrical configuration; this is in accordance with \fref{fig:levitation} because for $T\to T_c$, the levitation minimum
of the potential moves to large $D$ ($D_{0,s}=\Theta_{0,s}\xi_+\to\infty$) and disappears at $T=T_c$.
\par
In summary, as function of $\lambda$ there are three distinct levitation regimes:
\begin{itemize}
  \item[(i)] $\lambda>\lambda_0$ with $\lambda_0(d=3)\simeq0.88$ and $\lambda_0(d=4)=4/5$: There is no levitation
    and the critical Casimir force is attractive at all distances for any temperature.
  \item[(ii)] $\lambda_0>\lambda>\lambda_1$ with $\lambda_1(d=3)\simeq0.545$ and $\lambda_1(d=4)=1/2$: 
    Sufficiently close to $T_c$, i.e., for $\Psi=P/\sqrt{R\xi_+}<\Psi^*(\lambda)$ there is a local critical
    Casimir levitation minimum.
    Upon approaching $T_c$ its position $D_{0,s}=\Theta_{0,s}\xi_+$, with $\Theta_{0,s}(\xi_+\to\infty)$ finite,
    moves to macroscopic values proportional to the bulk correlation length.
  \item[(iii)] $\lambda_1>\lambda$: As in (ii) there is a local critical Casimir levitation minimum sufficiently
    close to $T_c$, i.e., for $\Psi<\Psi^*(\lambda)$.
    In general the onset of its appearance occurs further away from $T_c$ upon lowering $\lambda$.
    Upon approaching $T_c$ the position $D_{0,s}$ of this minimum diverges at a distinct nonzero reduced
    temperature given by $\Psi_0(\lambda)$, i.e., at $\xi_+=P^2/[R\Psi_0^2(\lambda)]$: $D_{0,s}=\Theta_{0,s}\xi_+$
    with $\Theta_{0,s}(\Psi\searrow\Psi_0(\lambda))\to\infty$.
\end{itemize}
\par
We note that, according to Figs.~\ref{fig:period_normal}, \ref{fig:normal_3d}, \ref{fig:kraft} and \ref{fig:comparesystem},
we expect that for $\Pi\lesssim2$ and $\Theta\lesssim4$ and for $\Pi\lesssim0.5$ and $\Theta\gtrsim4$, 
the DA does not provide a quantitatively reliable description of the actual behavior of 
$\Kperiod$ and therefore of $\Fperiod$; thus, for values of $\Psi\lesssim2\sqrt{\Theta_0}$ for $\Theta_0\lesssim4$, and 
$\Psi\lesssim0.5\sqrt{\Theta_0}$ for $\Theta_0\gtrsim4$,
we expect quantitative discrepancies between the actual behavior and the one predicted by the DA shown in \fref{fig:levitation}.
Nonetheless our results demonstrate that the geometric arrangement
of the chemical patterns allows one to design the normal critical Casimir force over a wide range.
\par
Figures~\ref{fig:levitation_example}(a) and (b) show the critical Casimir potential $\Phiperiod$ as a function of $D$
in $d=3$ within the DA based on Monte Carlo simulation data for the film geometry \cite{mcdata} for a variety of
specifically chosen values of the parameters $P$, $L_1$, $R$, and $\xi$.
The choice of these values is motivated by the typical experimental parameters which characterize recent investigations of 
the critical Casimir force acting on colloids immersed in binary liquid mixtures \cite{nellen:2009,soyka:2008,hertlein:2008,gambassi:2009}.
In particular, concerning the colloid radius we focus on the data of Ref.~\onlinecite{nellen:2009}, corresponding to $R=1.35\mu$m, 
while for the pattern we have chosen a periodicity $P=1\mu$m with $\lambda=0.4$ (i.e., $L_1=400$nm and $L_2=600$nm) 
[\fref{fig:levitation_example}(a)], or $P=0.4\mu$m with $\lambda=0.65$ (i.e., $L_1=260$nm and $L_2=140$nm) 
[\fref{fig:levitation_example}(b)].
A chemically patterned substrate with these characteristics appears to be realizable with presently available preparation
techniques \cite{soyka:2008,vogt:2009a,vogt:2009}.
[We note that $\Phiperiod$ as shown in \fref{fig:levitation_example}(a) and (b) is expected to describe the actual interaction
potential in the scaling regime characterized by values of $D$ and $\xi_+$ much larger than microscopic length scales
(such as $\xi_0^+\simeq0.2$nm \cite{hertlein:2008,gambassi:2009}) so that this prediction for $\Phiperiod$ is valid only 
for $D,\xi_+\gtrsim5$nm.]
With this choice of parameters we have calculated $\Phiperiod$ for various values of $\xi_+$ within an experimentally
accessible range \cite{nellen:2009,soyka:2008,hertlein:2008,gambassi:2009}.
From Figs.~\ref{fig:levitation_example}(a) and \ref{fig:levitation_example}(b) one can infer that for 
small values of $\xi_+$ (corresponding to large values of $\Psi>\Psi^*(\lambda)$) the critical Casimir potential is always attractive 
with a monotonic dependence on $D$ [see also \fref{fig:levitation}].
Upon approaching criticality, i.e., for increasing values of $\xi_+$ and decreasing values of $\Psi<\Psi^*(\lambda)$, a local maximum 
and a local minimum of the potential develop, so that for very small as well as for large $D$ the colloid is attracted to the 
patterned substrate, whereas within an intermediate range of values for $D$ it is repelled from it [see also \fref{fig:levitation}].
Thus, the colloid stably levitates at a distance $D_{0,s}$ corresponding to a local minimum of the potential.
The depth of this minimum ranges between a few $k_BT$ [\fref{fig:levitation_example}(a)] up to several $k_BT$ 
[\fref{fig:levitation_example}(b)].
Upon increasing $\xi_+$, $D_{0,s}$ increases as well, i.e., the colloid position is shifted away from the patterned substrate
with the potential minimum becoming more shallow.
In \fref{fig:levitation_example}(a) $\lambda=0.4$ and we find $\Psi^*(\lambda=0.4)\simeq4.65$ and $\Psi_0(\lambda=0.4)\simeq3.71$ 
[see \fref{fig:levitation}(b)] so that for $\Psi<\Psi_0(\lambda=0.4)$, i.e., for $\xi_+\gtrsim53.5$nm [\fref{fig:levitation_example}(a)]
the colloid does not exhibit stable levitation and the critical Casimir potential has a local maximum only.
The levitation minimum moves to macroscopic values of $D$ upon approaching the temperature corresponding to $\xi_+\simeq53.5$nm. 
In \fref{fig:levitation_example}(b) $\lambda=0.65$ and one has  $\Psi^*(\lambda=0.65)\simeq2.63$;
here $\Theta_{0,s}$ remains finite for $\Psi\to0$ in contrast to the case $\lambda<0.545$ [\fref{fig:levitation}(b)].
Thus, within the DA, for the case shown in \fref{fig:levitation_example}(b) stable levitation of the colloid is preserved 
for all finite values of $\xi_+>P^2/[R\,\,(\Psi^*(\lambda=0.65))^2]\simeq17$nm.
In this case upon approaching $T_c$ the levitation minimum moves to macroscopic values of $D$ proportional to the
bulk correlation length $\xi_+$.
\par
The discussion above focuses on the position of mechanical equilibrium of the colloid, corresponding to the point 
at which the forces acting on the particle vanish and the associated potential $\Phi$ has a local minimum 
$\Phi_{\rm min}$.
However, due to the thermal fluctuations of the surrounding near-critical fluid at temperature $T$, 
the colloid undergoes a Brownian diffusion which allows it to explore randomly such regions in space where the 
potential $\Phi$ is typically larger than $\Phi_{\rm min}$ for at most few $k_BT$. 
As a result, a position of mechanical equilibrium is stable against the effect of thermal fluctuations only if the 
potential depth of the minimum is larger than few $k_BT$.
In particular, if the potential barrier $\Phi(L_1,P,0,D=D_{0,u},R,T) - \Phi(L_1,P,0,D=D_{0,s},R,T)$, which separates 
the position of the local minimum at distance $D=D_{0,s}$ (levitation) from the global one at $D=0$ (stiction), is 
not sufficiently large [see, e.g., the curves corresponding to $\xi_+ = 36$nm in \fref{fig:levitation_example}(a) or 
corresponding to $\xi_+\lesssim 18$nm in \fref{fig:levitation_example}(b)], a de facto irreversible transition from levitation to 
stiction may occur as a consequence of thermal fluctuations. 
\par
In \fref{fig:levitation_example}(c) we show the resulting total potential of the forces acting on the colloid 
in the presence of an additional electrostatic repulsion which is experimentally practically unavoidable, in order 
to study its effect on critical Casimir levitation.
We assume that the electrostatic repulsion is laterally homogeneous and that it can be simply added to
the critical Casimir potential \cite{troendle:2009,gambassi:2009,nellen:2009} [see also Sec.~\ref{sec:summary} below].
Concerning the spatial dependence of the electrostatic repulsion we consider the one of Ref.~\onlinecite{nellen:2009}, which
corresponds to a colloid of radius $R=1.35\mu$m immersed in a near-critical water-lutidine mixture and 
close to a substrate exhibiting critical adsorption of water or lutidine \cite{nellen:2009}:
\begin{equation} 
  \label{eq:phitot}
  \Phi_{\textrm{el}}(D)/k_BT = \exp\{-\kappa(D-D_0)\},
\end{equation} 
where $D_0=88$nm and $\kappa^{-1}=11$nm \cite{nellen:2009}.
(Formally, $\Phi_{\textrm{el}}$ in \eref{eq:phitot} is finite for $D\to0$, and thus $\Phiperiod+\Phi_{\textrm{el}}$
is negative for $D\lesssim2$nm and has a global minimum at $D=0$ because $\Phiperiod\to-\infty$ for $D\to0$. 
However, \eref{eq:phitot} is actually the asymptotic form of the electrostatic interaction which is valid for distances larger than
the electrostatic screening length, i.e., $D\gg\kappa^{-1}$.
The corresponding total potential $\Phiperiod+\Phi_{\textrm{el}}$ is therefore not accurate for small
values  of $D$ and is reported in \fref{fig:levitation_example}(c) for $D>50$nm only.)
As in \fref{fig:levitation_example}(b) we choose $P=0.4\mu$m, $\lambda=0.65$, and experimentally 
accessible values of $\xi_+$.
Figure~\ref{fig:levitation_example}(c) provides a realistic comparison of the critical Casimir potential
with other forces as they typically occur in actual experimental systems.
One can infer from the graph reported in \fref{fig:levitation_example}(c) that for this choice of parameters the critical Casimir
levitation exhibited by the colloid is rather pronounced even in the presence of electrostatic interaction.
Far from the critical point ($\xi_+=10$nm) the interaction of the colloid with the substrate is completely dominated by
electrostatic repulsion.
Upon approaching criticality ($10$nm $\lesssim\xi_+\lesssim35$nm) a minimum in the total potential develops
and becomes deeper due to the increasing critical Casimir attraction working against the electrostatic repulsion.
For this latter range of values of $\xi_+$ the local minimum of the critical Casimir potential corresponding to levitation
is located at distances $D_{0,s}\lesssim60$nm at which the electrostatic repulsion still strongly contributes to
the resulting total potential [see \fref{fig:levitation_example}(c)].
Closer to the critical point ($\xi_+\gtrsim45$nm) the levitation minimum of the critical
Casimir potential occurs at distances $D_{0,s}\gtrsim100$nm [see \fref{fig:levitation_example}(b)]
at which the electrostatic force acting on the colloid is weak.
Thus, here the critical Casimir effect dominates and the position of the minimum of the total potential
increases with increasing values of $\xi_+$, which allows for measurements of the critical Casimir potential
for distances at which the precise form of $\Phi_{\textrm{el}}$ is not important.
Moreover, the depth of the minimum \emph{decreases} upon approaching criticality and the minimum becomes more 
shallow.
This behavior of the levitation minimum is distinct from the critical Casimir effect acting on a colloid close to
a \emph{homogeneous} substrate:
a local minimum also occurs in the latter case if the critical Casimir force is purely attractive ($\lambda=1$, $(-,-)$ BC) and works against
the electrostatic repulsion \cite{hertlein:2008,gambassi:2009}, due to the competition of different forces with opposite sign.
(We note that the critical Casimir levitation described above emerges from the critical Casimir force alone, i.e., it is a
feature of a \emph{single} force contribution.)
However, in this homogeneous case the preferred colloid position $D_{0,(-,-)}$ depends crucially on the 
form of the electrostatic interaction and is almost constant ($50$nm $<D_{0,(-,-)}< 75$nm).
Moreover, the depths of these latter minima monotonically increase as a function of of $\xi_+$ and become much larger 
than those shown in \fref{fig:levitation_example}(c) (see, e.g., 
Fig.~2(a) and Fig.~2(c) in Ref.~\onlinecite{hertlein:2008} and Fig.~3 in Ref.~\onlinecite{nellen:2009}).
In \fref{fig:levitation_example}(c) this is indicated by the shaded area and the shaded arrow, which corresponds to
the area of the graph within which minima of the total potential in the homogeneous case $\lambda=1$ occur for
$14$nm $<\xi_+<75$nm corresponding to potential depths of $0.5k_BT$ up to $70k_BT$.
On the other hand, the colloid position $D_{0,s}$ due to critical Casimir levitation can be much larger, can
reach values of several $\xi_+$, and can be tuned by temperature according to $D_{0,s}=\Theta_{0,s}\xi_+$.
In conclusion, the examples presented in \fref{fig:levitation_example} strongly suggest  that the 
critical Casimir levitation of a colloid close to a patterned substrate is experimentally accessible.
\par
By patterning the substrate, one introduces an additional (\emph{lateral}) length scale into the system, which, according to our results
presented above, can finally lead to stable levitation.
Introducing an additional length scale along the \emph{normal} direction by stacking different materials on top of
each other may lead to levitation due to \emph{quantum-electrodynamic} Casimir forces \cite{rodriguez:2010}.
The behavior of the stable levitation distance shows a bifurcation and irreversible transitions from separation
to stiction \cite{rodriguez:2010} similarly to the ones described above [see \fref{fig:levitation}].
In that context great importance has been given to the temperature dependence of the position $D_{0,s}$ of stable quantum
Casimir levitation \cite{rodriguez:2010}, which is quantified by the value of $\frac{d}{dT}D_{0,s}$.
In the critical Casimir case presented here, for an estimate of $\frac{d}{dT}D_{0,s}$ we pick as an example 
the stable levitation positions for $\xi_+=18$nm and $\xi_+=60$nm as
reported in \fref{fig:levitation_example} (a different choice would lead to similar results).
The results reported in \fref{fig:levitation_example} correspond to the experimentally relevant water-lutidine mixture with
$\xi_0^+=0.2$nm and $T_c\simeq307$K \cite{nellen:2009,hertlein:2008,gambassi:2009}.
Therefore, according to $\xi_+/\xi_0^+=|(T-T_c)/T_c|^{-\nu}$, the difference in temperature required to move from 
$\xi_+=18$nm to $\xi_+=60$nm is $\Delta T\simeq0.2$K.
Thus we find $\frac{d}{dT}D_{0,s}\simeq560\text{nm K}^{-1}$ for the average temperature dependence of critical Casimir levitation
[\fref{fig:levitation_example}(b)], and $\frac{d}{dT}D_{0,s}\simeq230\text{nm K}^{-1}$ by additionally taking electrostatics 
into account [\fref{fig:levitation_example}(c)].
We note that in the present critical case $\frac{d}{dT}D_{0,s}$ can become arbitrarily large at temperatures corresponding to 
the transition from
separation to stiction and the emergence of the local minimum and the local maximum of the critical Casimir potential
[see \fref{fig:levitation} and the curves for $\xi_+=34$nm and $\xi_+=36$nm in \fref{fig:levitation_example}(a)].
This shows that the critical Casimir levitation is strongly temperature dependent, even near room temperature, with the 
variation of stable separation $\frac{d}{dT}D_{0,s}$ being two orders of magnitude larger than the one 
predicted for the \emph{quantum-electrodynamic} Casimir effect in Ref.~\onlinecite{rodriguez:2010}.
In general the colloid will not only be exposed to the critical Casimir force and to an electrostatic force but also
to gravity and to laser tweezers, which generate a linearly increasing potential contribution.
This attractive contribution tends to reduce the potential barriers shown in \fref{fig:levitation_example}
and can eliminate small barriers altogether.
Thus these external forces can be used to switch levitation on and off (compare a similar discussion related to the
quantum-electrodynamic Casimir levitation in Ref.~\onlinecite{rodriguez:2010}).
%
\section{Cylinder \label{sec:cylinder}}
%
Currently, there is an increasing experimental interest in \emph{elongated} colloidal particles 
which have a typical diameter of up to several $100$~nm and a much larger length
(see, e.g., Refs.~\onlinecite{kondrat:2009,hoffmann:2008} and references therein).
These types of colloids resemble \emph{cylinders} rather than spheres.
The description of their behavior in confined critical solvents calls for a natural extension of the studies 
presented in Secs.~\ref{sec:homog}--\ref{sec:period}.
Hence, in the present section we consider the case of a $3d$ cylinder with $\MBC$ BC which is adjacent and parallel aligned 
to a periodically chemically patterned substrate consisting of alternating $\MBC$ and $\PBC$ stripes
as the ones discussed in Sec.~\ref{sec:period}.
Accordingly, the axis of rotational invariance of the cylinder is perpendicular to both the $x$ direction 
[\fref{fig:sketch}] and the direction normal to the substrate, and it is parallel to the direction of 
spatial translational invariance of the chemical stripes forming the pattern.
As compared with the case of the sphere the analysis for the cylinder is technically simpler because the system 
as a whole is invariant along all directions but two, the lateral one, $x$, and the one normal to the substrate. 
(For the sphere its finite extension in the second lateral direction, which is normal to the $x$-axis,
matters and thus leads to a basically three-dimensional problem.
Accordingly, here we do not consider short cylinders, for which this finite length matters, too.)
This reduction of the number of relevant dimensions allows us to perform numerical calculations of adequate 
precision for a range of various pattern geometries which is wider than in the case
of the sphere.
(Here, we do not consider a cylindrical colloid which is not perfectly aligned with the pattern and which 
would, therefore, experience a critical Casimir torque \cite{kondrat:2009}.)
Even though the expressions derived in Appendix~\ref{app:cylinder} can be used to study the case of a 
cylinder having its axis laterally displaced by an arbitrary amount $X$ from the chemical step, 
our numerical calculations for the case of a chemical stripe address only the case $X=0$. 
This corresponds to a lateral position of the symmetry axis of the cylinder which coincides 
with the center of an attractive $\MBC$ stripe.
\par
In Appendix~\ref{app:cylinder} we briefly derive the scaling behavior of the normal critical Casimir force acting
on the cylinder and compare it with the case of a sphere.
Then, we adapt the Derjaguin approximation appropriate for the geometry 
of the cylinder.
On this basis, we have calculated the scaling function of the normal
critical Casimir force acting on the cylinder in $d=3$ and $d=4$ on the basis of the Monte Carlo simulation data
for the film geometry \cite{mcdata} and of the analytic MFT expression for the critical
Casimir force for the film geometry \cite{krech:1997}, respectively.
In addition, within the same approach as the one of Sec.~\ref{sec:MFT} we have calculated numerically the MFT 
scaling functions corresponding to $\Delta\ne 0$, in order to assess the performance of the DA.
\par
\begin{figure}
  \ifTwocolumn
    \includegraphics[width=7.8cm]{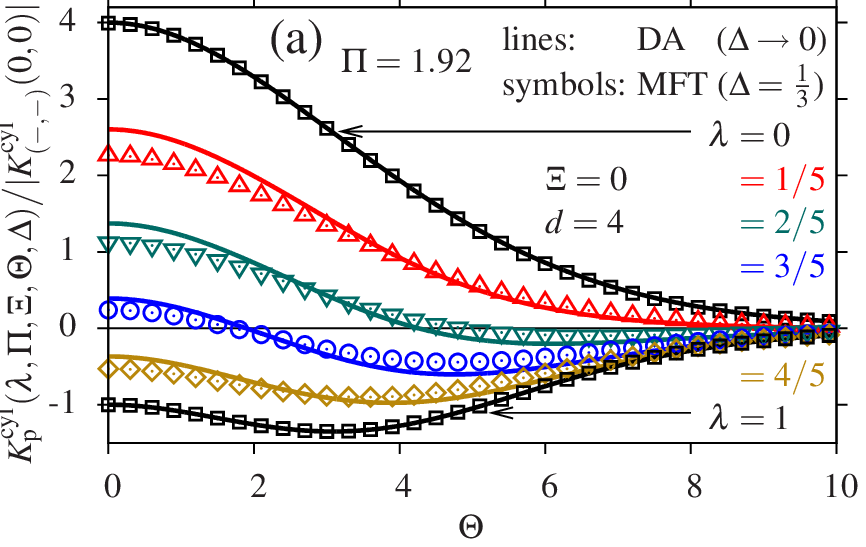}\\
    \includegraphics[width=7.8cm]{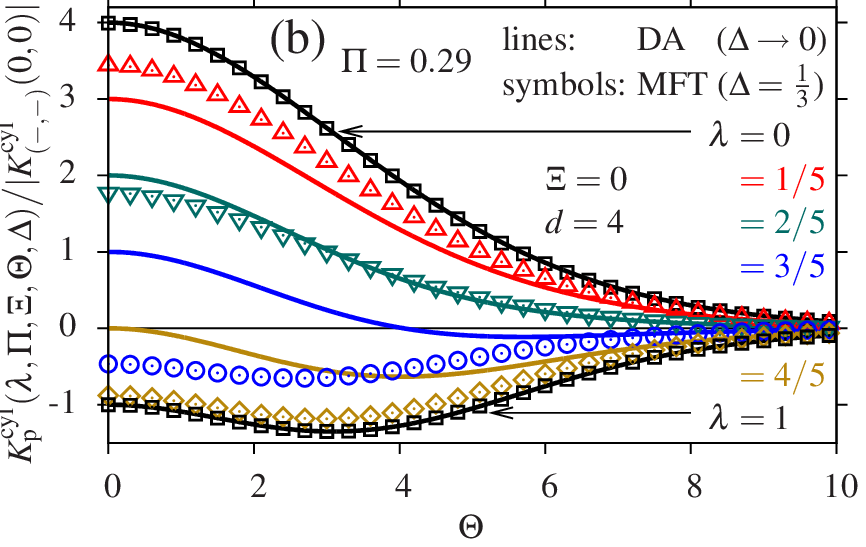}\\
    \includegraphics[width=7.8cm]{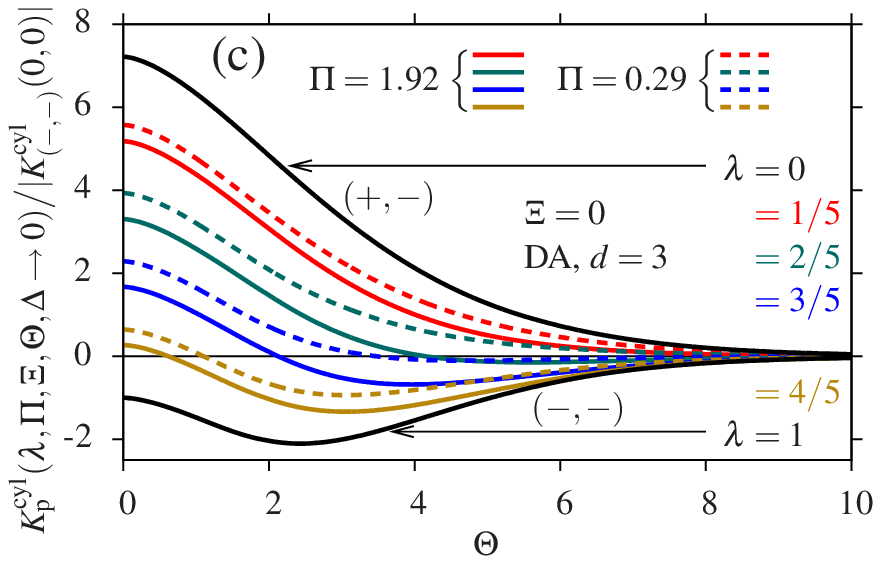}
  \else
    \linespread{1.2}
    \includegraphics{compare_cylinder_1_92}\\
    \includegraphics{compare_cylinder_0_29}\\
    \includegraphics{cylinder_3d}
  \fi
  \caption{%
    Normalized scaling function $\Kperiod^{\cyl}$ [see Appendix~\ref{app:cylinder}, including expressions for 
    $\Kmm^\cyl(0,0)$] of the normal critical Casimir force acting on a 
    \emph{cylindrical} colloid 
    close to and parallel to a periodically patterned substrate.
    The cylinder axis is aligned with the striped pattern and positioned above the center of a $(-)$ stripe which has 
    the same adsorption preference as the cylinder (analogous to \fref{fig:comparesystem} for a spherical colloid).
    In (a) for $\Pi=1.92$ the appropriate DA describes the actual MFT data rather well, and for $0.3\lesssim\lambda\lesssim0.7$ there
    is a change of sign of the force.
    In (b), instead, apart from the limiting homogeneous cases $\lambda=0$ and $\lambda=1$, for $\Pi=0.29$ the DA fails to describe 
    quantitatively the actual behavior [see the main text].
    In (c) $\Kperiod^{\cyl}$ is shown for $d=3$ within the DA based on the Monte Carlo simulation data for the film geometry
    \cite{mcdata} for the two cases $\Pi=0.29$ and $\Pi=1.92$.
    We expect that also in $d=3$ the DA for $\Pi=0.29$ is not quantitatively reliable.
    }
  \label{fig:cylinder}
\end{figure}
%
%
Here we focus on the comparison between the DA appropriate for the cylinder  
and the full numerical MFT data for the scaling function $\Kperiod^{\cyl}(\lambda,\Pi,\Xi=0,\Theta,\Delta)$
which characterizes the normal critical Casimir force in the presence of a periodically patterned substrate;  
$\lambda$, $\Pi$, $\Xi$, $\Theta$, and $\Delta$ are defined as in the case of the sphere
[see Sec.~\ref{sec:period} and Appendix~\ref{app:cylinder}].
Figure \ref{fig:cylinder} shows the scaling function of the normal critical Casimir force acting on
a cylinder as a function of $\Theta$ as obtained from the DA ($\Delta\to0$) in $d=4$ and from the full numerical MFT calculations for $\Delta=1/3$.
Besides the quantitative differences in the scaling function as a function of $\Theta$,
the \emph{qualitative} features of the behavior of the force acting on a cylinder,
which is reported in \fref{fig:cylinder} for various values of $\lambda$, 
are similar to the ones for the sphere [compare \fref{fig:comparesystem}].
For $\Pi=1.92$ [\fref{fig:cylinder}(a)] the DA describes the actual behavior of the critical
Casimir force rather well, in particular for $\Theta\gtrsim2$, even for most values of $\lambda$.
As in \fref{fig:comparesystem}, for a certain range of values of $\lambda$ the normal critical Casimir force changes
sign at $\Theta_0^\cyl(\Pi,\lambda,\Xi=0,\Delta)$.
On the other hand for small periodicities ($\Pi=0.29$ in \fref{fig:cylinder}(b)) the DA in $d=4$ fails to 
describe quantitatively the actual behavior of the force as obtained from the full numerical MFT calculations.
These strong deviations from the DA [\fref{fig:cylinder}(b)] indicate the relevance
of effects caused by the actual non-additivity of critical Casimir forces.
\par
For $\lambda\gtrsim0.6$ the scaling function $\Kperiod^{\cyl}$ of the normal critical Casimir force obtained 
numerically and represented by symbols in \fref{fig:cylinder}(b) is very close (much closer than within the DA) to the one corresponding to the homogeneous case 
with $(-,-)$ BC (corresponding to $\lambda=1$) and does not show a change of sign.
This means that, even if the substrate is not homogeneous but chemically patterned -- but such that the
larger part of the surface still corresponds to $\MBC$ BC, i.e., $\lambda\gtrsim0.5$ -- the resulting critical 
Casimir force acting on the colloid with $\MBC$ BC resembles the behavior for laterally homogeneous $(-,-)$ BC.
This can be understood in terms of the fixed point Hamiltonian in \eref{eq:hamiltonian}
which penalizes spatial variations of the order parameter at short scales.
Thus the system tries to smooth out spatial inhomogeneities of the order parameter profile, biased by the 
preference of the colloidal particle.
If the pattern is very finely structured, i.e., $\Pi=(L_1+L_2)/\sqrt{RD}\ll1$, regions with a positive order 
parameter close to the narrow $\PBC$ stripes ($\lambda\simeq 1^-$, i.e., $L_2\ll L_1$) extent only very little 
into the direction normal to the substrate and the resulting order parameter profile at a distance 
from the substrate remains negative only \cite{sprenger:2005}, so that the force resembles the one 
corresponding to the homogeneous case.
(Note that within the DA, the corresponding order parameter profile would simply consist of a patchwork of the order parameter profiles corresponding to the film geometry, with no smoothing taking place at the edges of the various spatial regions.)
Similarly, but in a weaker manner due to the opposite order parameter preference at the colloid, the
curves in \fref{fig:cylinder}(b) for $\lambda\lesssim0.5$ approach the corresponding homogeneous one 
for the case $(+,-)$ (i.e., $\lambda=0$).
Thus, the fact that both in \fref{fig:cylinder}(a) and \fref{fig:cylinder}(b) the curves for $\lambda=1/5$ are less close to their
limiting ones for $\lambda=0$ than the curves for $\lambda=4/5$ are close to the ones for $\lambda=1$ -- although
the portions of the minority part of the surface are the same -- is due to the fact that an order parameter 
profile with $(+,-)$ boundary conditions is energetically less preferred than the one with $(-,-)$ boundary conditions
because in the $(+,-)$ case an interface emerges between the two phases.
For broad stripes, i.e., in contrast to the case $\Pi\to0$, the energy costs for a similar behavior are seemingly
larger: the full numerical MFT data for $\lambda=1/5$ and $\lambda=4/5$ are less close to the corresponding 
limiting homogeneous cases $\lambda=0$ and $\lambda=1$, respectively, for $\Pi=1.92$ than for $\Pi=0.29$.
%
%
\begin{figure} 
  \ifTwocolumn
  \includegraphics[width=8.4cm]{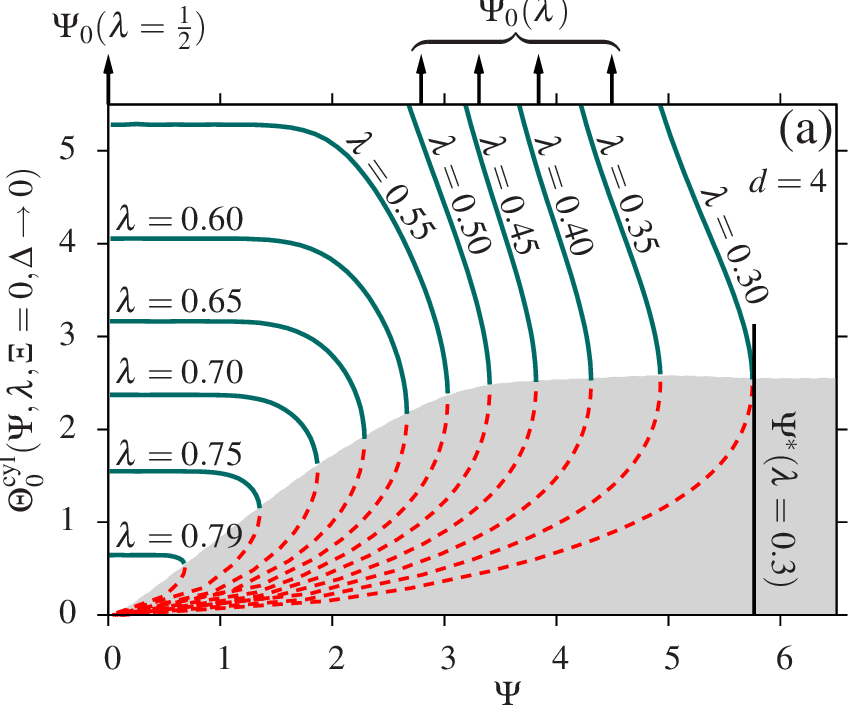}\\
  \includegraphics[width=8.4cm]{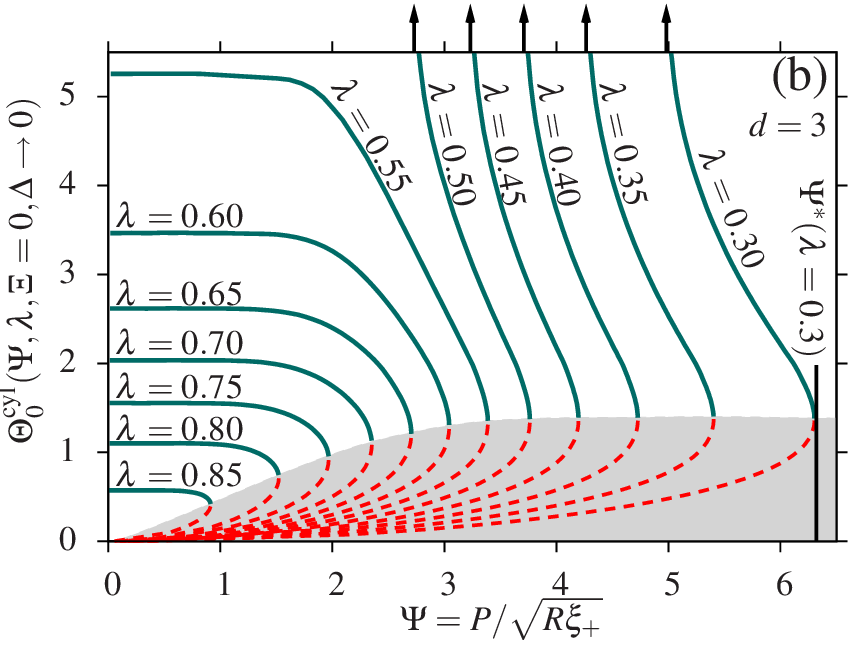}
  \else
  \includegraphics{levitation_4d_cylinder_c}\\
  \includegraphics{levitation_3d_cylinder_c}
  \fi
  \caption{%
    Values of the scaling variable $\Theta_0^\cyl$ at which the normal critical Casimir force $\Kperiod^\cyl$ acting
    on a cylinder close to a periodically patterned substrate vanishes as a function
    of $\Psi=P/\sqrt{R\xi_+}$ [compare \fref{fig:levitation} for the case of a sphere] within the DA.
    The region indicated by solid lines corresponds to the one in which the levitation of the  cylinder at a height 
    $D=D_0=\Theta_0\xi_+$ is stable against small perturbations of $D$, wheres in the shaded region indicated by dashed lines there is no such
    stable levitation although the normal critical Casimir force acting on the colloid vanishes.
    For $\lambda>\lambda_0$ with $\lambda_0(d=4)=4/5$ and $\lambda_0(d=3)\simeq0.88$, $\Theta_0^\cyl$ ceases
    to exist, i.e., $\Kperiod^\cyl$ does not exhibit a zero.
    For $\lambda<\lambda_1$ with $\lambda_1(d=4)=1/2$ and $\lambda_1(d=3)\simeq0.545$, $\Theta_0^\cyl(\Psi\searrow\Psi_0(\lambda))$
    diverges.
    (The values for $\Psi_0(\lambda)$ are indicated by upward arrows.)
    For any $\lambda<\lambda_0$, $\Theta_0^\cyl$ exists for $\Psi<\Psi^*(\lambda)$.
    We expect the DA to be quantitatively reliable only for $\Psi/\sqrt{\Theta_0}\gtrsim2$ for $\Theta_0\lesssim4$ and
    for $\Psi/\sqrt{\Theta_0}\gtrsim0.5$ for $\Theta_0\gtrsim4$.
  }   
  \label{fig:levitation_cylinder}
\end{figure}
%
%
%
\par
Figure~\ref{fig:cylinder}(c) shows the scaling function $\Kperiod^{\cyl}$ of the normal critical Casimir force for $d=3$ within the
DA as obtained by using Monte Carlo simulation data for the film geometry \cite{mcdata}.
One can infer from \fref{fig:cylinder}(c) that the qualitative features of the MFT scaling function as described above, such as the 
change of sign, are carried over to $d=3$.
\par

As discussed in the previous section, the vanishing of the normal critical Casimir force corresponds to a stable
levitation of the colloid at a distance $D_0$ from the substrate only if $\partial_D\Fperiod^\cyl|_{D=D_0}<0$.
Within the DA and at the laterally stable position $\Xi=0$ the sign of $\partial_D\Fperiod^\cyl$ is given by
\eref{eq:levitation-2} with $\Kperiod$ replaced by $\Kperiod^\cyl$.
The behavior of $\Theta_0^\cyl$ as a function of $\Psi$ and the demarcation of the regions where levitation
is stable against perturbations of $D$ is shown in \fref{fig:levitation_cylinder}, where the solid and the 
dashed lines correspond to stable and unstable levitation, respectively.
The behavior for the normal critical Casimir force acting on the cylinder is qualitatively similar to the one for the
sphere shown in \fref{fig:levitation}.
Analogously to the case of a sphere discussed in Sec.~\ref{sec:levitation}, no stable levitation is found at $T=T_c$
or for $\lambda>\lambda_0=\Dpm/(\Dpm-\Dmm)$, where $\lambda_0=0.80$ in $d=4$ and $\lambda_0\simeq0.88$ in $d=3$.
On the other hand, for $\Theta>0$, and $\lambda<\lambda_0$, it is always possible to find values of $P$ and $R$ such
that stable levitation of the cylinder occurs at a certain distance from the substrate.
The values of $\lambda_1$ below which one has a finite value $\Psi_0(\lambda)$ at which $\Theta_0$ diverges remain
the same as for the case of a sphere, i.e., $\lambda_1(d=4)=1/2$ and $\lambda_1(d=3)\simeq0.545$; also the corresponding
values of $\Psi_0(\lambda)$ remain the same [see \eref{eq:psi0}].
%
\section{Summary and conclusions \label{sec:summary}}
We have investigated the universal properties of the normal and lateral critical Casimir forces
acting on a spherical or cylindrical colloidal particle close to a chemically structured substrate with laterally varying
adsorption preferences for the species of a (near) critical classical binary liquid mixture (at its critical composition) 
in which the colloid is immersed.
Within the Derjaguin approximation (DA) [see \fref{fig:lat_color}] in spatial dimensions $d=3$ and $d=4$ 
we have derived analytic expressions for the corresponding universal scaling functions of the forces and the potentials 
for general fixed-point boundary conditions (BC) in terms of the scaling function of the critical Casimir force acting on
two parallel, homogeneous plates.
These expressions are given explicitly analytically at the bulk critical point $T=T_c$ and
-- for symmetry breaking boundary conditions -- far away from the critical point.
These relations enable one to obtain predictions for actual three-dimensional systems with a sphere-inhomogeneous plate geometry 
(for which currently computations are not possible) based on the scaling function for the parallel homogeneous plate geometry, 
for which, e.g., Monte Carlo simulation data in $d=3$ are available.
Moreover, results within mean-field theory (MFT, corresponding to $d=4$) and symmetry-breaking boundary conditions [Sec.~\ref{sec:MFT}]
have been obtained fully numerically and have been compared with the approximate results of the DA, which allows us
to explore the limits of validity of the latter.
We have studied several relevant situations [see \fref{fig:sketch}] and our main findings are the following:
\begin{enumerate}
  \item
First, we have studied a spherical colloid immersed in a binary liquid mixture close to a chemically
\emph{homogeneous} substrate which has, compared to the colloid, the same $\MBC$ or a different $\PBC$ adsorption
preference for one of the species of the mixture [Sec.~\ref{sec:homog}].
Close to the bulk critical point at $T=T_c$ the critical Casimir force induced by the confinement
of the order parameter (e.g., the concentration difference in a binary liquid mixture)
can be described in terms of universal scaling functions depending on the surface-to-surface distance $D$
of the colloid from the substrate scaled by the bulk correlation length, $\Theta=\sgn( (T-T_c)/T_c ) D/\xi_\pm$, 
and its ratio with the radius of the colloid, $\Delta=D/R$ [Eqs.~\eqref{eq:force-homog} and \eqref{eq:potential-homog}].
The scaling functions obtained within the DA [Eqs.~\eqref{eq:da-force} and \eqref{eq:derjaguinpot}] are
valid for $\Delta\to0$.
From the comparison with the full numerical MFT results [\fref{fig:homog}]
we find that in $d=4$ the DA describes the actual behavior quite well for $\Delta\lesssim0.4$.
Based on Monte Carlo simulation data for the scaling function of the critical Casimir force between parallel, homogeneous
plates and within the DA we have obtained also the scaling function for the critical Casimir force on a
spherical colloid close to a homogeneous substrate in $d=3$ [\fref{fig:homog}].
\item
The basic building block of a chemically patterned substrate is a \emph{chemical step}, which we have
studied in Sec.~\ref{sec:step}.
Due to the broken translational invariance in one lateral direction ($x$) the critical Casimir forces and potentials 
acquire a dependence on the additional scaling variable $\Xi=X/\sqrt{RD}$, which corresponds to the lateral 
distance $X$ of the center of the spherical colloid from the position of the chemical step along the plane
[Eqs.~\eqref{eq:step-force}, \eqref{eq:step-pot}, and \eqref{eq:step-lateral-force}].
Due to the different boundary conditions on both sides of the chemical step a \emph{lateral}
critical Casimir force emerges, which leads to a laterally varying potential for the colloid.
In the limit $\Delta\to0$ both the scaling function for the potential and for the lateral critical Casimir force 
as obtained within the DA are in agreement with the full numerical data [\fref{fig:lateral}].
We have derived the corresponding scaling functions within the DA also in $d=3$ by using Monte Carlo data for
the parallel plate geometry [\fref{fig:lateral}].
The preceding results have been partly presented in Ref.~\onlinecite{troendle:2009} as well as their
suitable comparison with corresponding experimental results \cite{soyka:2008}, which revealed that
the critical Casimir effect is rather sensitive to the geometrical details of the substrate patterns.
\item
Section~\ref{sec:stripe} deals with the critical Casimir forces and the corresponding potential
acting on a spherical colloid in front of a \emph{single chemical lane} of width $2L$, 
which additionally depends on a fourth scaling variable $\Lambda=L/\sqrt{RD}$ 
[Eqs.~\eqref{eq:stripe-force} and \eqref{eq:stripe-pot}].
It turns out that within the DA the scaling functions for the critical Casimir force and the critical
Casimir potential across a chemical lane can be expressed in terms of the ones for the chemical step [Eqs.~\eqref{eq:stripe-psi-da} 
and \eqref{eq:stripe-omega-da}].
For large values of $\Lambda$ the resulting potential can be described as a suitable superposition of chemical steps,
whereas for $\Lambda\lesssim 3$ one has explicitly to account for the finite width of the chemical stripe
[\fref{fig:stripe}].
Comparing the results of the DA with the ones obtained by a full numerical analysis, one finds that the DA
describes the actual behavior quite well for $\Delta\lesssim0.4$, even for small $\Lambda$.
Seemingly, in this respect, the nonlinearities inherent in the critical Casimir effect and edge effects do not considerably affect 
the resulting scaling functions [\fref{fig:omega_cl}].
\item
On the basis of the results of Sec.~\ref{sec:stripe}, in Sec.~\ref{sec:period} we have studied the universal scaling functions of the critical Casimir force and the corresponding potential
for a sphere opposite to a \emph{periodically patterned substrate} with laterally alternating chemical stripes of different
adsorption preferences [Sec.~\ref{sec:period}].
These scaling functions [Eqs.~\eqref{eq:period-force} and \eqref{eq:period-pot}] depend,
besides the scaling variables $\Theta$, $\Delta$, and $\Xi$, on two additional scaling variables $\Pi=P/\sqrt{RD}$ and
$\lambda=L_1/P$, which correspond to the period $P=L_1+L_2$ of the pattern and to the width $L_1\le P$ of the 
stripes with the same adsorption preference as the colloid.
The scaling function for the normal critical Casimir force obtained within the DA can be expressed in terms of
the one for the chemical step and describes the actual behavior well for $\Pi\gtrsim2$ [\eref{eq:period-psi-da}
and Figs.~\ref{fig:period_normal}, \ref{fig:normal_3d}(a) and \ref{fig:comparesystem}(a)].
However, for $\Pi\to0$ [\eref{eq:small-period-K}] the DA fails to capture quantitatively the numerically obtained
behavior within MFT, reflecting the importance of nonlinearities and edge effects in this context, which are not accounted for by the DA [Figs.~\ref{fig:period_normal},
\ref{fig:normal_3d}(a) and \ref{fig:comparesystem}(b)].
The failure of the DA in the limit $\Pi\to0$ can be traced back to the fact that for the \emph{film} geometry
of a patterned wall next to a laterally homogeneous flat wall, additivity of the critical Casimir forces
does not hold [\fref{fig:kraft}].
\item
The MFT scaling function of the normal critical Casimir force acting on a colloid close to a periodically
patterned substrate shows a remarkable behavior as a function of $\Theta=D/\xi_+$.
Within a certain range of values of $\Pi$ and $\lambda$ the critical Casimir force vanishes at $\Theta_0$ corresponding
to a distance $D=D_0$ between the colloid and the substrate.
We have analyzed the sign of the derivative of the critical Casimir force with respect to $D$ at $D_0$,
which is negative if for $D<D_0=D_{0,s}$ the colloid is repelled from the substrate whereas for $D>D_0=D_{0,s}$ it is
attracted to the substrate [\fref{fig:levitation}].
This means that in the absence of other forces the colloid can levitate above the substrate at a stable distance
which can be tuned by temperature.
Stable levitation points are found also in $d=3$, within the DA and on the basis of the 
Monte Carlo data
for the parallel plate geometry [Figs.~\ref{fig:normal_3d}(b), \ref{fig:comparesystem}(c), and \ref{fig:levitation}(b)].
Our analysis shows that \emph{at} the critical point $T=T_c$ levitation is not possible, whereas
off criticality a geometrical configuration leading to stable levitation can always be found.
For fixed geometrical parameters, the critical Casimir potential as a function of $D$
changes from a monotonic behavior to a non-monotonic one upon approaching criticality;
a local maximum and a local minimum, the latter corresponding to stable levitation, occur 
[\fref{fig:levitation_example}(a) and (b)].
Experimentally, this corresponds to a de facto irreversible transition from separation to stiction of a colloid and
a patterned substrate.
The depths of these potential minima can be up to several $k_BT$ so that the levitation is stable against Brownian motion
of the colloid.
The critical Casimir levitation can be rather pronounced and robust even
in the presence of electrostatic interactions [\fref{fig:levitation_example}(c)].
The levitation height is proportional to the bulk correlation length and thus can be tuned by varying temperature.
Depending on the geometric parameter $\lambda$ we have identified two distinct types of temperature dependences of the 
levitation height $D_{0,s}$.
In both cases it exhibits a high temperature sensitivity $\frac{d}{dT}D_{0,s}$ which, 
for realistic examples at room temperature, is of the order of
several $100\text{nm K}^{-1}$.
These results show that the periodic patterning of the substrate enables one to design critical 
Casimir forces over a wide range of properties.
\item
This behavior is also observed for a \emph{cylindrical} colloid which lies parallel to the substrate such that its axis is
aligned with the translationally invariant direction of the stripes [Sec.~\ref{sec:cylinder} and Appendix~\ref{app:cylinder}].
The main features of the scaling function for the corresponding normal critical Casimir force 
are similar to the ones for the spherical colloid: the DA describes well the actual behavior
as obtained from full numerical MFT calculations for large values of $\Pi$, but fails quantitatively for
$\Pi\lesssim2$ [\fref{fig:cylinder}].
The numerical studies for $\Pi\to0$ indicate that a substrate with a very fine pattern, dominated
by one of the two BC as far as the corresponding covered area is concerned, 
leads to a normal critical Casimir force which resembles the one for a
homogeneous substrate characterized by the dominating BC [\fref{fig:cylinder}(b)].
Based on Monte Carlo data for the parallel plate geometry we  calculated within the DA the critical Casimir
force acting on a cylinder in $d=3$ [\fref{fig:cylinder}(c)].
Above a chemically patterned substrate, also for a cylinder stable levitation is possible for a wide range
of parameters [\fref{fig:levitation_cylinder}].
\end{enumerate}
\par
%
Typically, in experiments with a colloidal suspension one has to consider also other forces, 
such as electrostatics, gravitation, and van der Waals forces which act on the colloidal particles in addition
to the critical Casimir forces.
The total force is approximately the sum of these contributions \cite{dantchev:2006,dantchev:2007} 
[see \fref{fig:levitation_example}(c)].
Upon approaching the critical point in the phase diagram, experiments \cite{soyka:2008,hertlein:2008,gambassi:2009} 
and theory (see, e.g., Refs.~\onlinecite{troendle:2009,gambassi:2009,sprenger:2006}) highlight the importance and the relevance 
of the critical Casimir effect in comparison with these other forces.
\par
The lateral critical Casimir forces occurring for patterned substrates as discussed here 
are highly sensitive to the details of the geometry of the pattern.
A detailed comparison with available experimental data \cite{soyka:2008} has to take 
this into account \cite{troendle:2009}.
This sensitivity even allows for an independent determination of the geometry of a 
chemically structured substrate by means of the critical Casimir effect.
This is useful in cases in which it is difficult to infer the geometry of the 
chemical pattern directly \cite{troendle:2009,soyka:2008}.
Concerning the comparison with experiments for chemically structured substrates, the theoretical 
predictions for the critical Casimir force are in agreement with the presently available data 
\cite{troendle:2009,soyka:2008}, for which the description in terms of independent chemical steps 
[Sec.~\ref{sec:step}] turns out to be sufficient \cite{troendle:2009}.
In order to test our specific predictions obtained for narrow single chemical lanes and for periodic 
chemical stripes, structures on the nanometer scale are needed.
Preliminary experimental data in this direction are encouraging \cite{vogt:2009,vogt:2009a}.
\par
In view of present basic research efforts and potential applications, it is important to study the effect of \emph{weak}
critical adsorption of the fluid at the confining surfaces, corresponding to \emph{finite} surface fields.
Such weak surface fields can be realized by applying suitable surface chemistry and they influence
the resulting behavior of the critical Casimir effect strongly \cite{mohry:2009,nellen:2009}.
Another approach to create an effective reduction of the surface adsorption is to create
fine periodic chemical patterns with  different (strong) adsorption preferences as discussed here.
However, our results for $\Pi\to0$ [Figs.~\ref{fig:comparesystem}(b) and \ref{fig:cylinder}(b)] show that a fine patterning
of the substrate with alternating boundary conditions does not necessarily lead to an
effective reduction of the surface adsorption at \emph{short} distances because in this range the critical Casimir 
force for a inhomogeneous
adsorption preference resembles the one for a \emph{homogeneous} substrate corresponding to strong adsorption.
On the other hand, at \emph{large} distances a periodically patterned substrate does lead to an effective BC
corresponding to a weak adsorption preference, and for $\lambda=1/2$ the surface fields even cancel out, 
leading to an effective BC resembling the so-called ordinary BC \cite{sprenger:2006}.
This offers the interesting perspective to study, at least asymptotically, critical Casimir forces with Dirichlet BC
by using classical fluids instead of superfluid quantum fluids \cite{krech:9192all,garcia:9902all,ganshin:2006,maciolek:2007}.
\par
A patterning on the \emph{molecular} scale is \emph{not} captured by the continuous approach pursued here,
which gives the universal features of the critical Casimir effect.
Nonetheless, a molecular patterning of the substrates may provide another means for an effective reduction of 
the adsorption of the corresponding fluid at the surface.
However, on a molecular scale the patterning is more likely to lead to randomly distributed surface fields which opens
a new challenge in the context of critical Casimir forces.

\acknowledgments
%
S.~K. and L.~H. gratefully acknowledge support by grant HA~2935/4-1 of the Deutsche Forschungsgemeinschaft.
A.~G. is supported by MIUR within the program ``Incentivazione alla mobilit{\`a} 
di studiosi stranieri e italiani residenti all'estero''.

\appendix

\section{Derjaguin approximation for a chemical step\label{app:step}}
%
In this appendix we first calculate within the DA the normal critical Casimir force $\Fstep(X,D,R,T)$ [\eref{eq:step-force}] acting
on a spherical colloid of radius $R$ facing a chemical step by using the DA. 
(We cannot directly calculate the lateral critical Casimir force 
$\Fstep^\parallel(X,D,R,T)$ within the DA because for two parallel homogeneous plates such a force vanishes.)
In a second step we derive the critical Casimir potential $\Phistep(X,D,R,T)=\int_D^\infty\upd z\; \Fstep(X,z,R,T)$ 
by integrating this result for the normal critical Casimir force.
In a third step the lateral critical Casimir force is obtained as $\Fstep^\parallel(X,D,R,T)=-\partial_X\Phistep(X,D,R,T)=
-\int_D^\infty\upd z\; \partial_X\Fstep(X,z,R,T)$ [see Sec.~\ref{sec:lateral}].
\par
\begin{figure} 
  \ifTwocolumn
  \includegraphics[width=7cm]{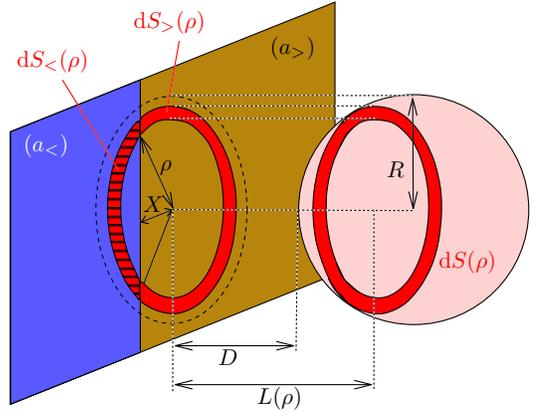}
  \else
  \includegraphics{lateraltest}
  \fi
  \caption{%
    Sketch concerning the Derjaguin approximation for the critical Casimir force acting
    on a colloid opposite to a chemical step.
    The critical Casimir force is subdivided into contributions from rings parallel to the substrate.
    The projection of the area $\upd S(\rho)$ of a ring onto the substrate is separated into the areal contributions
    $\upd S_<$ and $\upd S_>$ which emerge as the intersection of the projected ring with the half-planes 
    carrying $(a_<)$ and $(a_>)$ BC, respectively [see the main text].
    The sphere has a surface-to-surface distance $D$ from the substrate and its center has a lateral
    distance $X$ from the chemical step.
    }   
    \label{fig:lat_color}
\end{figure}

In the spirit of the DA, the surface of the spherical colloid with $(b)$ BC is thought of as being made of a pile of (infinitely thin)
rings parallel to the opposing substrate and with an area $\upd S(\rho)=2\pi\rho\upd\rho$, where $\rho$ is the radius of the
ring.
Each of these rings is partly facing (in normal direction) the surface with $(a_<)$ BC, with an extension $\upd S_<(\rho)$, 
and partly facing the surface with $(a_>)$ BC on the other side of the chemical step [\fref{fig:lat_color}],
with an extension $\upd S_>(\rho)$, such that $\upd S(\rho)=\upd S_<(\rho)+\upd S_>(\rho)$.
For an assigned $\rho$, $\upd S_\gtrless(\rho)$ depend, inter alia, on the lateral position $X$ of the colloid.
Using the \emph{assumption of additivity} of the forces underlying  the DA 
we suppose that the contribution $\upd \Fstep(\rho)$
of the ring to the total critical Casimir force $\Fstep$ is given by the \emph{sum} of the
critical Casimir forces which would act, in a film, on portions of areas $\upd S_<$ and $\upd S_>$ in the presence of
$(a_<,b)$ and $(a_>,b)$ BC, respectively. 
According to \eref{eq:planar-force} this leads to the following expression
for the force acting on a single ring:
\ifTwocolumn
\begin{multline} 
  \label{eq:app-step-force-ring}
  \frac{\upd \Fstep(\rho)}{k_BT}=
  \frac{\upd S_<(\rho)}{L^{d}(\rho)} k_\Asb(\sgn(t)\, L(\rho)/\xi_\pm)\\
    +
    \frac{\upd S_>(\rho)}{L^{d}(\rho)} k_\Alb(\sgn(t)\, L(\rho)/\xi_\pm),
\end{multline}
\else
\begin{equation} 
  \label{eq:app-step-force-ring}
  \frac{\upd \Fstep(\rho)}{k_BT}=
  \frac{\upd S_<(\rho)}{L^{d}(\rho)} k_\Asb(\sgn(t)\, L(\rho)/\xi_\pm)
    +
    \frac{\upd S_>(\rho)}{L^{d}(\rho)} k_\Alb(\sgn(t)\, L(\rho)/\xi_\pm),
\end{equation} 
\fi
where $L(\rho)$ is the substrate-ring distance [\fref{fig:lat_color}] as given in \eref{eq:da-L}, and $k_{(a_\gtrless,b)}$
are the scaling functions of the critical Casimir force in the film geometry with 
$(a_>,b)$ and $(a_<,b)$ BC, respectively [see \eref{eq:planar-force}].
This assumption neglects all edge effects along the boundary between the areas $\upd S_>(\rho)$ and 
$\upd S_<(\rho)$, which might actually be relevant in view of the spatial variation of the
order parameter profile.
It is therefore important to test the validity of this assumption at least in some relevant cases.
This is carried out in Sec.~\ref{sec:step} for $d=4$, i.e., within MFT.
\par
%
Without loss of generality in the following we assume $X>0$, i.e., that the normal projection of the 
center of the sphere falls on the part of the substrate with $(a_>)$ BC [Figs.~\ref{fig:sketch} and \ref{fig:lat_color}].
The results for $X<0$ are obtained by exchanging in the formulas below $a_< \leftrightarrow a_>$ and $X \leftrightarrow -X$.
Taking into account that $\upd S(\rho)=\upd S_<(\rho)+\upd S_>(\rho)$ one can rewrite \eref{eq:app-step-force-ring}
as 
\ifTwocolumn
\begin{multline} 
  \label{eq:app-step-force-ring-separation}
  \frac{\upd \Fstep(\rho)}{k_BT}=
  \frac{\upd S(\rho)}{L^{d}(\rho)} k_\Alb(\sgn(t)\,{L(\rho)}/{\xi_\pm})
    \\+
    \frac{\upd S_<(\rho)}{L^{d}(\rho)} \Delta k(\sgn(t)\,{L(\rho)}/{\xi_\pm}),
\end{multline} 
\else
\begin{equation} 
  \label{eq:app-step-force-ring-separation}
  \frac{\upd \Fstep(\rho)}{k_BT}=
  \frac{\upd S(\rho)}{L^{d}(\rho)} k_\Alb(\sgn(t)\,{L(\rho)}/{\xi_\pm})
    +
    \frac{\upd S_<(\rho)}{L^{d}(\rho)} \Delta k(\sgn(t)\,{L(\rho)}/{\xi_\pm}),
\end{equation} 
\fi
where $\Delta k(\Theta) = k_\Asb(\Theta) - k_\Alb(\Theta)$.
Summing up all force contributions from the rings of different radii $\rho$,
one finds for the \emph{total} normal force acting on the sphere
\begin{equation} 
  \label{eq:app-step-force-separation}
  \Fstep(X,D,R,T)=F_\Alb(D,R,T)+\Delta F(X,D,R,T),
\end{equation} 
where $F_\Alb$ is the force acting on a sphere close to a \emph{homogeneous} substrate with $(a_>)$ BC and
is given by \eref{eq:da-def} or by Eqs.~\eqref{eq:force-homog} and \eqref{eq:da-force}.
This term does not contribute to the \emph{lateral} critical Casimir force experienced by the 
colloid near the chemical step, because it does not depend on the lateral coordinate of the colloid.
The second term $\Delta F$ in \eref{eq:app-step-force-separation} corresponds to the integration
of the force differences $\Delta k$ in the region of overlap between the projection of the sphere 
onto the substrate plane and that part of the substrate with $(a_<)$ BC.
For each ring this area is given by [see \fref{fig:lat_color}]
\begin{equation} 
  \label{eq:}
  \upd S_<(\rho)=
  \begin{cases}
    0, &\rho<X,\\ 
    2 \arccos(X/\rho)\rho\upd \rho,&X\le\rho\le R.
  \end{cases}
\end{equation} 
This leads to
\begin{equation} 
  \label{eq:app-step-delta-f-int}
  \frac{\Delta F (X,D,R,T)}{k_BT} = 
  2\int\limits_X^R\upd\rho\,\rho\,\arccos\left(\frac{X}{\rho}\right)
    \,\frac{\Delta k(\sgn(t)\, L(\rho)/\xi_\pm)}{ L^{d}(\rho)}.
\end{equation} 
In the spirit of the DA, the radius of the sphere is taken to be large compared to its distance to the substrate, i.e.,
$\Delta=D/R\ll1$, and the contributions from the rings closest to the substrate dominate.
Therefore, it is well justified and in accordance with the DA to assume $X/R\ll1$ because the
contributions of rings with large radii do not change the behavior of the force in the Derjaguin limit.
Within these two limits we can use the \emph{parabolic} approximation for the distance of the rings to the substrate
[\eref{eq:da-L}], $L(\rho)\simeq D\alpha$, with $\alpha=1+{\rho^2}/{2RD}$.
Changing the integration variable in \eref{eq:app-step-delta-f-int}
we directly find
\begin{equation} 
  \label{eq:app-step-delta-f-int-rewritten-2}
  \Delta F (X,D,R,T) = k_B T\frac{R}{D^{d-1}}\Delta K(\Xi,\Theta,\Delta),
\end{equation} 
where $\Delta K$ is a universal scaling function given by
\begin{equation} 
  \label{eq:app-step-delta-K}
  \Delta K(\Xi,\Theta,\Delta\to0) = 2\!\!\!\!
                \int\limits_{1+\Xi^2/2}^{\infty}\!\!\!\!\upd\alpha\;
                \alpha^{-d}\arccos\left(\tfrac{\Xi}{\sqrt{2(\alpha-1)}}\right)
                  \Delta k( \alpha\Theta).
\end{equation} 
Note that the relevant scaling variable $\Xi=X/\sqrt{RD}$ can take on arbitrary values, irrespective of
the two assumptions $D/R\ll1$ and $X/R\ll1$.
From \eref{eq:app-step-delta-K} one finds with Eqs.~\eqref{eq:da-force} and \eqref{eq:step-K} directly
the expression for the scaling function $\psi_{(a_<|a_>,b)}$ given in \eref{eq:step-psi-da}.
\par
The critical Casimir potential  $\Phistep(X,D,R,T)=\int_D^\infty\upd l \Fstep(X,l,R,T)$ can be separated
analogously to \eref{eq:app-step-force-separation}, i.e.,
\begin{equation} 
  \label{eq:app-step-pot-separation}
  \Phistep(X,D,R,T)=\Phi_\Alb(D,R,T)+\Delta\Phi(X,R,D,T)
\end{equation} 
with
\begin{equation} 
  \label{eq:app-step-pot-delta}
\Delta\Phi(X,R,D,T)=
\int\limits_D^\infty\upd l \Delta F(X,l,R,T)\rdef k_BT\frac{R}{D^{d-2}}\Delta\vartheta(\Xi,\Theta,\Delta).
\end{equation}
\ifTwocolumn
\begin{widetext}
\fi
Using \eref{eq:app-step-delta-K}, the scaling function $\Delta\vartheta$ is given by
\begin{equation} 
  \label{eq:app-step-delta-vartheta-1}
  \Delta\vartheta(\Xi,\Theta,\Delta)=
  2\int\limits_1^\infty\upd y \frac{1}{y^{d-1}}
  \int\limits_{1+\Xi^2/(2y)}^{\infty}\!\!\!\!\upd\alpha\;
  \frac{1}{\alpha^{d}}\arccos\left(\frac{\Xi}{\sqrt{2y(\alpha-1)}}\right)
                                    \Delta k( y\alpha \Theta).
\end{equation} 
By changing the integration variable $\alpha\mapsto z\ddef 2y(\alpha-1)/\Xi^2$ followed by 
$y\mapsto v\ddef y+\Xi^2z/2$ one obtains
\begin{equation} 
  \label{eq:app-step-delta-vartheta-2}
  \Delta\vartheta(\Xi,\Theta,\Delta)=
  \Xi^2\int_1^\infty\upd z \int_{1+z\Xi^2/2}^\infty\upd v\frac{1}{v^d}\arccos(1/\sqrt{z})\Delta k(v\Theta).
\end{equation} 
After changing the order of integration
\begin{equation} 
  \label{eq:app-step-delta-vartheta-3}
  \int_1^\infty\upd z
  \int_{1+z\Xi^2/2}^\infty\upd v
  =
  \int_{1+\Xi^2/2}^\infty\upd v
  \int_1^{2(v-1)/\Xi^2}\upd z,
\end{equation} 
and using the primitive \cite{note:1} 
\begin{equation} 
  \label{eq:app-step-delta-vartheta-3b}
  \int \upd z \arccos(1/\sqrt{z})=z\arccos(1/\sqrt{z})-\sqrt{z-1}+c,
\end{equation} 
one obtains after a final change of variables $v\mapsto w\ddef2(v-1)/\Xi^2$
\begin{equation} 
  \label{eq:app-step-delta-vartheta-4}
  \Delta\vartheta(\Xi,\Theta,\Delta)=
  \frac{\Xi^4}{2}\int_1^\infty\upd s \frac{1}{(1+\Xi^2s/2)^d}\left[s\arccos(s^{-1/2})-\sqrt{s-1}\right]
  \Delta k (\Theta[1+\Xi^2s/2]).
\end{equation} 
From \eref{eq:app-step-delta-vartheta-4} together with \eref{eq:derjaguinpot} one obtains the final expression
for the scaling function of the critical Casimir potential as given in \eref{eq:step-omega-da}.
\subsection{Bulk critical point: $\Theta=0$ \label{app:step-crit}}
In order to calculate the critical Casimir force acting on the colloid at the \emph{bulk critical point} 
one inserts \eref{eq:delta-ab} into \eref{eq:app-step-delta-K} and obtains
\begin{eqnarray} 
  \label{eq:app-step-delta-K-crit}
  \Delta K(\Xi,\Theta=0,\Delta) &=& 2\left(\Delta_\Asb-\Delta_\Alb\right)
                \int_{1+\Xi^2/2}^{\infty}\!\!\!\!\upd\alpha\;
                \alpha^{-d}\arccos\left(\frac{\Xi}{\sqrt{2\alpha-2}}\right)\\
  \label{eq:app-step-I-def}
                &\rdef& \Xi^2\left(\Delta_\Asb-\Delta_\Alb\right)\; I_d(\Xi^2/2),\nonumber
\end{eqnarray}
\ifTwocolumn
\end{widetext}
\fi
where $\Delta_{(a,b)} = k_{(a,b)}(0)$ [see \eref{eq:delta-ab}], and with the substitution $\alpha\mapsto z=\Xi/\sqrt{2(\alpha-1)}$ for $d>1$,
\begin{equation} 
  \label{eq:app-step-I}
  I_d(a)=2\int_0^1\upd z \frac{z^{2d-3}}{(z^2+a)^d}\arccos(z).
\end{equation} 
For $I_d(a)$ the recursion relation
\begin{equation} 
  \label{eq:app-step-I-recursion}
  I_{d+1}(a)=\frac{1}{d}a^{1-d}\frac{\upd}{\upd a}[a^d I_d(a)]
\end{equation} 
holds, so that $I_4$ and $I_3$ can be expressed in terms of $I_2$.
Performing the integration we find \cite{note:2} 
%
\begin{equation} 
  \label{eq:app-step-I-2}
  I_2(a)=\frac{\pi}{2a}\left[1-\frac{a^{1/2}}{(1+a)^{1/2}}\right],
\end{equation} 
and therefore with \eref{eq:app-step-I-recursion}
\begin{equation} 
  \label{eq:app-step-I-3}
  I_3(a)
  =\frac{\pi}{4a}\left[1-\frac{\frac{3}{2}a^{1/2}+a^{3/2}}{(1+a)^{3/2}}\right],
\end{equation} 
and
\begin{equation} 
  \label{eq:app-step-I-4}
  I_4(a)=\frac{\pi}{6a}\left[1-\frac{\frac{15}{8}a^{1/2}+\frac{5}{2}a^{3/2}+a^{5/2}}{(1+a)^{5/2}}\right].
\end{equation} 
Thus, from Eqs.~\eqref{eq:app-step-I-def}, \eqref{eq:app-step-I-3}, and \eqref{eq:app-step-I-4} together with
the expression for $K_{(a_\gtrless,b)}(0,0)=2\pi\Delta_{(a_\gtrless,b)}/(d-1)$ [Sec.~\ref{sec:homog-da}] and 
\eref{eq:step-K}, one finds the expression for the scaling function 
$\psi_{(a_<|a_>,b)}$ given in \eref{eq:step-psi-da-crit}.
The critical Casimir potential at $\Theta=0$ for $d=3$ and $4$ can be found from \eref{eq:app-step-delta-vartheta-1}
together with \eref{eq:app-step-I-def}:
\begin{equation} 
  \label{eq:app-step-pot-1a}
  \Delta\vartheta(\Xi,\Theta=0,\Delta)=\Xi^2\left(\Delta_\Asb-\Delta_\Alb\right)
        \int_1^{\infty}\upd y\;y^{-d} I_d\left(\tfrac{\Xi^2}{2y}\right),
\end{equation}
and from a change of variable $y\mapsto a=\Xi^2/(2y)$ one finds
\begin{equation} 
  \label{eq:app-step-pot-1}
  \Delta\vartheta(\Xi,0,\Delta)=\frac{2^{d-2}}{\Xi^{2d-4}}\left(\Delta_\Asb-\Delta_\Alb\right)
        \int_0^{\Xi^2/2}\upd a\;a^{d-2} I_d(a).
\end{equation}
Using \eref{eq:app-step-I-recursion} and the limiting behavior $I_d(a\to0)=\pi/( 2(d-1) a)$, we find 
\begin{equation} 
  \label{eq:app-step-pot-2}
  \Delta\vartheta(\Xi,0,\Delta)=\frac{\Xi^{2}}{2(d-1)}\left(\Delta_\Asb-\Delta_\Alb\right)
  I_{d-1}(\Xi^2/2).
\end{equation} 
From Eqs.~\eqref{eq:app-step-I-2}, \eqref{eq:app-step-I-3}, and \eqref{eq:app-step-pot-2} together with 
$\vartheta(0,0)$ as given in Sec.~\ref{sec:homog-da} one obtains \eref{eq:step-omega-da-crit}
for the scaling function of the critical Casimir potential at $T_c$.
\ifTwocolumn
\begin{widetext}
\fi
%
\subsection{Far from criticality: $\Theta\gg1$ \label{app:step-gg}}
%
Far from the critical point, i.e., for $\Theta\gg1$, and for \emph{symmetry breaking} boundary conditions
$(a_<)=(+)$, $(a_>)=(-)$, and $(b)=(-)$ \eref{eq:exponential-decay} holds
and the integrals in Eqs.~\eqref{eq:app-step-delta-K} and \eqref{eq:app-step-delta-vartheta-4} can be
calculated analytically.
For $\Theta\gg1$ \eref{eq:app-step-delta-K} turns into 
\begin{equation} 
  \label{eq:app-step-K-gg-1}
  \Delta K(\Xi,\Theta\gg1,\Delta)=
  2(A_- - A_+)\Theta^d\int_{1+\Xi^2/2}^\infty
  \upd \alpha \arccos\left(\tfrac{\Xi}{\sqrt{2(\alpha-1)}}\right)
  e^{-\alpha\Theta}.
\end{equation} 
Substituting $\alpha\mapsto \beta=2(\alpha-2)/\Xi^2$ one has
\begin{equation} 
  \label{eq:app-step-K-gg-2}
  \Delta K(\Xi,\Theta\gg1,\Delta)=
  \Xi^2(A_- -A_+)\Theta^de^{-\Theta}\int_1^\infty
  \upd \beta \arccos(\beta^{-1})
  e^{-\Xi^2\Theta\beta/2}.
\end{equation} 
Integrating by parts leads to
\begin{equation} 
  \label{eq:app-step-K-gg-3}
  \Delta K(\Xi,\Theta\gg1,\Delta)=
  (A_- -A_+)\Theta^{d-1} e^{-\Theta}
  \int_1^\infty
  \upd \beta \frac{1}{\beta\sqrt{\beta-1}}
  e^{-\Xi^2\Theta\beta/2}.
\end{equation} 
By using the relation \cite{note:3}
%
\begin{equation} 
  \label{eq:app-erfc}
  \int_1^\infty
  \upd \beta \frac{1}{\beta\sqrt{\beta-1}}
  e^{-a^2\beta}=\pi\erfc(a),
\end{equation} 
where $a>0$ and $\erfc(a)=1-\erf(a)=2\pi^{-1/2}\int_a^\infty\upd t \exp(-t^2)$ is the complementary error function, we finally arrive at
\begin{equation} 
  \label{eq:app-step-K-gg-4}
  \Delta K(\Xi,\Theta\gg1,\Delta)=
  \pi(A_- -A_+)\Theta^{d-1} e^{-\Theta}
  \erfc(\Xi\sqrt{\Theta/2}).
\end{equation} 
The scaling function $\Kmpm$ for $\Theta\gg1$ in the homogeneous case [Sec.~\ref{sec:homog}]
is given by \cite{gambassi:2009}
\begin{equation} 
  \label{eq:app-step-K-homog}
  \Kmpm(\Theta\gg1,\Delta\to0)=2\pi A_\pm \Theta^{d-1}e^{-\Theta}
\end{equation}
and from Eqs.~\eqref{eq:step-K}, \eqref{eq:app-step-K-gg-4}, and \eqref{eq:app-step-K-homog} one obtains
the expression for $\psi_{(-|+,-)}$ as given in \eref{eq:erf}.
Similarly, after rewriting \eref{eq:app-step-delta-vartheta-4} for $\Theta\gg1$ as
\begin{equation} 
  \label{eq:app-step-vartheta-gg-1}
  \Delta \vartheta(\Xi,\Theta\gg1,\Delta)=
  (A_- -A_+)\Theta^{d} e^{-\Theta}\frac{\Xi^4}{2}
  \int_1^\infty
  \upd s \left(s\arccos(s^{-1/2})-\sqrt{s-1}\right)
  e^{-\Xi^2\Theta s/2},
\end{equation} 
one can integrate by parts, which yields
\begin{equation} 
  \label{eq:app-step-vartheta-gg-2}
  \Delta \vartheta(\Xi,\Theta\gg1,\Delta)=
  (A_- -A_+)\Theta^{d-2} e^{-\Theta}
  \int_1^\infty
  \upd s \frac{1}{\sqrt{s-1}}\left[\frac{1}{s}+\frac{\Theta\Xi^2}{2}\left(1+\Theta\Xi^2\right)-\frac{\Theta^2\Xi^4}{2}s\right]
  e^{-\Xi^2\Theta s/2}.
\end{equation} 
Using \eref{eq:app-erfc} and the relations
[which follow from taking successive derivatives $-d/d(a^2)$ of \eref{eq:app-erfc}]
\begin{equation} 
  \label{eq:app-step-vartheta-gg-3}
  \int_1^\infty\upd s\frac{1}{\sqrt{s-1}}e^{-a^2s}=\frac{\sqrt{\pi}}{a}e^{-a^2},\qquad
  \int_1^\infty\upd s\frac{s}{\sqrt{s-1}}e^{-a^2s}=\frac{\sqrt{\pi}}{2a^3}\left(1+2a^2\right)e^{-a^2},
\end{equation}
one ends up with
\begin{equation} 
  \label{eq:app-step-vartheta-gg-4}
  \Delta \vartheta(\Xi,\Theta\gg1,\Delta)=
  \pi (A_- -A_+)\Theta^{d-2} e^{-\Theta}
  \erfc(\Xi\sqrt{\Theta/2}).
\end{equation} 
Together with the expression for the homogeneous case [see Sec.~\ref{sec:homog} and Ref.~\onlinecite{gambassi:2009}],
\begin{equation} 
  \label{eq:app-step-vartheta-homog}
  \varthetampm(\Theta\gg1,\Delta\to0)=2\pi A_\pm \Theta^{d-2}e^{-\Theta},
\end{equation}
one obtains the expression for $\omega_{(-|+,-)}$ given in \eref{eq:erf}.
\section{Derjaguin approximation for a single chemical lane\label{app:stripe}}
%
Based on the assumption of additivity which underlies the Derjaguin approximation one can use the 
results presented in Sec.~\ref{sec:step} for a chemical step in order to study a chemical lane.
The chemical lane configuration can be regarded as the superposition of two chemical steps,
one $(A)$ being a chemical step located at $x=-L$ with $(a|a_{\stripe})$ BC, and the other one
$(B)$ being a chemical step located at $x=L$ with $(a_{\stripe}|a)$ BC.
This superposition overcounts a contribution corresponding to a homogeneous substrate with
$(a_{\stripe})$ BC which must be subtracted [see \eref{eq:step-K}]:
\begin{alignat}{8} 
  \label{eq:testdia}
  \left.
    \begin{aligned}
     (A):\;\;\stackrel{(a)}{\pprule}\!\!\!\!\!\!
     \mathop{|}_{-L}\!\!\!\!\!\!
      \stackrel{(a_{\stripe})}{\photon\photon}
      \\
  \fbox{$+$}\qquad\qquad
  \\
     (B):\;\; \stackrel{(a_{\stripe})}{\photon\photon}\!\!\!\!
     \mathop{|}_{L}\!\!\!\!
      \stackrel{(a)}{\pprule}
  \end{aligned}
  \right\}&\fbox{$-$}&\;\;
  \photon\!\!
  \stackrel{(a_{\stripe})}{\photon}\!\!
  \photon
  &=&\;\;
    \stackrel{(a)}{\pprule}\!\!\!\!\!\!
    \mathop{|}_{-L}\!\!\!\!\!\!
    \stackrel{(a_{\stripe})}{\photon}\!\!\!\!
    \mathop{|}_{L}\!\!\!\!
    \stackrel{(a)}{\pprule}
    \nonumber\\
    \Kstep^{(A)}\,+\;\Kstep^{(B)}\qquad
    &-&
    K_{(a_{\stripe},b)}\qquad
    &=&
    \Kstripe,\qquad
\end{alignat} 
where
\begin{equation} 
  \label{eq:configA}
  \Kstep^{(A)} (\Lambda,\Xi,\Theta,\Delta) 
  =  \frac{K_{(a,b)}+K_{(a_{\stripe},b)}}{2}\\
                                         +\frac{K_{(a,b)}-K_{(a_{\stripe},b)}}{2}
                                         \psi_{(a|a_{\stripe},b)}(\Xi+\Lambda,\Theta,\Delta)
\end{equation} 
and
\begin{equation} 
  \label{eq:configB}
  \Kstep^{(B)} (\Lambda,\Xi,\Theta,\Delta) =  \frac{K_{(a,b)}+K_{(a_{\stripe},b)}}{2}\\
                                         +\frac{K_{(a_{\stripe},b)}-K_{(a,b)}}{2}
                                    \psi_{(a_{\stripe}|a,b)}(\Xi-\Lambda,\Theta,\Delta).
\end{equation} 
Since within the DA $\psi_{(a_{\stripe}|a,b)}=\psi_{(a|a_{\stripe},b)}$, Eqs.~\eqref{eq:testdia}--\eqref{eq:configB}
and \eref{eq:stripe-psi} lead directly to \eref{eq:stripe-psi-da}.
The procedure for calculating the critical Casimir potential is analogous to the one discussed here for the force and leads to \eref{eq:stripe-omega-da}.
%
\section{Derjaguin approximation for periodic chemical patterns\label{app:period}}
%
In order to obtain the scaling function for the critical Casimir force and the potential of a sphere
close to a periodic chemical pattern one can follow a procedure analogous to the one
presented in Appendix~\ref{app:stripe}.
Indeed, in order to form a lane $\stripe'$ with $(a_1)$ BC on an otherwise homogeneous 
\emph{portion} of a substrate with $(a_2)$ BC and lateral extension $P$, one can proceed as follows:
\begin{itemize}
\item[(A):] superimpose onto the substrate the single chemical lane $\stripe$ studied in Sec.~\ref{sec:stripe}, 
  with $a_\ell=a_1$, $a=a_2$, suitably positioned in space such that it coincides with the lane $\ell'$ to be formed.
\item[(B):] subtract the contribution of a homogeneous substrate with $(a_2)$ BC, which is overcounted 
  in the previous superposition. 
  After this subtraction, the contribution to the force resulting from that part -- marked by $(?)$ in \eref{eq:diaper12} -- 
  of the original substrate which is not affected by the formation of the extra lane is unchanged.
\end{itemize}
\begin{alignat}{8}
  \label{eq:diaper12}
  \begin{aligned}
    \phantom{(A):\;\fbox{$+$}\,}
    \stackrel{(?)}{\gluon}\!\!{\overbrace{\stackrel{(a_2)}{\pprule\pprule\pprule}}^{P}}\!\!\stackrel{(?)}{\gluon}
    \\[3mm]
    (A):\;\fbox{$+$}\,%
    \stackrel{(a_2)}{\pprule\pprule}\!\!\!\!\!\!\!\!\!\!\!\!
    \mathop{|}_{X'-\frac{L_1}{2}}   \!\!\!\!\!\!\!\!\!\!\!
    \stackrel{(a_1)}{\photon}       \!\!\!\!\!\!\!\!\!\!\!\!
    \mathop{|}_{X'+\frac{L_1}{2}}   \!\!\!\!\!\!\!\!\!\!\!\!
    \stackrel{(a_2)}{\pprule\pprule}
    \\[1mm]
    (B):\;\fbox{$-$}\,
    \stackrel{(a_2)}{\pprule\pprule\pprule\pprule\pprule}
    \\[2mm]
    \phantom{(A):\;}\;
    =
    \;\,\stackrel{(?)}{\gluon}
    \stackrel{(a_2)}{\pprule}\!\!\!\!\!\!\!\!\!\!\!\!
    \mathop{|}_{X'-\frac{L_1}{2}}\!\!\!\!\!\!\!\!\!\!\!
    \stackrel{(a_1)}{\photon}\!\!\!\!\!\!\!\!\!\!\!\!
    \mathop{|}_{X'+\frac{L_1}{2}}\!\!\!\!\!\!\!\!\!\!\!\!
    \stackrel{(a_2)}{\pprule}
    \stackrel{(?)}{\gluon}
\ifTwocolumn
    \\[-1mm]
\else
    \\[-3mm]
\fi
    \qquad\qquad\qquad\text{\small lane $\ell'$}\qquad\qquad\;\;\;\;\;\;\,
  \end{aligned}
\end{alignat}
\par
The contribution $\Delta F$ to the critical Casimir force experienced by a colloid close to such a substrate and due to the addition of the lane is characterized by the scaling function [see~\eref{eq:stripe-psi}]
\begin{multline}
\label{eq:pippo}
\Delta K(\lambda,\Pi,\Xi-\Xi',\Theta,\Delta\rightarrow 0) = K_\stripe(\Pi\tfrac{\lambda}{2},\Xi-\Xi',\Theta) - K_{(a_2,b)}
=\frac{K_{(a_2,b)}-K_{(a_1,b)}}{2}\\
\times   \left[\psi_{(a_1|a,b)}(\Xi-\Xi' + \Pi \tfrac{\lambda}{2},\Theta,\Delta\rightarrow 0)-\psi_{(a_1|a,b)}(\Xi-\Xi' - \Pi \tfrac{\lambda}{2},
\Theta,\Delta\rightarrow 0)\right]
\end{multline}
where we have used the relation $(L_1/2)/\sqrt{RD} = \Pi \lambda/2$ and have introduced $\Xi' \equiv X'/\sqrt{RD}$, with $X'$ 
as the position of the center of the added lane $\stripe'$.
The force resulting from a periodic pattern can now be obtained by starting out with a homogeneous substrate with $(a_2)$ BC and by iterating the procedure discussed above which adds progressively displaced lanes at positions $X' = n P$, i.e., $\Xi' = n\Pi$, with $n\in \mathbb{Z}$. The resulting force is characterized by the scaling function
\begin{equation}
\Kperiod(\lambda,\Pi,\Xi,\Theta,\Delta\to0) = K_{(a_2,b)} + \sum_{n=-\infty}^{+\infty} \Delta K(\lambda,\Pi,\Xi-n\Pi,\Theta,\Delta\to0)
\end{equation}
which, together with \eref{eq:period-psi}, yields immediately \eref{eq:period-psi-da} for $\psiperiod$.
\par
For $\lambda=0$ or $\lambda=1$ one recovers from \eref{eq:period-psi-da} the homogeneous cases
with $(a_2,b)$ BC or $(a_1,b)$ BC, respectively.
Obviously, for $\lambda=0$, the sum in \eref{eq:period-psi-da} vanishes, and one is left
with $\psiperiod(\lambda=0,\Pi,\Xi,\Theta,\Delta\to0)=1$, corresponding to $(a_2,b)$ BC.
On the other hand for $\lambda=1$, the sum in \eref{eq:period-psi-da} can be easily evaluated
[see \eref{eq:step-psi-da} for $|\Xi|\to\infty$]:
\begin{multline} 
  \label{eq:app-period-sum-1}
\lim_{M,N\to\infty} \sum_{n=-M}^{N}
  \left\{\psi_{(a_1|a_2,b)}(\Xi+\Pi(n+\tfrac{1}{2}),\Theta,\Delta)
  -\psi_{(a_1|a_2,b)}(\Xi+\Pi(n-\tfrac{1}{2}),\Theta,\Delta)\right\}\\
=
\lim_{M,N\to\infty}
  \left\{\psi_{(a_1|a_2,b)}(\Xi+\Pi(N+\tfrac{1}{2}),\Theta,\Delta)
  -\psi_{(a_1|a_2,b)}(\Xi+\Pi(-M-\tfrac{1}{2}),\Theta,\Delta)\right\}=-2,
\end{multline} 
where we have used the fact that $\psi_{(a_1|a_2,b)}(\Xi=\pm\infty,\Theta,\Delta)=\mp1$.
Accordingly, $\psiperiod(\lambda=1,\Pi,\Xi,\Theta,\Delta\to0)=-1$, which corresponds to the homogeneous
case with $(a_1,b)$ BC.
\par
In the limit $\Pi\to0$ (i.e., for very fine patterns compared with $\sqrt{RD}$), the sum
in \eref{eq:period-psi-da} turns into an integral:
\begin{multline} 
  \label{eq:app-period-sum-2}
  \sum_{n=-\infty}^{\infty}
  \left\{\psi_{(a_1|a_2,b)}(\Xi+\Pi(n+\tfrac{\lambda}{2}),\Theta,\Delta)
  -\psi_{(a_1|a_2,b)}(\Xi+\Pi(n-\tfrac{\lambda}{2}),\Theta,\Delta)\right\}\\
  \xrightarrow[\Pi\to0]{}
  \frac{1}{\Pi}
  \int_{-\infty}^{\infty}\upd \eta
  \left\{\psi_{(a_1|a_2,b)}(\Xi+\eta+\tfrac{\Pi\lambda}{2},\Theta,\Delta)
  -\psi_{(a_1|a_2,b)}(\Xi+\eta-\tfrac{\Pi\lambda}{2},\Theta,\Delta)\right\}\\
  =
  \int_{-\infty}^\infty \upd \eta\;\lambda\;\frac{\upd}{\upd\eta}
  \psi_{(a_1|a_2,b)}(\Xi+\eta,\Theta,\Delta)\\
  =
       \lambda \left\{\psi_{(a_1|a_2,b)}(+\infty,\Theta,\Delta)-
            \psi_{(a_1|a_2,b)}(-\infty,\Theta,\Delta)\right\}=-2\lambda,
\end{multline}
and finally one finds \eref{eq:small-period}.
\par
For completeness, we provide the corresponding expression for the scaling function of the critical
Casimir potential $\omegaperiod$ within the DA:
\ifTwocolumn
\begin{equation}
  \label{eq:period-omega-da}
  \omegaperiod(\lambda,\Pi,\Xi,\Theta,\Delta\to0)=
  1+\sum_{n=-\infty}^{\infty} \left\{\omega_{(a_1|a_2,b)}(\Xi+\Pi(n+\tfrac{\lambda}{2}),\Theta,\Delta\to0)
                              -\omega_{(a_1|a_2,b)}(\Xi+\Pi(n-\tfrac{\lambda}{2}),\Theta,\Delta\to0)\right\}.
\end{equation} 
\else
\begin{multline}
  \label{eq:period-omega-da}
  \omegaperiod(\lambda,\Pi,\Xi,\Theta,\Delta\to0)=\\
  1+\sum_{n=-\infty}^{\infty} \left\{\omega_{(a_1|a_2,b)}(\Xi+\Pi(n+\tfrac{\lambda}{2}),\Theta,\Delta\to0)
                              -\omega_{(a_1|a_2,b)}(\Xi+\Pi(n-\tfrac{\lambda}{2}),\Theta,\Delta\to0)\right\}.
\end{multline} 
\fi
In the limit $\Pi\to0$, $\omegaperiod$ reduces to
\begin{equation} 
  \label{eq:small-period-omega}
  \omegaperiod(\lambda,\Pi\to0,\Xi,\Theta,\Delta\to0)=1-2\lambda.
\end{equation} 
Accordingly, within the DA and in the limit $\Pi\to0$ the critical Casimir potential is the average 
of the ones corresponding to the two boundary conditions, weighted with the corresponding relative stripe width:
\begin{equation} 
  \label{eq:small-period-vartheta}
  \varthetaperiod(\lambda,\Pi\to0,\Xi,\Theta,\Delta\to0)=
    \lambda \vartheta_{(a_1,b)}(\Theta,\Delta\to0)+(1-\lambda)\vartheta_{(a_2,b)}(\Theta,\Delta\to0).
\end{equation} 

\ifTwocolumn
\end{widetext}
\fi

\section{Cylinder close to a patterned substrate\label{app:cylinder}}
%
\subsection{Derjaguin approximation for a homogeneous substrate}
Similarly to the case of a sphere discussed before, the critical Casimir force
$\Fab^\cyl$ per unit length acting on a (three-dimensional) cylinder of radius $R$ with $(b)$ BC close to and parallel 
to a substrate with $(a)$ BC at a surface-to-surface distance $D$ can be expressed in terms of 
a universal scaling function $K^\cyl$:
\begin{equation} 
  \label{eq:app-cyl-force}
  \Fab^\cyl(D,R,T)=k_BT\frac{R^{1/2}}{D^{d-1/2}}\Kab^\cyl(\Theta,\Delta),
\end{equation} 
with $\Theta=\sgn(t)\,D/\xi_\pm$ and $\Delta=D/R$ as before.
Equation~\eqref{eq:app-cyl-force} describes a force divided by a length and per $D^{d-3}$ which for 
$d=4$ corresponds to considering $\Fab^\cyl$ per length of its axis and per length of the
\emph{extra} translationally invariant direction of a hypercylinder [compare \eref{eq:force-homog}].
The geometric prefactor in \eref{eq:app-cyl-force}, however,  differs from the one for the sphere [\eref{eq:force-homog}]
because it is chosen such that within the 
DA ($\Delta\to0$) the scaling function $\Kab^\cyl$ attains a nonzero and finite limit, as discussed before.
The DA can be implemented along the lines of Sec.~\ref{sec:homog-da} for the sphere.
Here the surface of the cylindrical colloid is decomposed into pairs of infinitely narrow stripes of
combined area $\upd S=2 M \upd \rho$, positioned parallel to the substrate at a distance $L(\rho)$ from it
[\eref{eq:da-L}] and each at a distance $\rho$ from the symmetry plane of the configuration.
$M$ is the length of the cylinder and drops out from $\Fab^\cyl$ which follows analogously from
Eqs.~\eqref{eq:da-dF} and \eqref{eq:da-def}:
\begin{equation} 
  \label{eq:app-cyl-force-da-def}
  \Fab^\cyl(D,R,T)/k_BT\simeq 2  \int_0^R \upd\rho  \left[L(\rho)\right]^{-d} \kab(\sgn(t)\, L(\rho)/\xi_\pm),
\end{equation} 
where $L(\rho)$ is given in \eref{eq:da-L}.
Finally, in the limit $\Delta\to0$ we obtain 
\begin{equation} 
  \label{eq:app-cyl-force-da}
  \Kab^\cyl(\Theta,\Delta\to0)=\sqrt{2}\int\limits_1^\infty\upd\alpha\,(\alpha-1)^{-\frac{1}{2}}\,\alpha^{-d}\,\kab(\Theta\alpha).
\end{equation} 
At the bulk critical point $\Theta=0$ one finds $\Kab^\cyl(0,0)=\sqrt{2\pi}[\Gamma(d-\frac{1}{2})/\Gamma(d)]\Dab$ so that
$\Kab^\cyl(0,0)=[3\pi/(4\sqrt{2})]\Dab \simeq 1.66 \times\Dab$ for $d=3$ and 
$\Kab^\cyl(0,0)=[5\pi/(8\sqrt{2})]\Dab \simeq 1.38 \times\Dab$ for $d=4$.
\subsection{Derjaguin approximation for a chemical step}
Here, we assume that the axis of the cylinder is parallel to the chemical step,
i.e., perpendicular to the $x$ direction [\fref{fig:sketch}], as well as parallel to the substrate.
The projection of the position of the axis of the cylinder with respect to the $x$ direction
is denoted by $X$, so that at $X=0$ the cylinder is positioned directly above the chemical step [\fref{fig:sketch}].
Accordingly, the problem is effectively two-dimensional and the corresponding DA can be performed much easier
than in Appendix~\ref{app:step}.
Following an approach analogous to the one adopted for the sphere in Sec.~\ref{sec:step} and in Appendix~\ref{app:step},  we rewrite the normal critical Casimir
force per unit length acting on the cylinder as in  \eref{eq:app-step-force-separation}: 
\begin{equation} 
  \label{eq:app-cyl-F-step}
  \Fstep^\cyl(X,D,R,T)=F_\Alb^\cyl(D,R,T)+\Delta F^\cyl(X,D,R,T).
\end{equation} 
Within the DA we find for $\Delta\to0$ [compare \eref{eq:app-step-delta-f-int-rewritten-2}]
\begin{equation} 
  \label{eq:app-cyl-delta-F}
  \Delta F^\cyl(X,D,R,T)=k_BT\frac{R^{1/2}}{D^{d-1/2}}\Delta K^\cyl(\Xi,\Theta,\Delta\to0),
\end{equation} 
where [compare \eref{eq:app-step-delta-K}]
\begin{equation} 
  \label{eq:app-cyl-delta-K}
  \Delta K^\cyl(\Xi,\Theta,\Delta\to0)=
  \sqrt{2}\int_{1+\Xi^2/2}^\infty\upd\alpha\,(\alpha-1)^{-\frac{1}{2}}\,\alpha^{-d}\Delta k(\Theta\alpha).
\end{equation} 
Using \eref{eq:app-cyl-delta-K} and \eref{eq:app-cyl-force-da} we find for the whole range of values of $\Xi$ 
the scaling function $\psi_{(a_<|a_>,b)}^\cyl$ which is defined completely analogous to \eref{eq:step-K} 
[compare \eref{eq:step-psi-da}]:
\ifTwocolumn
\begin{multline} 
  \label{eq:app-cyl-psi-step}
  \psi_{(a_<|a_>,b)}^\cyl(\Xi\gtrless0,\Theta,\Delta\to0)=
    \mp 1\\
    \pm
    \frac{\sqrt{2}
    \int_{1+\Xi^2/2}^\infty\upd\alpha\,(\alpha-1)^{-\frac{1}{2}}\,\alpha^{-d}\Delta k(\Theta\alpha)}
    {\Kasb^\cyl(\Theta,\Delta\to0)-\Kalb^\cyl(\Theta,\Delta\to0)}.
\end{multline} 
\else
\begin{equation} 
  \label{eq:app-cyl-psi-step}
  \psi_{(a_<|a_>,b)}^\cyl(\Xi\gtrless0,\Theta,\Delta\to0)=
    \mp 1
    \pm
    \frac{\sqrt{2}
    \int_{1+\Xi^2/2}^\infty\upd\alpha\,(\alpha-1)^{-\frac{1}{2}}\,\alpha^{-d}\Delta k(\Theta\alpha)}
    {\Kasb^\cyl(\Theta,\Delta\to0)-\Kalb^\cyl(\Theta,\Delta\to0)}.
\end{equation} 
\fi
\subsection{Derjaguin approximation for a periodic chemical pattern}
The derivation of the scaling function for the critical Casimir force
acting on the cylinder close to and aligned with a periodic chemical pattern as studied
in Sec.~\ref{sec:cylinder} is analogous to the one for the sphere described in 
Appendix~\ref{app:period}.
The final formula for $\psiperiod^\cyl$ is the same as  in \eref{eq:period-psi-da}
with $\psi_{(a_1|a_2,b)}$ replaced by $\psi_{(a_1|a_2,b)}^\cyl$ given by \eref{eq:app-cyl-psi-step}.
This renders the critical Casimir force per unit length 
\begin{equation}
  \Fperiod^\cyl(L_1,P,X,D,R,T)=k_BT\frac{R^{1/2}}{D^{d-1/2}}\Kperiod^\cyl(\lambda,\Pi,\Xi,\Theta,\Delta)
\end{equation}
where $\Kperiod^\cyl$ is defined as in \eref{eq:period-psi} with 
$K_{(a_1,b)}$ and $K_{(a_2,b)}$
replaced by 
$K_{(a_1,b)}^\cyl$ and $K_{(a_2,b)}^\cyl$, respectively, which are given by \eref{eq:app-cyl-force-da},
and with $\psiperiod$ replaced by $\psiperiod^\cyl$.
The corresponding results are shown in \fref{fig:cylinder}.

%
%
%

\ifTwocolumn
\clearpage
\fi


\begin{thebibliography}{64}
\expandafter\ifx\csname natexlab\endcsname\relax\def\natexlab#1{#1}\fi
\expandafter\ifx\csname bibnamefont\endcsname\relax
  \def\bibnamefont#1{#1}\fi
\expandafter\ifx\csname bibfnamefont\endcsname\relax
  \def\bibfnamefont#1{#1}\fi
\expandafter\ifx\csname citenamefont\endcsname\relax
  \def\citenamefont#1{#1}\fi
\expandafter\ifx\csname url\endcsname\relax
  \def\url#1{\texttt{#1}}\fi
\expandafter\ifx\csname urlprefix\endcsname\relax\def\urlprefix{URL }\fi
\providecommand{\bibinfo}[2]{#2}
\providecommand{\eprint}[2][]{\url{#2}}

\bibitem[{\citenamefont{Casimir}(1948)}]{casimir:1948}
\bibinfo{author}{\bibfnamefont{H.~G.~B.} \bibnamefont{Casimir}},
  \bibinfo{journal}{Proc. K. Ned. Akad. Wet.} \textbf{\bibinfo{Volume}{51}},
  \bibinfo{pages}{793} (\bibinfo{year}{1948}).

\bibitem[{\citenamefont{Kardar and Golestanian}(1999)}]{kardar:1999}
\bibinfo{author}{\bibfnamefont{M.}~\bibnamefont{Kardar}} \bibnamefont{and}
  \bibinfo{author}{\bibfnamefont{R.}~\bibnamefont{Golestanian}},
  \bibinfo{journal}{Rev. Mod. Phys.} \textbf{\bibinfo{Volume}{71}},
  \bibinfo{pages}{1233} (\bibinfo{year}{1999}).

\bibitem[{\citenamefont{Fisher and de~Gennes}(1978)}]{fisher:1978}
\bibinfo{author}{\bibfnamefont{M.~E.} \bibnamefont{Fisher}} \bibnamefont{and}
  \bibinfo{author}{\bibfnamefont{P.~G.} \bibnamefont{de~Gennes}},
  \bibinfo{journal}{C. R. Acad. Sci., Paris, Ser. B}
  \textbf{\bibinfo{Volume}{287}}, \bibinfo{pages}{207} (\bibinfo{year}{1978}).

\bibitem[{\citenamefont{Gambassi}(2009)}]{gambassi:2009conf}
\bibinfo{author}{\bibfnamefont{A.}~\bibnamefont{Gambassi}},
  \bibinfo{journal}{J. Phys.: Conf. Ser.} \textbf{\bibinfo{Volume}{161}},
  \bibinfo{pages}{012037} (\bibinfo{year}{2009}).

\bibitem[{\citenamefont{Gambassi
  et~al.}(2009{\natexlab{a}})\citenamefont{Gambassi, Hertlein, Helden,
  Bechinger, and Dietrich}}]{gambassi:2009news}
\bibinfo{author}{\bibfnamefont{A.}~\bibnamefont{Gambassi}},
  \bibinfo{author}{\bibfnamefont{C.}~\bibnamefont{Hertlein}},
  \bibinfo{author}{\bibfnamefont{L.}~\bibnamefont{Helden}},
  \bibinfo{author}{\bibfnamefont{C.}~\bibnamefont{Bechinger}},
  \bibnamefont{and} \bibinfo{author}{\bibfnamefont{S.}~\bibnamefont{Dietrich}},
  \bibinfo{journal}{Europhysics News} \textbf{\bibinfo{Volume}{40/1}},
  \bibinfo{pages}{18} (\bibinfo{year}{2009}{\natexlab{a}}).

\bibitem[{\citenamefont{Hertlein et~al.}(2008)\citenamefont{Hertlein, Helden,
  Gambassi, Dietrich, and Bechinger}}]{hertlein:2008}
\bibinfo{author}{\bibfnamefont{C.}~\bibnamefont{Hertlein}},
  \bibinfo{author}{\bibfnamefont{L.}~\bibnamefont{Helden}},
  \bibinfo{author}{\bibfnamefont{A.}~\bibnamefont{Gambassi}},
  \bibinfo{author}{\bibfnamefont{S.}~\bibnamefont{Dietrich}}, \bibnamefont{and}
  \bibinfo{author}{\bibfnamefont{C.}~\bibnamefont{Bechinger}},
  \bibinfo{journal}{Nature} \textbf{\bibinfo{Volume}{451}},
  \bibinfo{pages}{172} (\bibinfo{year}{2008}).

\bibitem[{\citenamefont{Gambassi
  et~al.}(2009{\natexlab{b}})\citenamefont{Gambassi, Macio{\l}ek, Hertlein,
  Nellen, Helden, Bechinger, and Dietrich}}]{gambassi:2009}
\bibinfo{author}{\bibfnamefont{A.}~\bibnamefont{Gambassi}},
  \bibinfo{author}{\bibfnamefont{A.}~\bibnamefont{Macio{\l}ek}},
  \bibinfo{author}{\bibfnamefont{C.}~\bibnamefont{Hertlein}},
  \bibinfo{author}{\bibfnamefont{U.}~\bibnamefont{Nellen}},
  \bibinfo{author}{\bibfnamefont{L.}~\bibnamefont{Helden}},
  \bibinfo{author}{\bibfnamefont{C.}~\bibnamefont{Bechinger}},
  \bibnamefont{and} \bibinfo{author}{\bibfnamefont{S.}~\bibnamefont{Dietrich}},
  \bibinfo{journal}{Phys. Rev. E} \textbf{\bibinfo{Volume}{80}},
  \bibinfo{pages}{061143} (\bibinfo{year}{2009}{\natexlab{b}}).

\bibitem[{\citenamefont{Krech}(1994)}]{krech:book}
\bibinfo{author}{\bibfnamefont{M.}~\bibnamefont{Krech}},
  \emph{\bibinfo{title}{The Casimir Effect in Critical Systems}}
  (\bibinfo{publisher}{World Scientific, Singapore}, \bibinfo{year}{1994}).

\bibitem[{\citenamefont{Brankov et~al.}(2000)\citenamefont{Brankov, Danchev,
  and Tonchev}}]{brankov:book}
\bibinfo{author}{\bibfnamefont{J.~G.} \bibnamefont{Brankov}},
  \bibinfo{author}{\bibfnamefont{D.~M.} \bibnamefont{Danchev}},
  \bibnamefont{and} \bibinfo{author}{\bibfnamefont{N.~S.}
  \bibnamefont{Tonchev}}, \emph{\bibinfo{title}{Theory of critical phenomena in
  finite size systems}} (\bibinfo{publisher}{World Scientific, Singapore},
  \bibinfo{year}{2000}).

\bibitem[{\citenamefont{Krech and Dietrich}(1991)}]{krech:9192all}
\bibinfo{author}{\bibfnamefont{M.}~\bibnamefont{Krech}} \bibnamefont{and}
  \bibinfo{author}{\bibfnamefont{S.}~\bibnamefont{Dietrich}},
  \bibinfo{journal}{Phys. Rev. Lett.} \textbf{\bibinfo{Volume}{66}},
  \bibinfo{pages}{345} (\bibinfo{year}{1991}); \bibinfo{note}{{Phys.} Rev. A
  \textbf{46}, 1886 (1992); ibid 1922 (1992)}.

\bibitem[{\citenamefont{Evans and Stecki}(1994)}]{evans:1994}
\bibinfo{author}{\bibfnamefont{R.}~\bibnamefont{Evans}} \bibnamefont{and}
  \bibinfo{author}{\bibfnamefont{J.}~\bibnamefont{Stecki}},
  \bibinfo{journal}{Phys. Rev. B} \textbf{\bibinfo{Volume}{49}},
  \bibinfo{pages}{8842} (\bibinfo{year}{1994}).

\bibitem[{\citenamefont{Diehl et~al.}(2006)\citenamefont{Diehl, Gr{\"u}neberg,
  and Shpot}}]{diehl:2006}
\bibinfo{author}{\bibfnamefont{H.~W.} \bibnamefont{Diehl}},
  \bibinfo{author}{\bibfnamefont{D.}~\bibnamefont{Gr{\"u}neberg}},
  \bibnamefont{and} \bibinfo{author}{\bibfnamefont{M.~A.} \bibnamefont{Shpot}},
  \bibinfo{journal}{EPL} \textbf{\bibinfo{Volume}{75}}, \bibinfo{pages}{241}
  (\bibinfo{year}{2006}).

\bibitem[{\citenamefont{Zandi et~al.}(2007)\citenamefont{Zandi, Shackell,
  Rudnick, Kardar, and Chayes}}]{zandi:2007}
\bibinfo{author}{\bibfnamefont{R.}~\bibnamefont{Zandi}},
  \bibinfo{author}{\bibfnamefont{A.}~\bibnamefont{Shackell}},
  \bibinfo{author}{\bibfnamefont{J.}~\bibnamefont{Rudnick}},
  \bibinfo{author}{\bibfnamefont{M.}~\bibnamefont{Kardar}}, \bibnamefont{and}
  \bibinfo{author}{\bibfnamefont{L.~P.} \bibnamefont{Chayes}},
  \bibinfo{journal}{Phys. Rev. E} \textbf{\bibinfo{Volume}{76}},
  \bibinfo{pages}{030601} (\bibinfo{year}{2007}).

\bibitem[{\citenamefont{Schmidt and Diehl}(2008)}]{schmidt:2008}
\bibinfo{author}{\bibfnamefont{F.~M.} \bibnamefont{Schmidt}} \bibnamefont{and}
  \bibinfo{author}{\bibfnamefont{H.~W.} \bibnamefont{Diehl}},
  \bibinfo{journal}{Phys. Rev. Lett.} \textbf{\bibinfo{Volume}{101}},
  \bibinfo{pages}{100601} (\bibinfo{year}{2008}).

\bibitem[{\citenamefont{Mohry et~al.}(2008)\citenamefont{Mohry, Macio{\l}ek,
  and Dietrich}}]{mohry:2009}
\bibinfo{author}{\bibfnamefont{T.~F.} \bibnamefont{Mohry}},
  \bibinfo{author}{\bibfnamefont{A.}~\bibnamefont{Macio{\l}ek}},
  \bibnamefont{and} \bibinfo{author}{\bibfnamefont{S.}~\bibnamefont{Dietrich}},
  \bibinfo{journal}{arXiv:1004.0112 (2010); {T.~F.~Mohry}, diploma thesis,
  University of Stuttgart}  (\bibinfo{year}{2008}).

\bibitem[{\citenamefont{Munday et~al.}(2009)\citenamefont{Munday, Capasso, and
  Parsegian}}]{Munday:2009}
\bibinfo{author}{\bibfnamefont{J.~N.} \bibnamefont{Munday}},
  \bibinfo{author}{\bibfnamefont{F.}~\bibnamefont{Capasso}}, \bibnamefont{and}
  \bibinfo{author}{\bibfnamefont{V.~A.} \bibnamefont{Parsegian}},
  \bibinfo{journal}{Nature} \textbf{\bibinfo{Volume}{457}},
  \bibinfo{pages}{170} (\bibinfo{year}{2009}).

\bibitem[{\citenamefont{Garcia and Chan}(1999)}]{garcia:9902all}
\bibinfo{author}{\bibfnamefont{R.}~\bibnamefont{Garcia}} \bibnamefont{and}
  \bibinfo{author}{\bibfnamefont{M.~H.~W.} \bibnamefont{Chan}},
  \bibinfo{journal}{Phys. Rev. Lett.} \textbf{\bibinfo{Volume}{83}},
  \bibinfo{pages}{1187} (\bibinfo{year}{1999}); \bibinfo{note}{{ibid}
  \textbf{88}, 086101 (2002)}.

\bibitem[{\citenamefont{Ganshin et~al.}(2006)\citenamefont{Ganshin,
  Scheidemantel, Garcia, and Chan}}]{ganshin:2006}
\bibinfo{author}{\bibfnamefont{A.}~\bibnamefont{Ganshin}},
  \bibinfo{author}{\bibfnamefont{S.}~\bibnamefont{Scheidemantel}},
  \bibinfo{author}{\bibfnamefont{R.}~\bibnamefont{Garcia}}, \bibnamefont{and}
  \bibinfo{author}{\bibfnamefont{M.~H.~W.} \bibnamefont{Chan}},
  \bibinfo{journal}{Phys. Rev. Lett.} \textbf{\bibinfo{Volume}{97}},
  \bibinfo{pages}{075301} (\bibinfo{year}{2006}).

\bibitem[{\citenamefont{Fukuto et~al.}(2005)\citenamefont{Fukuto, Yano, and
  Pershan}}]{fukuto:2005}
\bibinfo{author}{\bibfnamefont{M.}~\bibnamefont{Fukuto}},
  \bibinfo{author}{\bibfnamefont{Y.~F.} \bibnamefont{Yano}}, \bibnamefont{and}
  \bibinfo{author}{\bibfnamefont{P.~S.} \bibnamefont{Pershan}},
  \bibinfo{journal}{Phys. Rev. Lett.} \textbf{\bibinfo{Volume}{94}},
  \bibinfo{pages}{135702} (\bibinfo{year}{2005}).

\bibitem[{\citenamefont{Rafai et~al.}(2007)\citenamefont{Rafai, Bonn, and
  Meunier}}]{rafai:2007}
\bibinfo{author}{\bibfnamefont{S.}~\bibnamefont{Rafai}},
  \bibinfo{author}{\bibfnamefont{D.}~\bibnamefont{Bonn}}, \bibnamefont{and}
  \bibinfo{author}{\bibfnamefont{J.}~\bibnamefont{Meunier}},
  \bibinfo{journal}{Physica A} \textbf{\bibinfo{Volume}{386}},
  \bibinfo{pages}{31} (\bibinfo{year}{2007}).

\bibitem[{\citenamefont{Hucht}(2007)}]{hucht:2007}
\bibinfo{author}{\bibfnamefont{A.}~\bibnamefont{Hucht}},
  \bibinfo{journal}{Phys. Rev. Lett.} \textbf{\bibinfo{Volume}{99}},
  \bibinfo{pages}{185301} (\bibinfo{year}{2007}).

\bibitem[{\citenamefont{Vasilyev et~al.}(2007)\citenamefont{Vasilyev, Gambassi,
  Macio{\l}ek, and Dietrich}}]{vasilyev:2007}
\bibinfo{author}{\bibfnamefont{O.}~\bibnamefont{Vasilyev}},
  \bibinfo{author}{\bibfnamefont{A.}~\bibnamefont{Gambassi}},
  \bibinfo{author}{\bibfnamefont{A.}~\bibnamefont{Macio{\l}ek}},
  \bibnamefont{and} \bibinfo{author}{\bibfnamefont{S.}~\bibnamefont{Dietrich}},
  \bibinfo{journal}{EPL} \textbf{\bibinfo{Volume}{80}}, \bibinfo{pages}{60009}
  (\bibinfo{year}{2007}).

\bibitem[{\citenamefont{Vasilyev et~al.}(2009)\citenamefont{Vasilyev, Gambassi,
  Macio{\l}ek, and Dietrich}}]{vasilyev:2009}
\bibinfo{author}{\bibfnamefont{O.}~\bibnamefont{Vasilyev}},
  \bibinfo{author}{\bibfnamefont{A.}~\bibnamefont{Gambassi}},
  \bibinfo{author}{\bibfnamefont{A.}~\bibnamefont{Macio{\l}ek}},
  \bibnamefont{and} \bibinfo{author}{\bibfnamefont{S.}~\bibnamefont{Dietrich}},
  \bibinfo{journal}{Phys. Rev. E} \textbf{\bibinfo{Volume}{79}},
  \bibinfo{pages}{041142} (\bibinfo{year}{2009}); \bibinfo{note}{{ibid}
  \textbf{80}, 039902(E) (2009)}.

\bibitem[{\citenamefont{Hasenbusch}(2009)}]{Hasenbusch:2009}
\bibinfo{author}{\bibfnamefont{M.}~\bibnamefont{Hasenbusch}},
  \bibinfo{journal}{J. Stat. Mech., P07031 (2009); Phys. Rev. E \textbf{80},
  061120 (2009); arXiv:0907.2847}  (\bibinfo{year}{2009}).

\bibitem[{\citenamefont{Burkhardt and Eisenriegler}(1995)}]{Burkhardt:1995}
\bibinfo{author}{\bibfnamefont{T.~W.} \bibnamefont{Burkhardt}}
  \bibnamefont{and}
  \bibinfo{author}{\bibfnamefont{E.}~\bibnamefont{Eisenriegler}},
  \bibinfo{journal}{Phys. Rev. Lett.} \textbf{\bibinfo{Volume}{74}},
  \bibinfo{pages}{3189} (\bibinfo{year}{1995}); \bibinfo{note}{ibid
  {\textbf{78}}, 2867 (1997)}.

\bibitem[{\citenamefont{Eisenriegler and Ritschel}(1995)}]{Eisenriegler:1995}
\bibinfo{author}{\bibfnamefont{E.}~\bibnamefont{Eisenriegler}}
  \bibnamefont{and} \bibinfo{author}{\bibfnamefont{U.}~\bibnamefont{Ritschel}},
  \bibinfo{journal}{Phys. Rev. B} \textbf{\bibinfo{Volume}{51}},
  \bibinfo{pages}{13717} (\bibinfo{year}{1995}).

\bibitem[{\citenamefont{Hanke et~al.}(1998)\citenamefont{Hanke, Schlesener,
  Eisenriegler, and Dietrich}}]{hanke:1998}
\bibinfo{author}{\bibfnamefont{A.}~\bibnamefont{Hanke}},
  \bibinfo{author}{\bibfnamefont{F.}~\bibnamefont{Schlesener}},
  \bibinfo{author}{\bibfnamefont{E.}~\bibnamefont{Eisenriegler}},
  \bibnamefont{and} \bibinfo{author}{\bibfnamefont{S.}~\bibnamefont{Dietrich}},
  \bibinfo{journal}{Phys. Rev. Lett.} \textbf{\bibinfo{Volume}{81}},
  \bibinfo{pages}{1885} (\bibinfo{year}{1998}).

\bibitem[{\citenamefont{Schlesener et~al.}(2003)\citenamefont{Schlesener,
  Hanke, and Dietrich}}]{Schlesener:2003}
\bibinfo{author}{\bibfnamefont{F.}~\bibnamefont{Schlesener}},
  \bibinfo{author}{\bibfnamefont{A.}~\bibnamefont{Hanke}}, \bibnamefont{and}
  \bibinfo{author}{\bibfnamefont{S.}~\bibnamefont{Dietrich}},
  \bibinfo{journal}{J. Stat. Phys.} \textbf{\bibinfo{Volume}{110}},
  \bibinfo{pages}{981} (\bibinfo{year}{2003}).

\bibitem[{\citenamefont{Eisenriegler}(2004)}]{Eisenriegler:2004}
\bibinfo{author}{\bibfnamefont{E.}~\bibnamefont{Eisenriegler}},
  \bibinfo{journal}{J. Chem. Phys.} \textbf{\bibinfo{Volume}{121}},
  \bibinfo{pages}{3299} (\bibinfo{year}{2004}).

\bibitem[{\citenamefont{Kondrat et~al.}(2009)\citenamefont{Kondrat, Harnau, and
  Dietrich}}]{kondrat:2009}
\bibinfo{author}{\bibfnamefont{S.}~\bibnamefont{Kondrat}},
  \bibinfo{author}{\bibfnamefont{L.}~\bibnamefont{Harnau}}, \bibnamefont{and}
  \bibinfo{author}{\bibfnamefont{S.}~\bibnamefont{Dietrich}},
  \bibinfo{journal}{J. Chem. Phys.} \textbf{\bibinfo{Volume}{131}},
  \bibinfo{pages}{204902} (\bibinfo{year}{2009}).

\bibitem[{\citenamefont{Tröndle et~al.}(2009)\citenamefont{Tröndle, Kondrat,
  Gambassi, Harnau, and Dietrich}}]{troendle:2009}
  \bibinfo{author}{\bibfnamefont{M.}~\bibnamefont{Tr{\"o}ndle}},
  \bibinfo{author}{\bibfnamefont{S.}~\bibnamefont{Kondrat}},
  \bibinfo{author}{\bibfnamefont{A.}~\bibnamefont{Gambassi}},
  \bibinfo{author}{\bibfnamefont{L.}~\bibnamefont{Harnau}}, \bibnamefont{and}
  \bibinfo{author}{\bibfnamefont{S.}~\bibnamefont{Dietrich}},
  \bibinfo{journal}{EPL} \textbf{\bibinfo{Volume}{88}}, \bibinfo{pages}{40004}
  (\bibinfo{year}{2009}).

\bibitem[{\citenamefont{Soyka et~al.}(2008)\citenamefont{Soyka, Zvyagolskaya,
  Hertlein, Helden, and Bechinger}}]{soyka:2008}
\bibinfo{author}{\bibfnamefont{F.}~\bibnamefont{Soyka}},
  \bibinfo{author}{\bibfnamefont{O.}~\bibnamefont{Zvyagolskaya}},
  \bibinfo{author}{\bibfnamefont{C.}~\bibnamefont{Hertlein}},
  \bibinfo{author}{\bibfnamefont{L.}~\bibnamefont{Helden}}, \bibnamefont{and}
  \bibinfo{author}{\bibfnamefont{C.}~\bibnamefont{Bechinger}},
  \bibinfo{journal}{Phys. Rev. Lett.} \textbf{\bibinfo{Volume}{101}},
  \bibinfo{pages}{208301} (\bibinfo{year}{2008}).

\bibitem[{\citenamefont{Sprenger et~al.}(2006)\citenamefont{Sprenger,
  Schlesener, and Dietrich}}]{sprenger:2006}
\bibinfo{author}{\bibfnamefont{M.}~\bibnamefont{Sprenger}},
  \bibinfo{author}{\bibfnamefont{F.}~\bibnamefont{Schlesener}},
  \bibnamefont{and} \bibinfo{author}{\bibfnamefont{S.}~\bibnamefont{Dietrich}},
  \bibinfo{journal}{J. Chem. Phys.} \textbf{\bibinfo{Volume}{124}},
  \bibinfo{pages}{134703} (\bibinfo{year}{2006}).

\bibitem[{\citenamefont{Tr{\"o}ndle et~al.}(2008)\citenamefont{Tr{\"o}ndle,
  Harnau, and Dietrich}}]{troendle:2008}
\bibinfo{author}{\bibfnamefont{M.}~\bibnamefont{Tr{\"o}ndle}},
  \bibinfo{author}{\bibfnamefont{L.}~\bibnamefont{Harnau}}, \bibnamefont{and}
  \bibinfo{author}{\bibfnamefont{S.}~\bibnamefont{Dietrich}},
  \bibinfo{journal}{J. Chem. Phys.} \textbf{\bibinfo{Volume}{129}},
  \bibinfo{pages}{124716} (\bibinfo{year}{2008}).

\bibitem[{\citenamefont{Krech}(1997)}]{krech:1997}
\bibinfo{author}{\bibfnamefont{M.}~\bibnamefont{Krech}},
  \bibinfo{journal}{Phys. Rev. E} \textbf{\bibinfo{Volume}{56}},
  \bibinfo{pages}{1642} (\bibinfo{year}{1997}).

\bibitem[{\citenamefont{Leonhardt and Philbin}(2007)}]{leonhardt:2007}
\bibinfo{author}{\bibfnamefont{U.}~\bibnamefont{Leonhardt}} \bibnamefont{and}
  \bibinfo{author}{\bibfnamefont{T.~G.} \bibnamefont{Philbin}},
  \bibinfo{journal}{New J. Phys.} \textbf{\bibinfo{Volume}{9}},
  \bibinfo{pages}{254} (\bibinfo{year}{2007}).

\bibitem[{\citenamefont{Rodriguez et~al.}(2008)\citenamefont{Rodriguez, Munday,
  Joannopoulos, Capasso, Dalvit, and Johnson}}]{rodriguez:2008}
\bibinfo{author}{\bibfnamefont{A.~W.} \bibnamefont{Rodriguez}},
  \bibinfo{author}{\bibfnamefont{J.~N.} \bibnamefont{Munday}},
  \bibinfo{author}{\bibfnamefont{J.~D.} \bibnamefont{Joannopoulos}},
  \bibinfo{author}{\bibfnamefont{F.}~\bibnamefont{Capasso}},
  \bibinfo{author}{\bibfnamefont{D.~A.~R.} \bibnamefont{Dalvit}},
  \bibnamefont{and} \bibinfo{author}{\bibfnamefont{S.~G.}
  \bibnamefont{Johnson}}, \bibinfo{journal}{Phys. Rev. Lett.}
  \textbf{\bibinfo{Volume}{101}}, \bibinfo{pages}{190404}
  (\bibinfo{year}{2008}).

\bibitem[{\citenamefont{Rodriguez
  et~al.}(2010{\natexlab{a}})\citenamefont{Rodriguez, McCauley, Woolf, Capasso,
  Joannopoulos, and Johnson}}]{rodriguez:2009}
\bibinfo{author}{\bibfnamefont{A.~W.} \bibnamefont{Rodriguez}},
  \bibinfo{author}{\bibfnamefont{A.~P.} \bibnamefont{McCauley}},
  \bibinfo{author}{\bibfnamefont{D.}~\bibnamefont{Woolf}},
  \bibinfo{author}{\bibfnamefont{F.}~\bibnamefont{Capasso}},
  \bibinfo{author}{\bibfnamefont{J.~D.} \bibnamefont{Joannopoulos}},
  \bibnamefont{and} \bibinfo{author}{\bibfnamefont{S.~G.}
  \bibnamefont{Johnson}}, \bibinfo{journal}{Phys. Rev. Lett.}
  \textbf{\bibinfo{Volume}{104}}, \bibinfo{pages}{160402}
  (\bibinfo{year}{2010}{\natexlab{a}}).

\bibitem[{\citenamefont{Rahi et~al.}(2009)\citenamefont{Rahi, Kardar, and
  Emig}}]{rahi:2009}
\bibinfo{author}{\bibfnamefont{S.~J.} \bibnamefont{Rahi}},
  \bibinfo{author}{\bibfnamefont{M.}~\bibnamefont{Kardar}}, \bibnamefont{and}
  \bibinfo{author}{\bibfnamefont{T.}~\bibnamefont{Emig}},
  \bibinfo{journal}{arXiv:0911.5364}  (\bibinfo{year}{2009}).

\bibitem[{\citenamefont{Rahi and Zaheer}(2009)}]{rahi:2009a}
\bibinfo{author}{\bibfnamefont{S.~J.} \bibnamefont{Rahi}} \bibnamefont{and}
  \bibinfo{author}{\bibfnamefont{S.}~\bibnamefont{Zaheer}},
  \bibinfo{journal}{arXiv:0909.4510}  (\bibinfo{year}{2009}).

\bibitem[{\citenamefont{Zhao et~al.}(2009)\citenamefont{Zhao, Zhou, Koschny,
  Economou, and Soukoulis}}]{zhao:2009}
\bibinfo{author}{\bibfnamefont{R.}~\bibnamefont{Zhao}},
  \bibinfo{author}{\bibfnamefont{J.}~\bibnamefont{Zhou}},
  \bibinfo{author}{\bibfnamefont{T.}~\bibnamefont{Koschny}},
  \bibinfo{author}{\bibfnamefont{E.~N.} \bibnamefont{Economou}},
  \bibnamefont{and} \bibinfo{author}{\bibfnamefont{C.~M.}
  \bibnamefont{Soukoulis}}, \bibinfo{journal}{Phys. Rev. Lett.}
  \textbf{\bibinfo{Volume}{103}}, \bibinfo{pages}{103602}
  (\bibinfo{year}{2009}).

\bibitem[{\citenamefont{Binder}(1983)}]{binder:1983}
\bibinfo{author}{\bibfnamefont{K.}~\bibnamefont{Binder}}, in
  \emph{\bibinfo{booktitle}{Phase Transitions and Critical Phenomena}}, edited
  by \bibinfo{editor}{\bibfnamefont{C.}~\bibnamefont{Domb}} \bibnamefont{and}
  \bibinfo{editor}{\bibfnamefont{J.~L.} \bibnamefont{Lebowitz}}
  (\bibinfo{publisher}{Academic, London}, \bibinfo{year}{1983}),
  Vol.~\bibinfo{volume}{8}, p.~\bibinfo{pages}{1}.

\bibitem[{\citenamefont{Diehl}(1986)}]{diehl:1986}
\bibinfo{author}{\bibfnamefont{H.~W.} \bibnamefont{Diehl}}, in
  \emph{\bibinfo{booktitle}{Phase Transitions and Critical Phenomena}}, edited
  by \bibinfo{editor}{\bibfnamefont{C.}~\bibnamefont{Domb}} \bibnamefont{and}
  \bibinfo{editor}{\bibfnamefont{J.~L.} \bibnamefont{Lebowitz}}
  (\bibinfo{publisher}{Academic, London}, \bibinfo{year}{1986}),
  Vol.~\bibinfo{volume}{10}, p.~\bibinfo{pages}{75}.

\bibitem[{\citenamefont{Diehl}(1997)}]{diehl:1997}
\bibinfo{author}{\bibfnamefont{H.~W.} \bibnamefont{Diehl}},
  \bibinfo{journal}{Int. J. Mod. Phys. B} \textbf{\bibinfo{Volume}{11}},
  \bibinfo{pages}{3503} (\bibinfo{year}{1997}).

\bibitem[{\citenamefont{Pelissetto and Vicari}(2002)}]{pelissetto:2002}
\bibinfo{author}{\bibfnamefont{A.}~\bibnamefont{Pelissetto}} \bibnamefont{and}
  \bibinfo{author}{\bibfnamefont{E.}~\bibnamefont{Vicari}},
  \bibinfo{journal}{Phys. Rep.} \textbf{\bibinfo{Volume}{368}},
  \bibinfo{pages}{549} (\bibinfo{year}{2002}).

\bibitem[{\citenamefont{Privman et~al.}(1991)\citenamefont{Privman, Hohenberg,
  and Aharony}}]{privman:1991}
\bibinfo{author}{\bibfnamefont{V.}~\bibnamefont{Privman}},
  \bibinfo{author}{\bibfnamefont{P.~C.} \bibnamefont{Hohenberg}},
  \bibnamefont{and} \bibinfo{author}{\bibfnamefont{A.}~\bibnamefont{Aharony}},
  in \emph{\bibinfo{booktitle}{Phase Transitions and Critical Phenomena}},
  edited by \bibinfo{editor}{\bibfnamefont{C.}~\bibnamefont{Domb}}
  \bibnamefont{and} \bibinfo{editor}{\bibfnamefont{J.~L.}
  \bibnamefont{Lebowitz}} (\bibinfo{publisher}{Academic, London},
  \bibinfo{year}{1991}), Vol.~\bibinfo{volume}{14}, p. \bibinfo{pages}{1 and p.
  364}.

\bibitem[{\citenamefont{Tarko and Fisher}(1973)}]{tarko:all}
\bibinfo{author}{\bibfnamefont{H.~B.} \bibnamefont{Tarko}} \bibnamefont{and}
  \bibinfo{author}{\bibfnamefont{M.~E.} \bibnamefont{Fisher}},
  \bibinfo{journal}{Phys. Rev. Lett.} \textbf{\bibinfo{Volume}{31}},
  \bibinfo{pages}{926} (\bibinfo{year}{1973}); \bibinfo{note}{{Phys.} Rev. B
  \textbf{11}, 1217 (1975)}.

\bibitem[{\citenamefont{Borjan and Upton}(2008)}]{borjan:2008}
\bibinfo{author}{\bibfnamefont{Z.}~\bibnamefont{Borjan}} \bibnamefont{and}
  \bibinfo{author}{\bibfnamefont{P.~J.} \bibnamefont{Upton}},
  \bibinfo{journal}{Phys. Rev. Lett.} \textbf{\bibinfo{Volume}{101}},
  \bibinfo{pages}{125702} (\bibinfo{year}{2008}).

\bibitem[{\citenamefont{Burkhardt and Diehl}(1994)}]{burkhardt:1994}
\bibinfo{author}{\bibfnamefont{T.~W.} \bibnamefont{Burkhardt}}
  \bibnamefont{and} \bibinfo{author}{\bibfnamefont{H.~W.} \bibnamefont{Diehl}},
  \bibinfo{journal}{Phys. Rev. B} \textbf{\bibinfo{Volume}{50}},
  \bibinfo{pages}{3894} (\bibinfo{year}{1994}).

\bibitem[{\citenamefont{Diehl and Smock}(1993)}]{diehl:1993}
\bibinfo{author}{\bibfnamefont{H.~W.} \bibnamefont{Diehl}} \bibnamefont{and}
  \bibinfo{author}{\bibfnamefont{M.}~\bibnamefont{Smock}},
  \bibinfo{journal}{Phys. Rev. B} \textbf{\bibinfo{Volume}{47}},
  \bibinfo{pages}{5841} (\bibinfo{year}{1993}); \bibinfo{note}{ibid
  \textbf{48}, 6740 (1993)}.

\bibitem[{\citenamefont{Derjaguin}(1934)}]{derjaguin:1934}
\bibinfo{author}{\bibfnamefont{B.}~\bibnamefont{Derjaguin}},
  \bibinfo{journal}{Kolloid Z.} \textbf{\bibinfo{Volume}{69}},
  \bibinfo{pages}{155} (\bibinfo{year}{1934}).

\bibitem[{mcd()}]{mcdata}
\bibinfo{note}{For the scaling function in $d=3$ of the critical Casimir force
  acting on two parallel planar walls with $(\pm)$ BC, we use the approximation
  denoted by $(i)$ in Figs. 9 and 10 of Ref.~\onlinecite{vasilyev:2009}. The
  uncertainty of the overall amplitude of the scaling functions is about 10\%
  to 20\% as indicated by the different results obtained by the various
  approximations used in Ref.~\onlinecite{vasilyev:2009}. Correspondingly, this
  uncertainty affects our predictions for the scaling functions $\Kpmm$,
  $\varthetapmm$, $\Kstep^\parallel$, $\Kperiod$, $\Phiperiod$, and
  $\Kperiod^\cyl$ based on such Monte Carlo simulation data. However, the
  normalized scaling functions $\omega_{(+|-,-)}$, $\omegastripe$,
  $\psiperiod$, and $\psiperiod^\cyl$ are affected less leading to an
  uncertainty of at most 3\%.}

\bibitem[{\citenamefont{Nellen et~al.}(2009)\citenamefont{Nellen, Helden, and
  Bechinger}}]{nellen:2009}
\bibinfo{author}{\bibfnamefont{U.}~\bibnamefont{Nellen}},
  \bibinfo{author}{\bibfnamefont{L.}~\bibnamefont{Helden}}, \bibnamefont{and}
  \bibinfo{author}{\bibfnamefont{C.}~\bibnamefont{Bechinger}},
  \bibinfo{journal}{EPL} \textbf{\bibinfo{Volume}{88}}, \bibinfo{pages}{26001}
  (\bibinfo{year}{2009}).

\bibitem[{\citenamefont{Vogt}(2009)}]{vogt:2009a}
\bibinfo{author}{\bibfnamefont{D.}~\bibnamefont{Vogt}},
  \bibinfo{journal}{diploma thesis, University of Stuttgart}
  (\bibinfo{year}{2009}).

\bibitem[{vog()}]{vogt:2009}
\bibinfo{note}{Private communication by D. Vogt, O. Zvyagolskaya, and C.
  Bechinger}.

\bibitem[{\citenamefont{Rodriguez
  et~al.}(2010{\natexlab{b}})\citenamefont{Rodriguez, Woolf, McCauley, Capasso,
  Joannopoulos, and Johnson}}]{rodriguez:2010}
\bibinfo{author}{\bibfnamefont{A.~W.} \bibnamefont{Rodriguez}},
  \bibinfo{author}{\bibfnamefont{D.}~\bibnamefont{Woolf}},
  \bibinfo{author}{\bibfnamefont{A.~P.} \bibnamefont{McCauley}},
  \bibinfo{author}{\bibfnamefont{F.}~\bibnamefont{Capasso}},
  \bibinfo{author}{\bibfnamefont{J.~D.} \bibnamefont{Joannopoulos}},
  \bibnamefont{and} \bibinfo{author}{\bibfnamefont{S.~G.}
  \bibnamefont{Johnson}}, \bibinfo{journal}{arXiv:1004.2733}
  (\bibinfo{year}{2010}{\natexlab{b}}).

\bibitem[{\citenamefont{Hoffmann et~al.}(2008)\citenamefont{Hoffmann, Lu,
  Schrinner, Ballauff, and Harnau}}]{hoffmann:2008}
\bibinfo{author}{\bibfnamefont{M.}~\bibnamefont{Hoffmann}},
  \bibinfo{author}{\bibfnamefont{Y.}~\bibnamefont{Lu}},
  \bibinfo{author}{\bibfnamefont{M.}~\bibnamefont{Schrinner}},
  \bibinfo{author}{\bibfnamefont{M.}~\bibnamefont{Ballauff}}, \bibnamefont{and}
  \bibinfo{author}{\bibfnamefont{L.}~\bibnamefont{Harnau}},
  \bibinfo{journal}{J. Phys. Chem. B} \textbf{\bibinfo{Volume}{112}},
  \bibinfo{pages}{14843} (\bibinfo{year}{2008}); \bibinfo{note}{{M.} Hoffmann,
  M. Siebenb{\"u}rger, L. Harnau, M. Hund, C. Hanske, Y. Lu, C.~S. Wagner, M.
  Drechsler, and M. Ballauff, Soft Matter \textbf{6}, 1125 (2010)}.

\bibitem[{\citenamefont{Sprenger et~al.}(2005)\citenamefont{Sprenger,
  Schlesener, and Dietrich}}]{sprenger:2005}
\bibinfo{author}{\bibfnamefont{M.}~\bibnamefont{Sprenger}},
  \bibinfo{author}{\bibfnamefont{F.}~\bibnamefont{Schlesener}},
  \bibnamefont{and} \bibinfo{author}{\bibfnamefont{S.}~\bibnamefont{Dietrich}},
  \bibinfo{journal}{Phys. Rev. E} \textbf{\bibinfo{Volume}{71}},
  \bibinfo{pages}{056125} (\bibinfo{year}{2005}).

\bibitem[{\citenamefont{Dantchev et~al.}(2006)\citenamefont{Dantchev, Diehl,
  and Gruneberg}}]{dantchev:2006}
\bibinfo{author}{\bibfnamefont{D.}~\bibnamefont{Dantchev}},
  \bibinfo{author}{\bibfnamefont{H.~W.} \bibnamefont{Diehl}}, \bibnamefont{and}
  \bibinfo{author}{\bibfnamefont{D.}~\bibnamefont{Gruneberg}},
  \bibinfo{journal}{Phys. Rev. E} \textbf{\bibinfo{Volume}{73}},
  \bibinfo{pages}{016131} (\bibinfo{year}{2006}).

\bibitem[{\citenamefont{Dantchev et~al.}(2007)\citenamefont{Dantchev,
  Schlesener, and Dietrich}}]{dantchev:2007}
\bibinfo{author}{\bibfnamefont{D.}~\bibnamefont{Dantchev}},
  \bibinfo{author}{\bibfnamefont{F.}~\bibnamefont{Schlesener}},
  \bibnamefont{and} \bibinfo{author}{\bibfnamefont{S.}~\bibnamefont{Dietrich}},
  \bibinfo{journal}{Phys. Rev. E} \textbf{\bibinfo{Volume}{76}},
  \bibinfo{pages}{011121} (\bibinfo{year}{2007}).

\bibitem[{\citenamefont{Macio{\l}ek et~al.}(2007)\citenamefont{Macio{\l}ek,
  Gambassi, and Dietrich}}]{maciolek:2007}
\bibinfo{author}{\bibfnamefont{A.}~\bibnamefont{Macio{\l}ek}},
  \bibinfo{author}{\bibfnamefont{A.}~\bibnamefont{Gambassi}}, \bibnamefont{and}
  \bibinfo{author}{\bibfnamefont{S.}~\bibnamefont{Dietrich}},
  \bibinfo{journal}{Phys. Rev. E} \textbf{\bibinfo{Volume}{76}},
  \bibinfo{pages}{031124} (\bibinfo{year}{2007}).

\bibitem{note:1}
  {See Eq.~$7.8.3$ on p.~$168$ in \textit{Tables of indefinite integrals},
edited by Y.~A. Brychkov, O.~I. Marichev, and A.~P. Prudnikov (Gordon and Breach, New York, 1989),
with the substitution $z\mapsto x=1/\sqrt{z}$.
  }

\bibitem{note:2}
  {See Eq.~(4) of Tab.~$234$ in \textit{Nouvelles tables d'int\'egrales d\'efinies}, 
edited by D.~B. De Haan (P. Engels, Leide, 1867);
note that there is a misprint in Eq.~$4.521.8$ in \textit{Table of Integrals, 
Series, and Products}, Sixth edition, edited by I.~S.~Gradshteyn and I.~M.~Ryzhik (Academic, London, 2000).
The correct expression is $\int_0^1\upd x\; x\big(\arccos(x)\big)/(1+qx^2)^2=\pi(\sqrt{1+q}-1)/(4q\sqrt{1+q})$ for $q>-1$.
}

\bibitem{note:3}
  {See Eq.~(26) on p.~$136$ in \textit{Tables of Integrals Transforms}, Vol. I, Bateman Manuscript
Project, edited by H. Erdelyi (McGraw-Hill, New York, 1954).
}

\end{thebibliography}
\end{document}